\documentclass[twocolumn]{aastex631} %
\usepackage{graphicx}
\usepackage[version=4]{mhchem}
\usepackage{mathtools} 
\usepackage{comment}
\usepackage{txfonts}
\NewDocumentCommand{\evalat}{sO{\big}mm}{%
  \IfBooleanTF{#1}
   {\mleft. #3 \mright|_{#4}}
   {#3#2|_{#4}}%
}

\begin{document}
\title{Global Chemical Transport on Hot Jupiters:\\Insights from 2D VULCAN photochemical model}

\author[0000-0002-8163-4608]{Shang-Min Tsai}
\affiliation{Department of Earth and Planetary Sciences, University of California, Riverside, CA, USA}
\author[0000-0001-9521-6258]{Vivien Parmentier}
\affiliation{Universit\'e C\^ote d'Azur, Observatoire de la C\^ote d'Azur, CNRS, Laboratoire Lagrange, Nice, France}
\author[0000-0002-6907-4476]{Jo\~{a}o M. Mendon\c{c}a}
\affiliation{National Space Institute, Technical University of Denmark, Elektrovej 328, DK-2800 Kgs. Lyngby, Denmark}
\author[0000-0003-2278-6932]{Xianyu Tan}
\affiliation{Tsung-Dao Lee Institute, Shanghai Jiao Tong University, 520 Shengrong Road, Shanghai, People’s Republic of China}
\affiliation{School of Physics and Astronomy, Shanghai Jiao Tong University, 800 Dongchuan Road, Shanghai, People’s Republic of China}
\author[0000-0001-9423-8121]{Russell Deitrick}
\affiliation{School of Earth and Ocean Sciences, University of Victoria, Victoria, British Columbia, Canada}
\author[0000-0002-6893-522X]{Mark Hammond}
\affiliation{Atmospheric, Oceanic and Planetary Physics, Department of Physics, University of Oxford, UK}
\author[0000-0002-2454-768X]{Arjun B. Savel}
\affiliation{Center for Computational Astrophysics, Flatiron Institute, New York, USA}
\affiliation{Astronomy Department, University of Maryland, College Park, 4296 Stadium Dr., College Park, USA}
\author[0000-0002-8706-6963]{Xi Zhang}
\affiliation{Department of Earth and Planetary Sciences, University of California Santa Cruz, Santa Cruz, CA, USA}
\author[0000-0002-5887-1197]{Raymond T. Pierrehumbert}
\affiliation{Atmospheric, Oceanic and Planetary Physics, Department of Physics, University of Oxford, UK}   
\author[0000-0002-2949-2163]{Edward W. Schwieterman}  
\affiliation{Department of Earth and Planetary Sciences, University of California, Riverside, CA, USA}



\begin{abstract}
The atmospheric dynamics of tidally-locked hot Jupiters is characterized by strong equatorial winds. Understanding the interaction between global circulation and chemistry is crucial in atmospheric studies and interpreting observations. Two-dimensional (2D) photochemical transport models shed light on how the atmospheric composition depends on circulation. In this paper, we introduce the 2D photochemical {\it (horizontal and vertical)} transport model, VULCAN 2D, which improves on the pseudo-2D approaches by allowing for non-uniform zonal winds. We extensively validate our VULCAN 2D with analytical solutions and benchmark comparisons. Applications to HD 189733 b and HD 209458 b reveal a transition in mixing regimes: horizontal transport predominates below $\sim$0.1 mbar while vertical mixing is more important at higher altitudes above 0.1 mbar. Motivated by the previously inferred carbon-rich atmosphere, we find that HD 209458 b with super-solar carbon-to-oxygen ratio (C/O) exhibits pronounced \ce{C2H4} absorption on the morning limb but not on the evening limb, owing to horizontal transport from the nightside. We discuss when a pseudo-2D approach is a valid assumption and its inherent limitations. Finally, we demonstrate the effect of horizontal transport in transmission observations and its impact on the morning-evening limb asymmetry with synthetic spectra, highlighting the need to consider global transport when interpreting exoplanet atmospheres.


\end{abstract}
   


%

\section{Introduction}
We are entering the era of detailed exoplanet atmosphere characterization. The atmospheric characterization has come a long way since the first transit observation with the Hubble Space Telescope (HST) \citep{Charbonneau2002}. Recent JWST spectral measurements revealed unprecedented details of gas giants \citep[e.g.,][]{ERF2023,Rustamkulov2023,Bell2023}. Transmission spectroscopy is central for probing atmospheric composition. To improve our ability to interpret observations through theoretical models, it is important to consider how the temperature and composition at the terminators and various regions are shaped by global circulation. This aspect cannot be addressed by traditional 1D models that omit thermal and chemical variations and transport processes in the horizontal direction.

Since transmission observations probe terminator regions that are composed of opposite sides of the planet, recent works have highlighted the need to account for spatial inhomogeneities. \cite{Espinoza2021} and \cite{Grant2023} introduced methods to separate the atmospheric components from the morning (leading) and evening (trailing) limbs. The advancement of high-resolution spectroscopy also expands our capacity to resolve the climate \citep[e.g., ][]{Louden2015} and chemical variations \citep[e.g., ][]{Ehrenreich2020,Wardenier2021} across the planet. Atmospheric retrieval frameworks are beginning to extend to 2D and 3D to prevent biases from fitting the data with 1D profiles \citep{MacDonald2020,Chubb2022,Pluriel2022}. As we gain more complete observations from transit, eclipse, and phase curves, we will soon have the ability to probe the composition distributions across the planet. It is essential to have models resolving global variations to support the progress of observations.


The atmospheric composition of gas giants is generally governed by thermochemistry, photochemistry, and mixing processes. The merit of 1D models is they can incorporate all the above processes as needed. However, the 1D structure intrinsically neglects the variations in 3D, and more importantly, how the global circulation impacts the local properties \citep{Mendonca2018,Drummond2020,Zamyatina2023}. 3D general circulation models (GCMs) can capture the intricate dynamic interactions across the planet and have become commonly applied to interpret observations \citep[e.g., ][]{Beltz2021,Kempton2023}. However, simplifications of the physical and chemical processes are usually required due to computation limitations. 
In addition to reducing the full Navier–Stokes dynamics into primitive equations \citep{Mayne2019}, 3D GCMs often make further simplifications to radiative transfer and chemical processes. The chemical processes in particular are often severely simplified in most applications, using the assumption of thermochemical equilibrium, which is valid only at high temperatures and pressures. Modeling efforts have been put into implementing reduced chemical schemes in a GCM to account for transport-induced disequilibrium \citep{Drummond2020,Tsai2022,Lee2023}. While Earth climate models have provided insights into the oxygen response of Earth-like atmospheres \citep{Chen2021,Cooke2022,Braam2023,Deitrick2023,Ji2023}, to date, the incorporation of photochemistry into 3D models for diverse non-Earth-like atmospheres has not been realized yet.

For most tidally-locked giant exoplanets, the stationary day-night heating drives large-scale equatorial waves. These waves interact with background flows and transport momentum from higher latitudes toward the equator, leading to an equatorial superrotating jet \citep{Showman2011,Tsai2014,Showman2020}. The strong jet plays a central role in global circulation, facilitating the transport of heat, aerosols, and gaseous species \citep{Komacek2016,Hammond2021,Steinrueck2021,Drummond2020}. For these planets, ``pseudo-2D" photochemical models \citep{Agundez2014,Venot2020,Baeyens2022,Moses2022} that employ a rotating 1D column to mimic a uniform jet in a Lagrangian frame have emerged as useful complementary tools. Their relatively fast computation enables them to incorporate the same detailed mechanisms as 1D models, but the simulated circulation is limited to uniform winds\footnote{Strictly speaking, it is the angular velocity held constant in a rotating 1D model. The top layer of the equivalent flow moves faster than the bottom by $H_{atm} \omega$, where $H_{atm}$ is the atmospheric depth in the model and $\omega$ is the assumed rotational frequency applied to simulate the equatorial jet.}. On the other hand, 2D models implemented with horizontal diffusion have been used to model the meridional plane for solar system gas giants \citep{Liang2005,Zhang2013,Hue2018} but have not yet been applied to exoplanets. In this work, we present the 2D (horizontal and vertical) photochemical-transport model, VULCAN 2D, which lifts this restriction on wind patterns while sharing the same advantage as pseudo-2D models. While VULCAN 2D is constructed with horizontal and vertical dimensions in general, here we will focus on the zonal and vertical directions.
   
In Section \ref{method}, we describe the construction of the VULCAN 2D model. In Section \ref{validation}, we validate the numerical scheme and modeling results through comparisons to an analytical solution, a pseudo-2D approach, and a 3D GCM, respectively. In Section \ref{results}, we delve into the applications to canonical hot Jupiters, HD 189733 b and HD 209458 b, including a super-solar C/O scenario for HD 209458 b. We use limiting cases to demonstrate the roles of horizontal and vertical transport and draw comparisons with previous works \citep{Agundez2014}. We then compare VULCAN 2D to pseudo-2D models in Section \ref{sec:pseudo2D}, exploring when the uniform wind assumption in pseudo-2D models may no longer hold. Finally, we discuss the observational implications and the morning-evening limb asymmetry arising from horizontal transport in Section \ref{sec:spectra}.   

\section{Configuration of the 2D chemical transport model}\label{method}
\subsection{The 2D Grid}
\begin{figure}
   \centering
   \includegraphics[width=0.4\columnwidth]{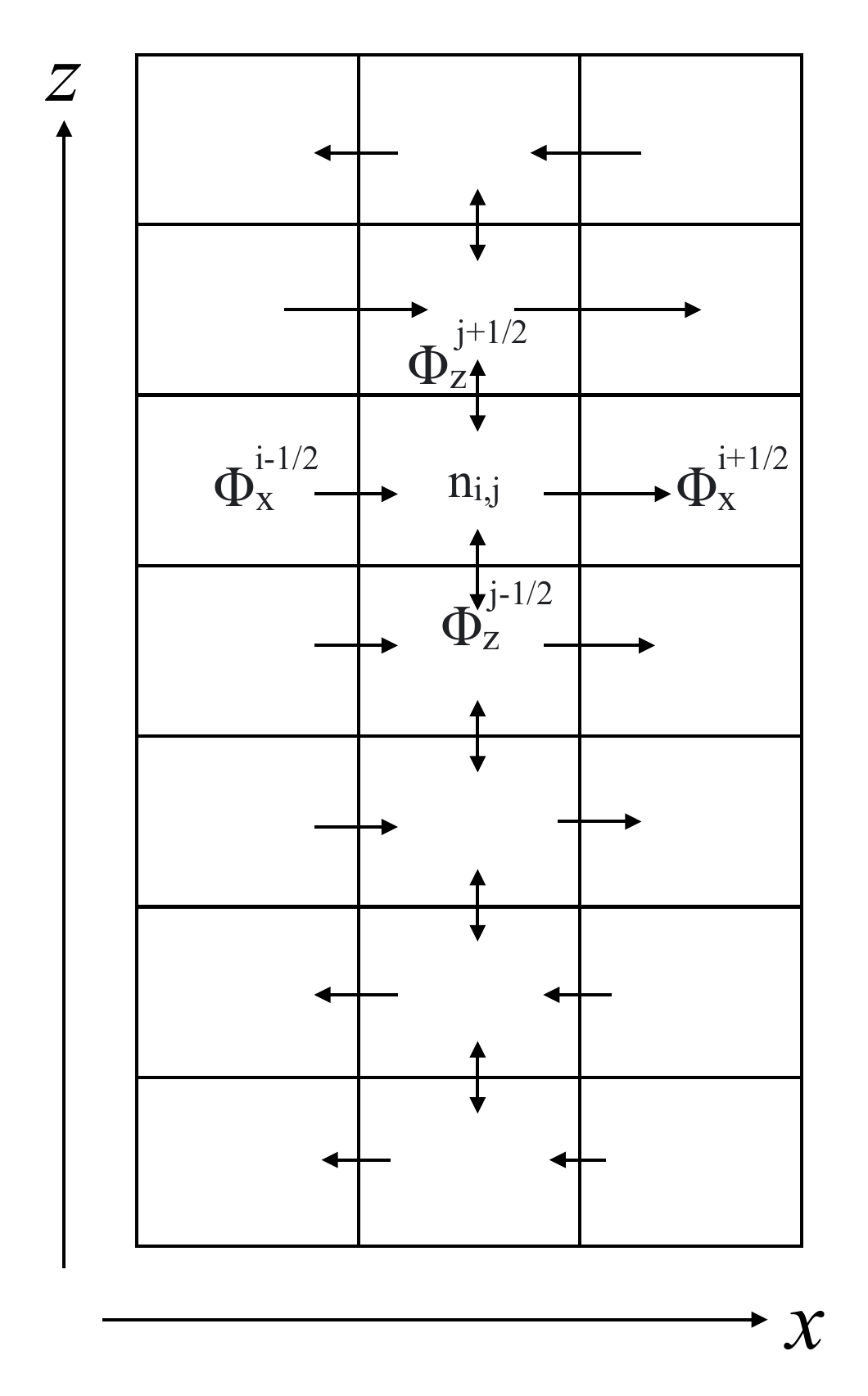}
   \includegraphics[width=0.55\columnwidth]{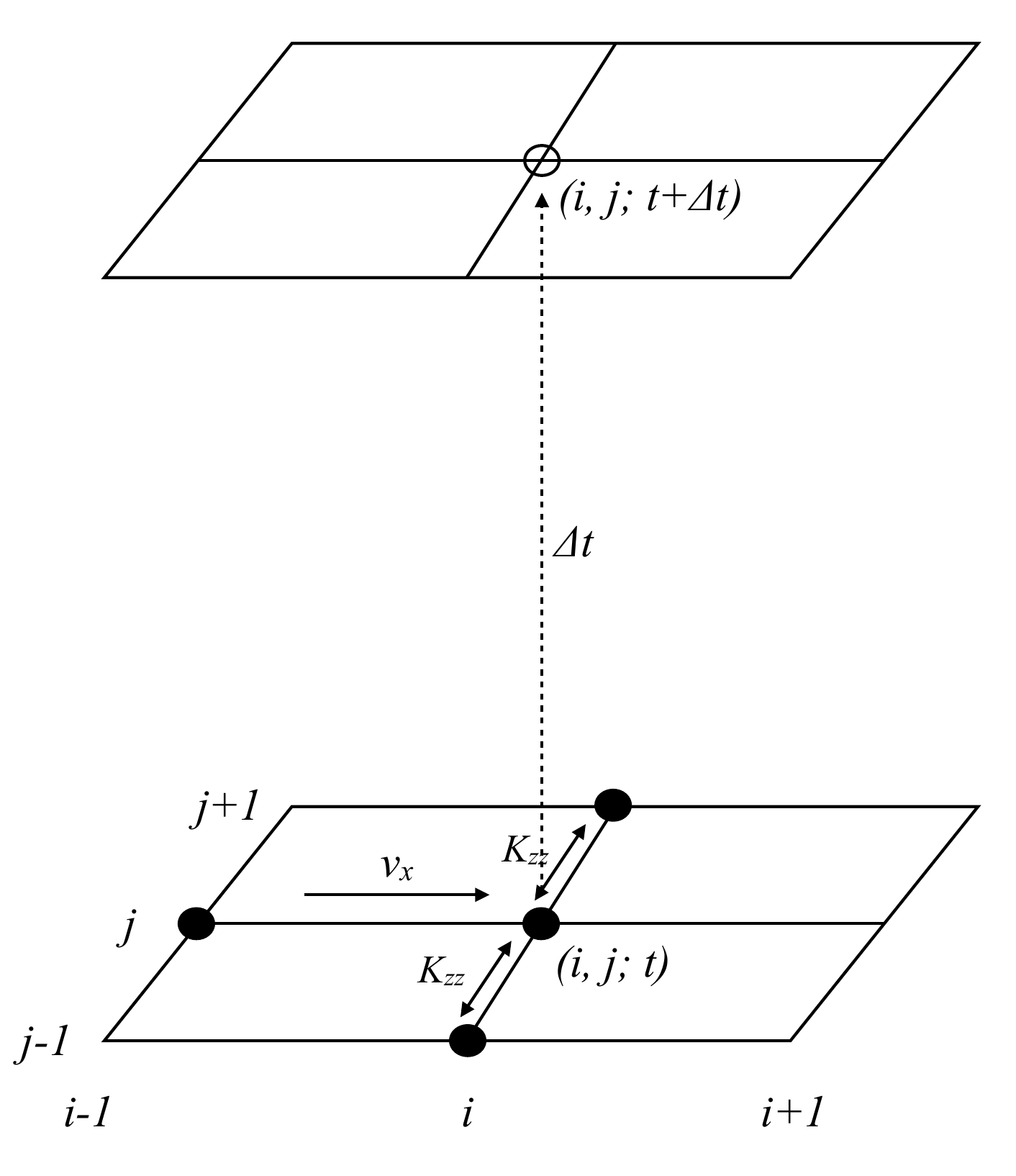}
\caption{The grid configuration of VULCAN 2D (left) and the stencil of the upwind scheme
for horizontal ($i$) advection and vertical ($j$) diffusion (right). The arrows illustrate the advection and diffusion transport between adjacent grids.}
\label{fig:grid_schematic} 
\end{figure}

A standard 1D photochemical kinetics model solves the continuity equation in the form of a set of coupled partial differential equations:
\begin{equation}
\frac{\partial n (z, t)}{\partial t} = {\cal P} - {\cal L} - \frac{\partial \phi}{\partial z},
\label{eq:1D}
\end{equation}
where $n$ is the number density (cm$^{-3}$) of each species and $t$ denotes the time. ${\cal P}$ and ${\cal L}$ are the chemical production and loss rates (cm$^{-3}$ s$^{-1}$) of the corresponding species at each vertical layer \citep{Tsai2017}. Equation (\ref{eq:1D}) can be readily generalized to a 2D Cartesian space ($x$, $z$) to include horizontal transport:
\begin{equation}
\frac{\partial n (x, z, t)}{\partial t} = {\cal P} - {\cal L} - \frac{\partial \phi_{z}}{\partial z} - \evalat[\bigg]{\frac{\partial \phi_{x}}{\partial x}}{P}
\label{eq:2D-master} 
\end{equation}
where $\phi_{z}$, $\phi_{x}$ are the vertical ($z$) and horizontal ($x$) transport flux, respectively. In general, $\phi_{z}$ encompasses vertical advection, eddy diffusion, and molecular diffusion \citep{Tsai2021}, whereas $\phi_{x}$ describes horizontal transport on isobaric surfaces. Using isobaric flux is motivated by the fact that previous analysis of 3D GCMs is traditionally done in isobaric coordinates and winds on the isobaric levels typically dominate heat and chemical transport on tidally-locked planets \citep[e.g.][]{Koll2018,Showman2020,Hammond2021}.
We note that we use log-pressure coordinates with the vertical coordinate defined as $z = - H \; \textrm{ln}(p/P_s)$, where $H$ is the pressure scale height that depends on the local temperature, mean molecular weight, and altitude-dependent gravity and $P_s$ is the reference pressure. While each vertical column uses the same pressure grid spaced uniformly in log-space, the corresponding $z$ differs between columns due to the horizontal temperature gradient. Therefore, the horizontal derivative is evaluated at the {\it same pressure level but not the same geometric height}.
The configuration of our 2D grid is illustrated in Figure \ref{fig:grid_schematic}.

The real challenge lies in numerically solving Equation (\ref{eq:2D-master}). Most of the numerical methods for stiff equations require evaluating the Jacobian matrix \citep{Brasseur2017}. In a 1D system, the Jacobian matrix is neatly constructed through nested looping each species within each vertical layer (See Figure 14. in \cite{Tsai2017}). For example, the Rosenbrock method with a band matrix solver is employed in VULCAN. However, this structure breaks down when extending to 2D. To take advantage of the established numerical solver built for 1D systems, we apply an asynchronous integrator within each timestep \citep{Small2013}. Specifically, the right-hand side of Equation (\ref{eq:2D-master}) is first discretized in $x$ as 
\begin{equation}\label{eq:discret}
\frac{d n}{d t} = {\cal P}- {\cal L} - \frac{\phi_z^{i,j+1/2} - \phi_z^{i,j-1/2}}{\Delta z} -  \frac{\phi_x^{i+1/2,j} - \phi_x^{i-1/2,j}}{\Delta x}    
\end{equation}
where $i$, $j$ denote the horizontal, vertical indices, respectively, and the +1/2 and -1/2 represent the interfaces enclosing that layer. Equation (\ref{eq:discret}) is evaluated for each species and each vertical column. Since the last term (i.e. the horizontal transport flux) on the right-hand side of Equation (\ref{eq:discret}) only contributes to the diagonal elements in the Jacobian matrix, Equation (\ref{eq:discret}) has the same numerical structure as that in a 1D system (Equation (5) in \cite{Tsai2017}). This allows us to evaluate each column, including horizontal transport, using the existing 1D solver. Specifically, Equation (\ref{eq:discret}) is computed for each $x$ column within each timestep. Although errors can arise from the asynchronous update of horizontal transport flux associated with each column, we have tested integrating each column in different orders and found the errors to be negligible.


For the horizontal advection, we adopt a first-order upwind difference scheme, where the local concentration is affected by the upwind cell only (same as the vertical advection described in \cite{Tsai2021}). The advective flux for the $k$ cell in the $x$ direction is 
\begin{equation}
\begin{aligned}
\phi_{x-1/2} = &\begin{cases}
v_{k-1/2} n_{k-1}, \text{for } v_{k-1/2} > 0 \\ 
v_{k-1/2} n_{k}, \text{for } v_{k-1/2} < 0  
\end{cases}\\
&\text{and}\\ 
\phi_{x+1/2} = &\begin{cases}
-v_{k+1/2} n_{k}, \text{for } v_{k+1/2} > 0 \\ 
-v_{k+1/2} n_{k+1}, \text{for } v_{k+1/2} < 0  
\end{cases}
\end{aligned}
\label{eq:phix}
\end{equation} 
where $v_{k-1/2}$ and $v_{k+1/2}$ are the zonal wind velocity at the left and right interfaces of the grid cell $k$, respectively. In addition to advective transport, horizontal diffusion contributes to meridional transport on Jupiter and Saturn \citep{Zhang2013,Hue2015,Hue2018} and is also implemented in VULCAN 2D. However, as horizontal winds can be directly obtained from GCM output, we focus only on advection for horizontal transport. For vertical transport, although the numerical functions include vertical advection, we restrict our model to vertical diffusion in this study and do not consider vertical advection to ensure numerical stability. We leave the full exploration of vertical advection to future work.

The advantage of our 2D spatial grid is it accommodates non-uniform two-dimensional flow patterns, such as the mean meridional or zonal circulation derived from the 3D GCM. In contrast, for pseudo-2D models that employ a Lagrangian rotating 1D column \citep{Agundez2014}, the horizontal transport is restricted to a uniform jet by design. Both our 2D model and the pseudo-2D model share a limitation known as the Courant–Friedrichs–Lewy (CFL) condition \citep{Courant1928}, as already pointed out by \cite{Zhang2013}. Given that we explicitly solve for each vertical column, the integration step size must be constrained by the time it takes for the horizontal flow to travel to neighbouring grid cells. This constraint becomes more stringent as the number of horizontal columns increases. In practice, to reduce the overall integration time to achieve convergence, we initially run our 2D model without horizontal transport (thereby removing the stepsize restriction) to attain the 1D steady state. We then run the full 2D VULCAN from this steady state (achieved without horizontal transport) with the stepsize limited by the CFL condition: $dt < min( \frac{dx}{v_x} )$ until the final steady state. Current VULCAN 2D evaluates each vertical column in sequence. Therefore, the simulation runtime scales with the number of columns linearly. However, there are opportunities to enhance code efficiency with parallel integration across the horizontal domain. Future development of VULCAN 2D will explore multiprocessing to speed up column-based computations.



\begin{figure}[ht!]
   \centering
   \includegraphics[width=\columnwidth]{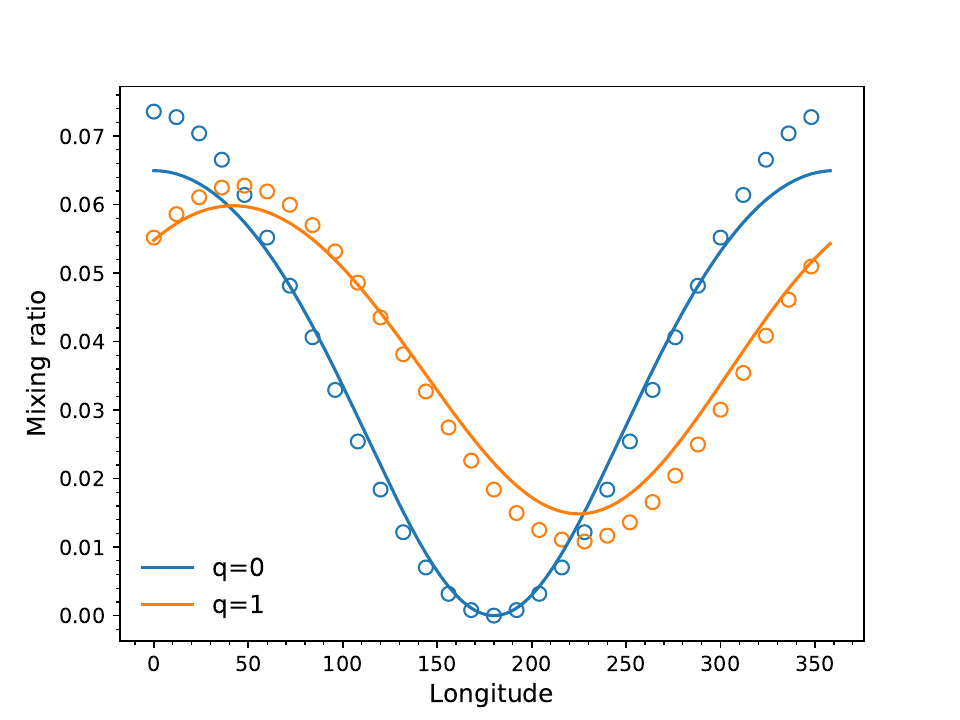}
\caption{Comparison of VULCAN 2D (solid curve) to analytical solutions (open circles), showing the volume mixing ratio at $\xi$ = 1 with q = 0 (no zonal transport) and q = 1 (eastward zonal transport). The q = 1 case represents zonal transport on a timescale comparable to the chemical timescale.}
\label{fig:ana}
\end{figure}

 \begin{figure*}[ht!]
   \centering
\includegraphics[width=\columnwidth]{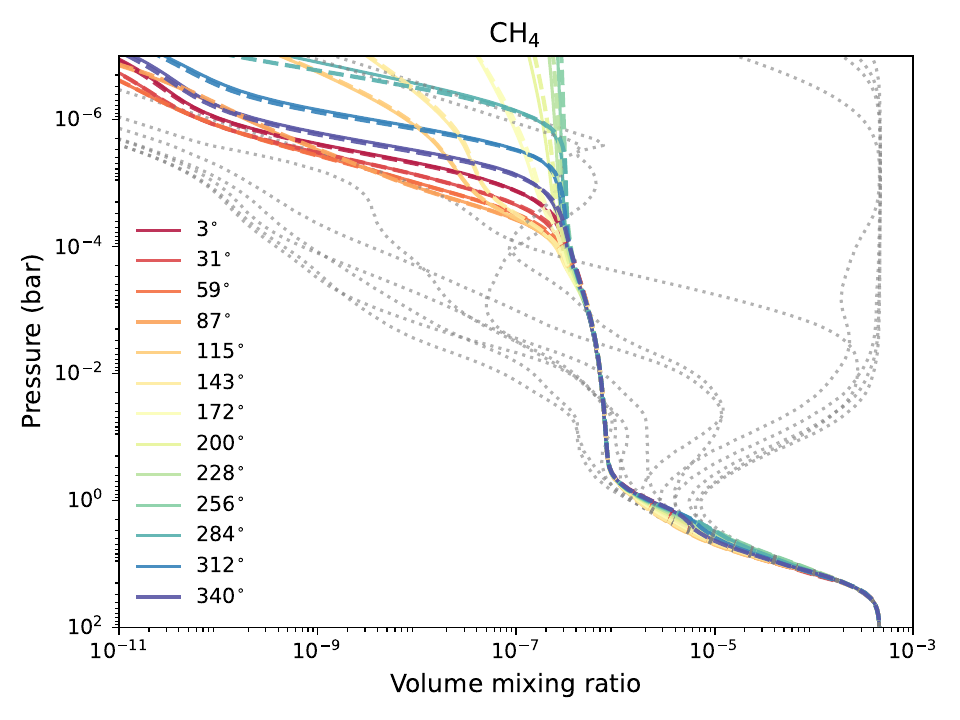}
\includegraphics[width=\columnwidth]{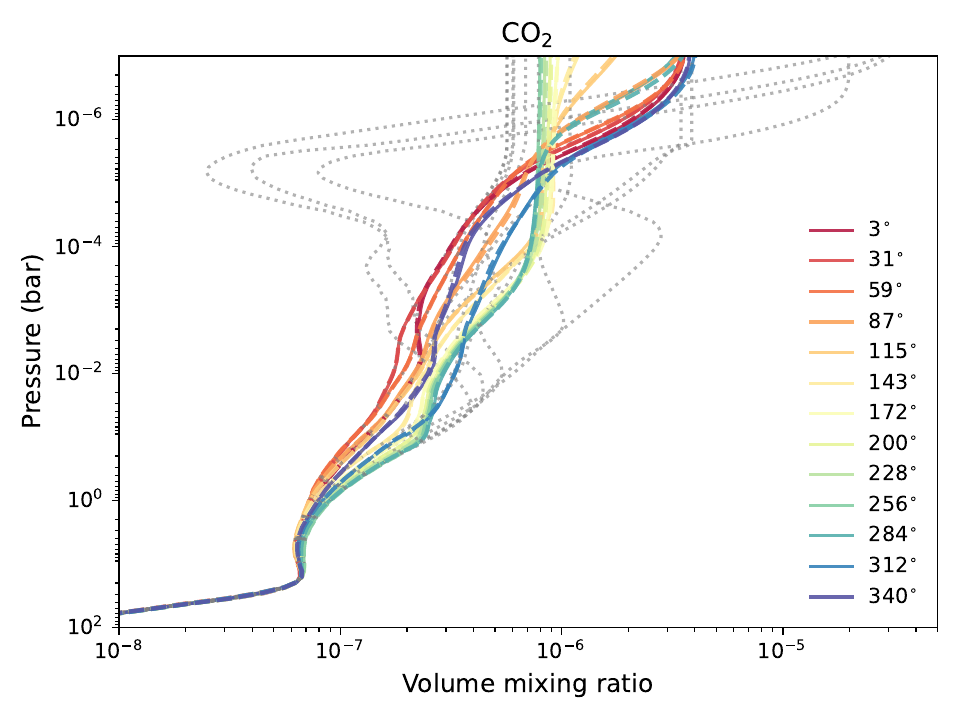}
\includegraphics[width=\columnwidth]{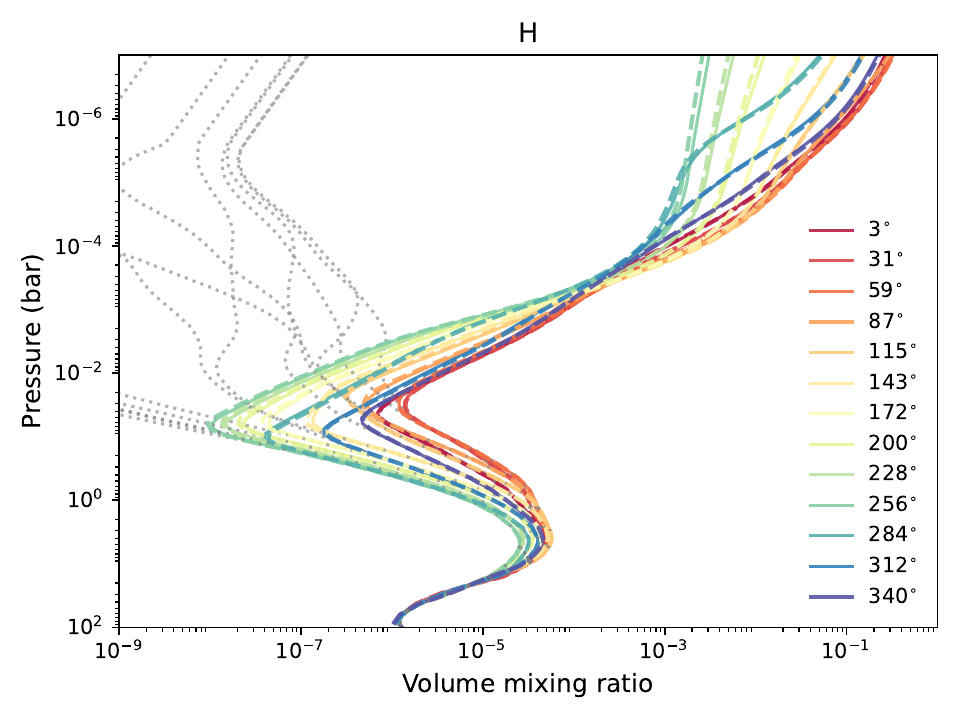}
\includegraphics[width=\columnwidth]{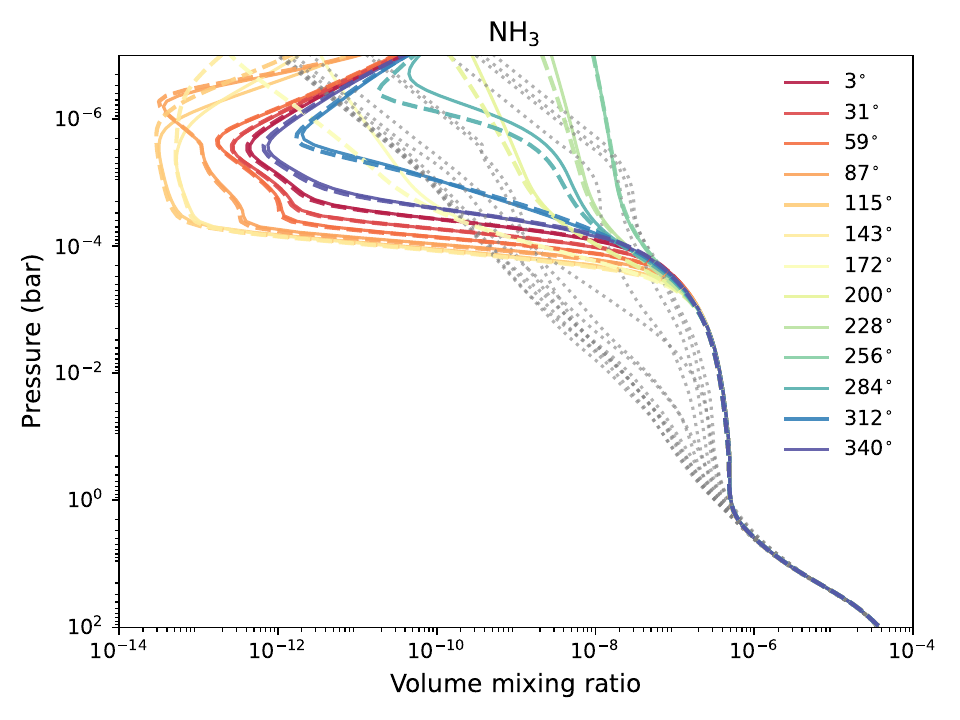}
\caption{Mixing ratio profiles at various longitudes of HD 189733 b using Lagrangian-1D VULCAN (dashed) compared to those using 2D VULCAN with equivalent zonal winds (solid; see text for details). The substellar point is at 0$^{\circ}$ longitude. Lagrangian-1D VULCAN has the same configuration as the pseudo-2D model \citep[e.g.,][]{Agundez2014,Baeyens2021}. Equilibrium abundances are indicated by the grey lines.} 
\label{fig:psuedo2D}
\end{figure*}

\section{Validation and comparison to previous work}\label{validation}
\subsection{Validation with analytical solutions}

\cite{Zhang2013} derived an analytical solution for a 2D advective-diffusive system with parameterized chemical sources and sinks. We apply their analytical solution for a 2D zonal plane to validate our numerical model. To briefly recap, Equation (13) in \cite{Zhang2013} describes the steady-state abundance as a function of longitude and altitude, governed by vertical diffusion, horizontal advection, and chemical sources and sinks. We assume vertically uniform diffusion to simplify the expression, i.e., with $\alpha$ = $\gamma$ = 0, the analytical solution for the volume mixing ratio under day-night sinusoidal chemical production ($k$ = 1) is
\begin{equation}    
\chi(\lambda, \xi) =  \frac{P_0}{L_0} \left[ 1 + \frac{1}{\sqrt{1+q^2}} \mbox{cos}(\lambda - \phi) \right] e^{-\xi}
\end{equation} 
where $\chi$ is the volume mixing ratio, $\lambda$ is longitude, $\phi$ is the phase shift, $\xi$ is the vertical coordinate, $P_0$ and $L_0$ are the production and loss rates, respectively, and q relates the ratio of advection to chemical timescales, following the same notation as \cite{Zhang2013}.

To compare with the analytical solution, we implemented a mock chemical network with only A and B as reactive species. The rate constant of \ce{A -> B} is given by $L_0N_0 e^{-\xi}$/[A] and that of \ce{B -> A} is given by $P_0N_0 (1 + \mbox{cos}k\lambda)$/[B], corresponding to the loss and production prescriptions in Section 5 of \cite{Zhang2013}. We assume constant zonal winds of 71.5 m/s and $R_p$ = $R_J$, together with $P_0N_0$ = 2 $\times$ 10$^{-7}$ cm$^{-3}$ s$^{-1}$ and $L_0N_0$ = 2 $\times$ 10$^{-7}$ cm$^{-3}$ s$^{-1}$ to yield $q$ = 1. The vertical diffusion is set to 10$^8$ cm$^{-2}$ s$^{-1}$ but the horizontal distribution does not directly depend on this value. 

Figure \ref{fig:ana} compares our 2D numerical model to the analytical solution. The impact of the horizontal wind is twofold: it both transports and homogenizes the chemical gradient. While our numerical results show somewhat lower peak amplitudes than the analytical solution, which can likely be attributed to numerical diffusion, they effectively capture the overall distribution shape and correctly reproduce the ``phase shift" due to eastward transport.

\subsection{Comparison with the Pseudo 2D approach \citep{Agundez2014}}
\cite{Agundez2014} apply a rotating 1D model to represent an air column moving with a uniform zonal jet in a Lagrangian frame. Their chemical kinetics scheme was based on \cite{Venot2012}. For a like-for-like comparison, we adopt the same rotating 1D-column as \cite{Agundez2014} (referred to as Lagrangian-1D VULCAN) and compare it against our 2D photochemical model (2D VULCAN) using the identical N-C-H-O chemical scheme in VULCAN. We chose the canonical hot Jupiter HD 189733 b for this comparison, employing the same GCM output as that in \cite{Agundez2014}. In Lagrangian-1D VULCAN, the 1D column is set to travel around the equator over 2.435 days, following \cite{Agundez2014}, i.e. with a constant angular velocity of 1.493$\times$10$^{-5}$ rad/s. To translate this rotation into the equivalent zonal-mean wind, we use a quasi-uniform zonal wind (faster at higher altitudes) in 2D VULCAN to match the same angular velocity. The zonal wind is 2430 m/s at 1 bar and scaled to each altitude according to the geometry. The equatorial plane is divided into 64 columns in longitude for both 2D VULCAN and Lagrangian-1D VULCAN. We used the same eddy diffusion coefficient ($K_{\textrm{zz}}$) for vertical mixing as \cite{Agundez2014}: $K_{\textrm{zz}}$ = 10$^7$ $\times$ ($P$/ 1 bar)$^{-0.65}$ cm$^2$s$^{-1}$. Other physical and chemical parameters were kept as much the same as possible between Lagrangian-1D and 2D VULCAN models. 

The comparisons between our Lagrangian-1D and 2D VULCAN for species displaying apparent longitudinal gradients are shown in Figure \ref{fig:psuedo2D}. 2D VULCAN and Lagrangian-1D VULCAN exhibit consistent results for species with either short (e.g. H) or long (e.g. \ce{CH4}) chemical timescales. This indicates that 2D VULCAN correctly captures the processes of vertical mixing and horizontal transport. We find minor deviations that appeared in trace species (volume mixing ratios $\lesssim$ 10$^{-6}$). For instance, \ce{NH3} exhibits the most notable differences above 10$^{-4}$ bar. The discrepancy is likely due to the numerical integration: the first-order upwind in 2D VULCAN has stronger numerical diffusion than the Rosenberg method with second-order convergence in time used in Lagrangian-1D VULCAN. Nevertheless, the agreements are generally within a factor of two in the region below 0.1 mbar level. Our 2D VULCAN, with the equivalent uniform wind, can successfully reproduce the results obtained by the Lagrangian-1D approach.

\begin{figure}[ht!]
   \centering
   \includegraphics[width=\columnwidth]{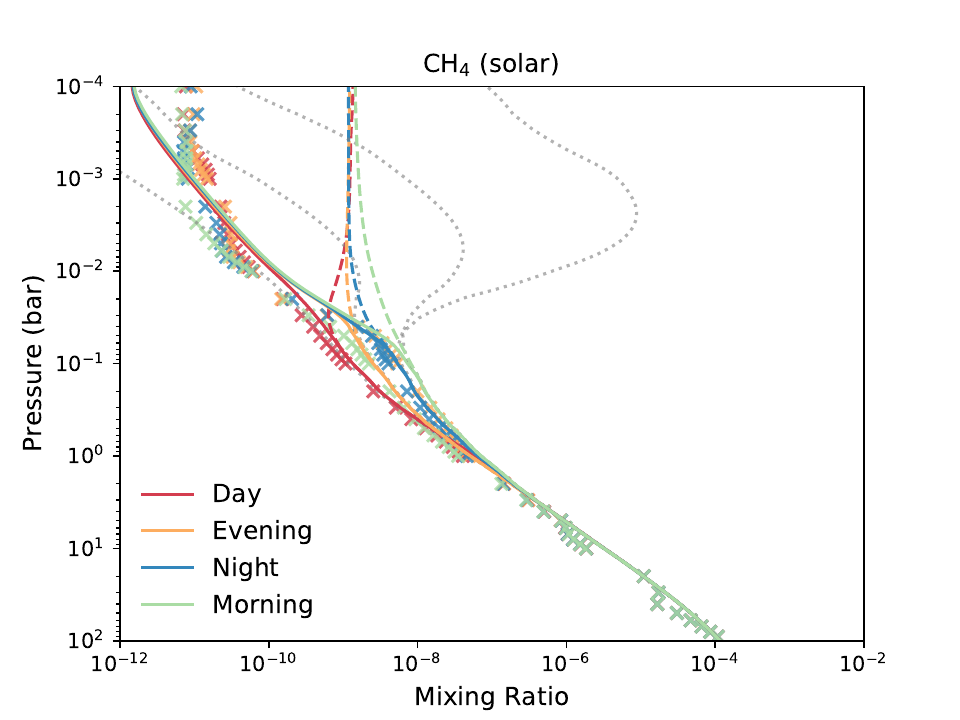}
   \includegraphics[width=\columnwidth]{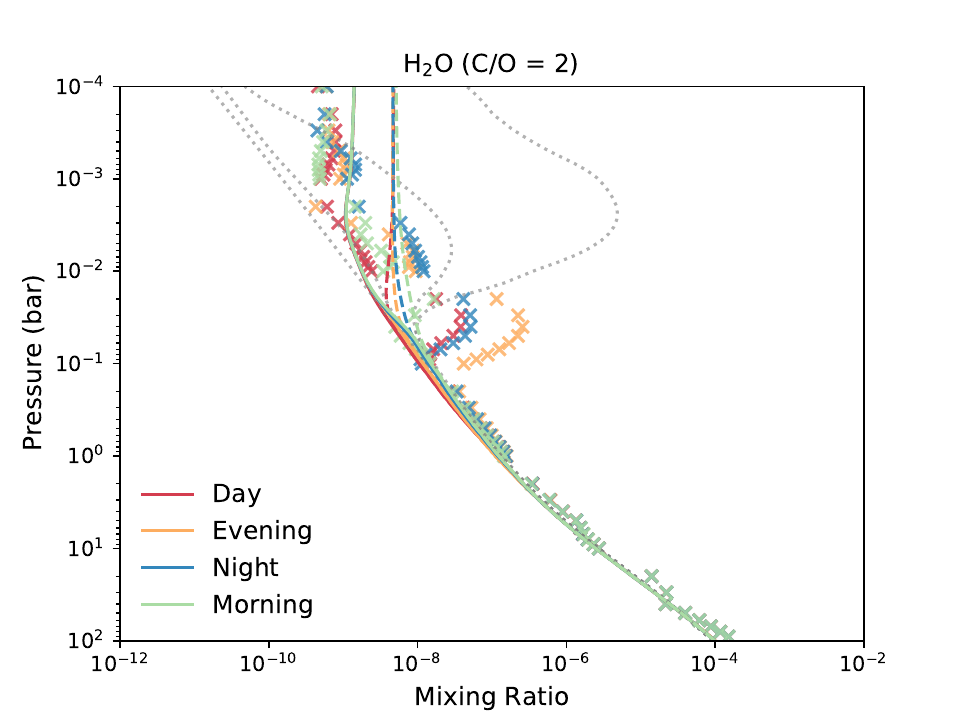}
\caption{\ce{CH4} (top) and \ce{H2O} (bottom) abundances computed by VULCAN 2D for WASP-43 b compared to those in \cite{Mendonca2018a} (crosses), showing the averaged mixing-ratio profiles with , eddy diffusion $K_{\textrm{zz}} = 0.01 \times w_{rms} H$ (solid) and $K_{\textrm{zz}} = w_{rms} H$ (dashed) on the dayside (315$^{\circ}$ -- 45$^{\circ}$), morning limb (225$^{\circ}$ -- 315$^{\circ}$), nightside (135$^{\circ}$ -- 225$^{\circ}$), and evening limb (45$^{\circ}$ -- 135$^{\circ}$). We present the cases that exhibit the most compositional gradient: \ce{CH4} from the solar composition simulation and \ce{H2O} from the C/O = 2 simulation. The resulting distribution profiles with  $K_{\textrm{zz}} = 0.01 \times w_{rms} H$ better represent those from the 3D GCM.}
\label{fig:relax}
\end{figure}

\subsection{Comparison with 3D transport (without photochemistry)} %
We have validated VULCAN 2D through an analytical solution and a pseudo-2D approach, to ensure the accuracy of our physical and numerical implementation within the 2D framework. In our next step, we will perform additional comparisons with a 3D GCM to see whether our 2D model effectively captures the primary transport process. 

 \cite{Mendonca2018a} previously investigated the global transport of chemically active tracers: \ce{H2O}, \ce{CH4}, CO, and \ce{CO2} in the 3D simulation of WASP-43b. The chemical relaxation scheme \citep{Tsai2018} implemented in \cite{Mendonca2018a} simplifies the thermochemical kinetics with a linear response according to the chemical timescale. We compare with the work by \cite{Mendonca2018a} because their chemical timescales are derived from the same chemical network implemented in VULCAN. Here, photochemistry in 2D VULCAN is switched off to have an equivalent comparison. The equatorial region ( $\pm$ 20$^{\circ}$) is divided into four quadrants in longitude: dayside (315$^{\circ}$ -- 45$^{\circ}$), morning limb (225$^{\circ}$ -- 315$^{\circ}$), nightside (135$^{\circ}$ -- 225$^{\circ}$), and evening limb (45$^{\circ}$ -- 135$^{\circ}$), following Figure 4. in \cite{Mendonca2018a}. We adopt the average temperature and zonal wind field in this equatorial region from the same GCM output in \cite{Mendonca2018,Mendonca2018a}. The eddy diffusion coefficients are often estimated from mixing length theory, using root-mean-square vertical velocity ($w_{rms}$) as the turbulent velocity and the scale height ($H$) as the mixing length. However, \cite{Parmentier2013} demonstrated that this approach tends to overestimate vertical mixing efficiency for hot Jupiters. Given this likely overestimation, we make two assumptions: $K_{\textrm{zz}} = w_{rms} H$ and $K_{\textrm{zz}} = 0.01 \times w_{rms} H$ to bracket the plausible range of $K_{\textrm{zz}}$. The average temperature from the GCM is used to compute the scale height $H = \frac{k_B T}{m g}$ ($k_B$: Boltzmann constant; g: altitude-dependent gravitational acceleration), but we fix the mean molecular weight $m$ to 2.2387, following the same setting in \cite{Mendonca2018,Mendonca2018a}.
 
Figure \ref{fig:relax} compares the abundance distribution of \ce{CH4} (for solar composition) and \ce{H2O} (for C/O = 2) in the four quadrants of the equatorial region computed by VULCAN 2D and \cite{Mendonca2018}. Our 2D model with $K_{\textrm{zz}} = 0.01 \times w_{rms} H$ predicts vertical quench levels close to those in the 3D results. The preference for a weaker $K_{\textrm{zz}}$ might be attributed to the high gravity of WASP-43b. The slight increase of \ce{H2O} between 0.1 and 0.01 in the 3D GCM is likely associated with meridional transport from higher latitude \citep{Drummond2018} not accounted for in the 2D model. The overall good agreement between VULCAN 2D and the 3D results reinforces our 2D model's representation of the mixing process in the equatorial region of a hot Jupiter.


\begin{figure*}[!htp]
   \centering
   \includegraphics[width=\columnwidth]{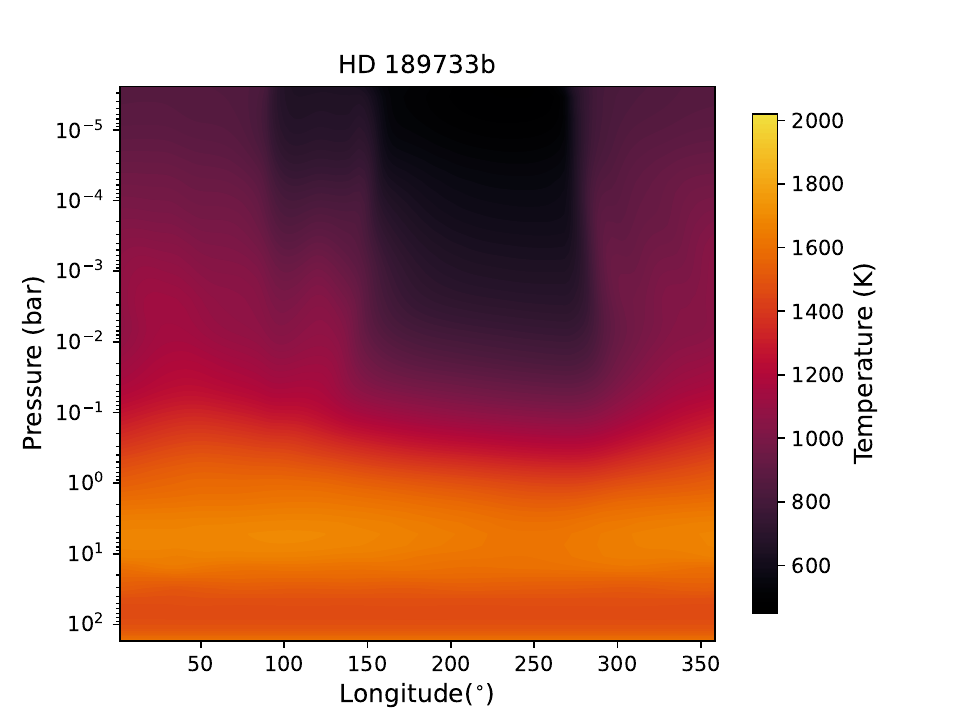}
   \includegraphics[width=\columnwidth]{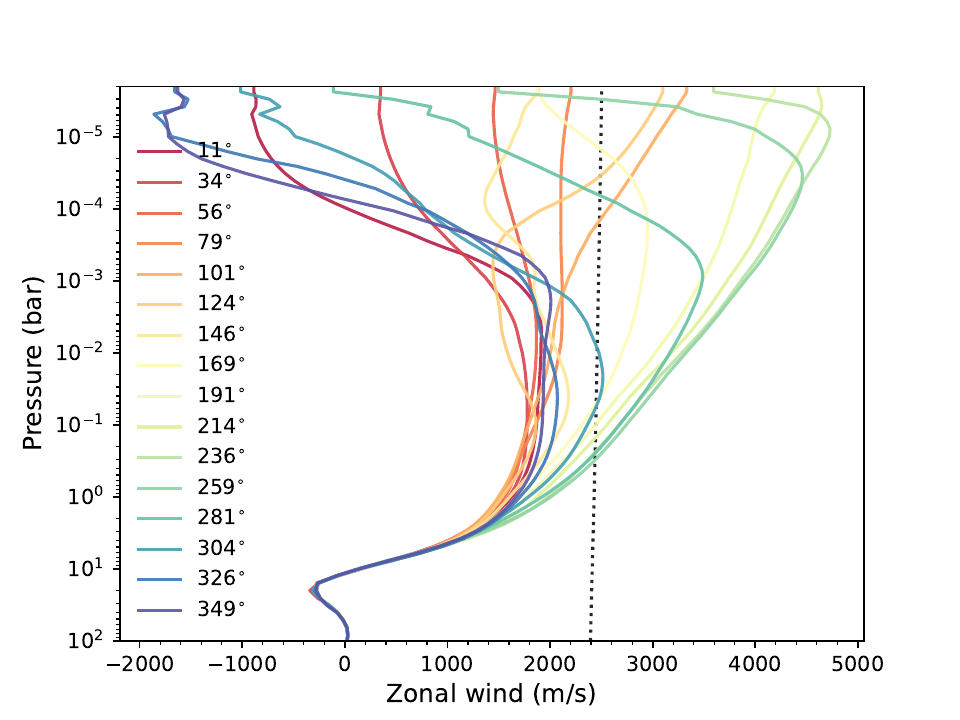}
   \includegraphics[width=\columnwidth]{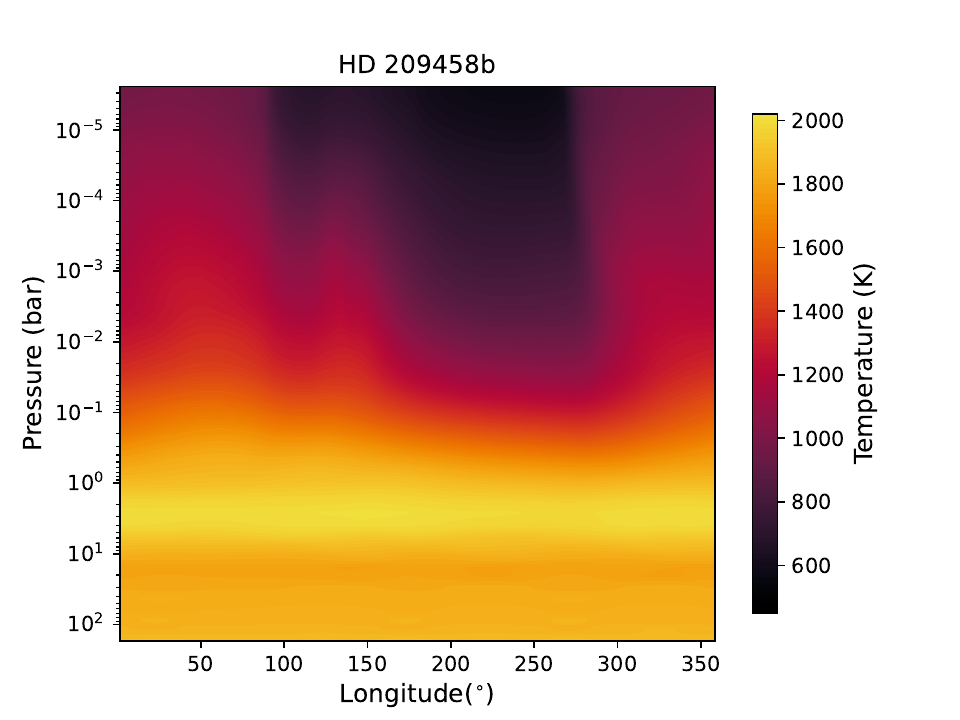}
   \includegraphics[width=\columnwidth]{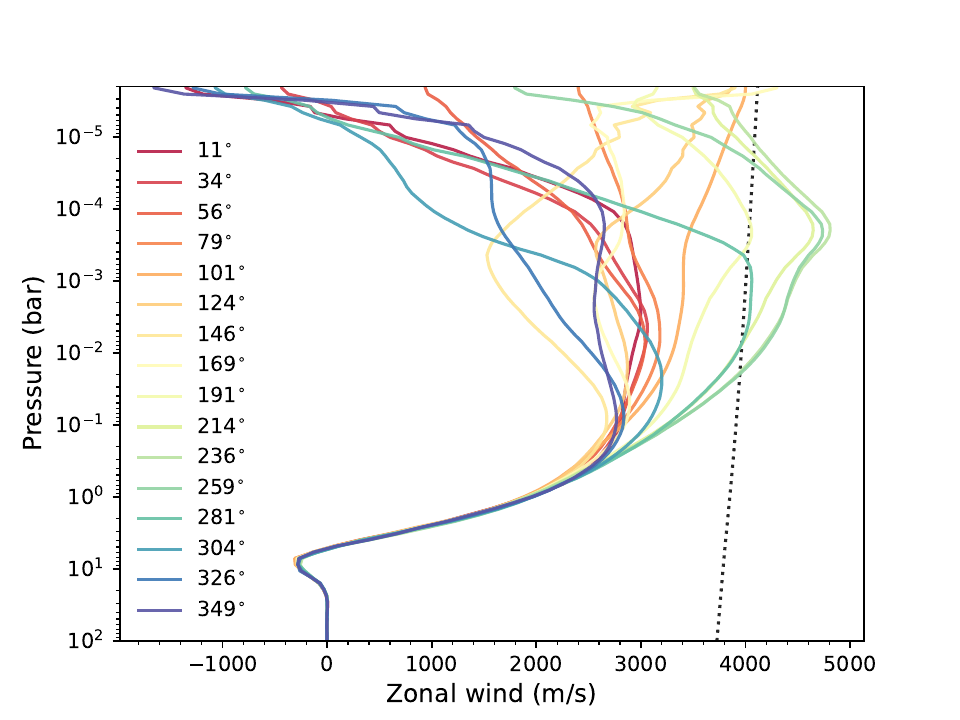}
\caption{The equatorial temperature (left) and zonal winds (right) from our GCMs of HD 209458 b and HD 189733b, averaged over $\pm$ 30$^{\circ}$ latitudes. The substellar point is located at 0$^{\circ}$ longitude. The uniform zonal winds (constant angular velocities with 2430 m/s for HD 189733 b and 3850 m/s for HD 209458 b at the 1 bar pressure level, respectively) assumed in \cite{Agundez2014} are illustrated in black dotted lines in the right panels for comparison.} 
\label{fig:GCMs}
\end{figure*}

\section{Results: Equatorial Chemical Distributions on HD 189733 b and HD 209458 b}\label{results}
We now first present an overview of the abundance distribution in the atmosphere of our fiducial simulations of HD 189733 b and HD 2094598 b. We briefly compare our results to previous work by \cite{Agundez2014} on the same planets. To gain insight into the effects of vertical and horizontal transport, we examine the limiting cases where individual transport processes are isolated following the approach of \cite{Agundez2014}. We then explore the sensitivity to the eddy diffusion coefficients converted from the GCM wind and discuss the results guided by the limiting cases. Lastly, we explore the global chemical transport when C/O $\gtrsim$ 1 with HD 209458 b, motivated by recent high-spectral-resolution observations.


\begin{figure*}[!ht]
   \centering
   \includegraphics[width=0.7\columnwidth]{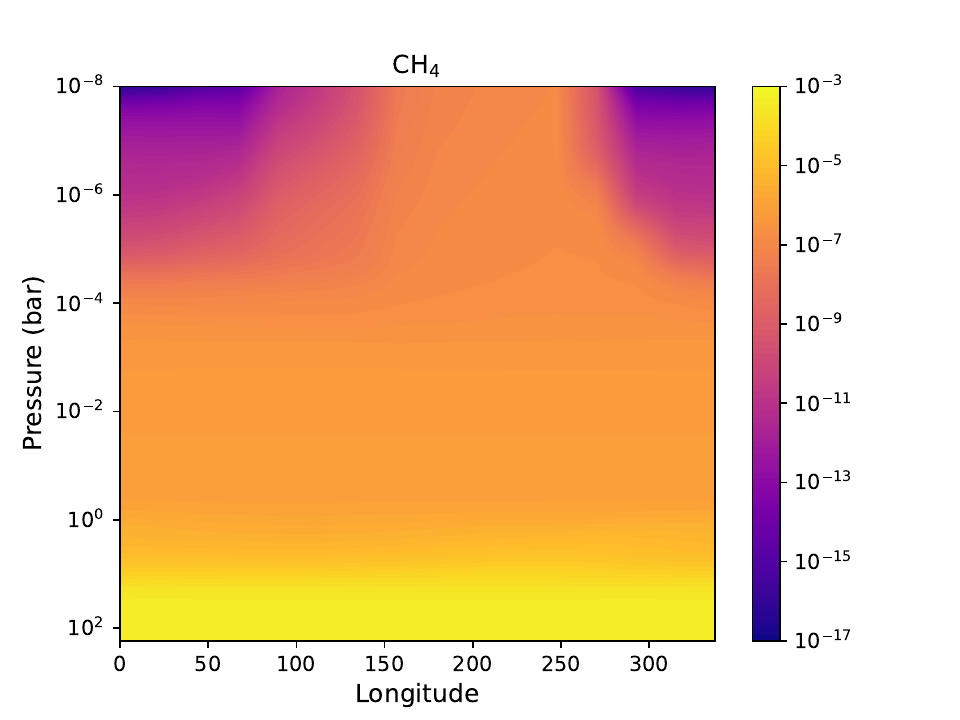}
   \includegraphics[width=0.7\columnwidth]{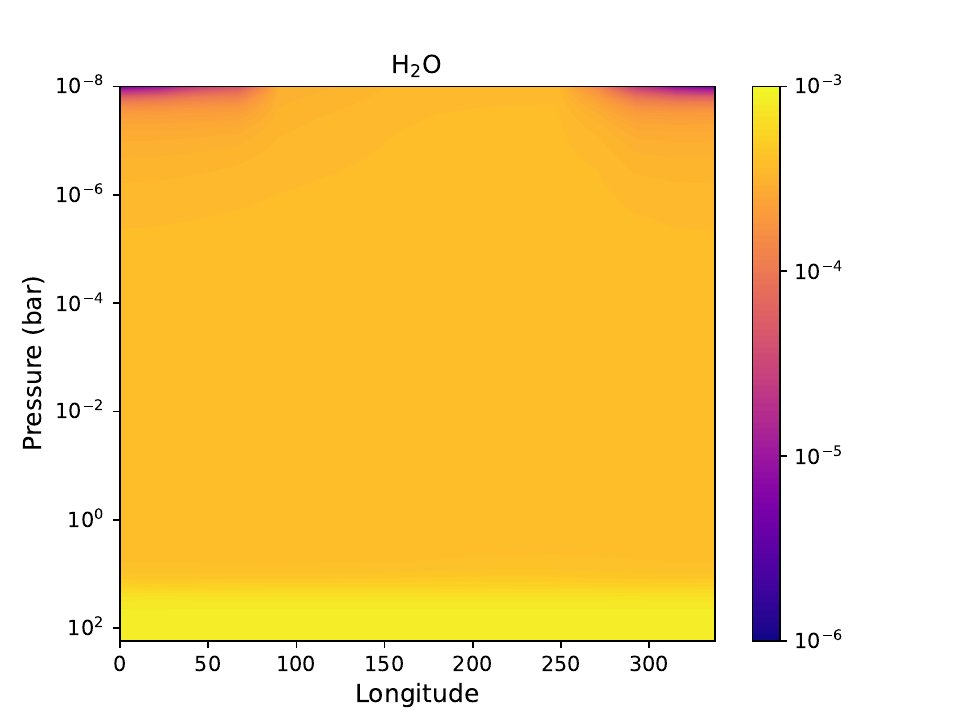}
   \includegraphics[width=0.7\columnwidth]{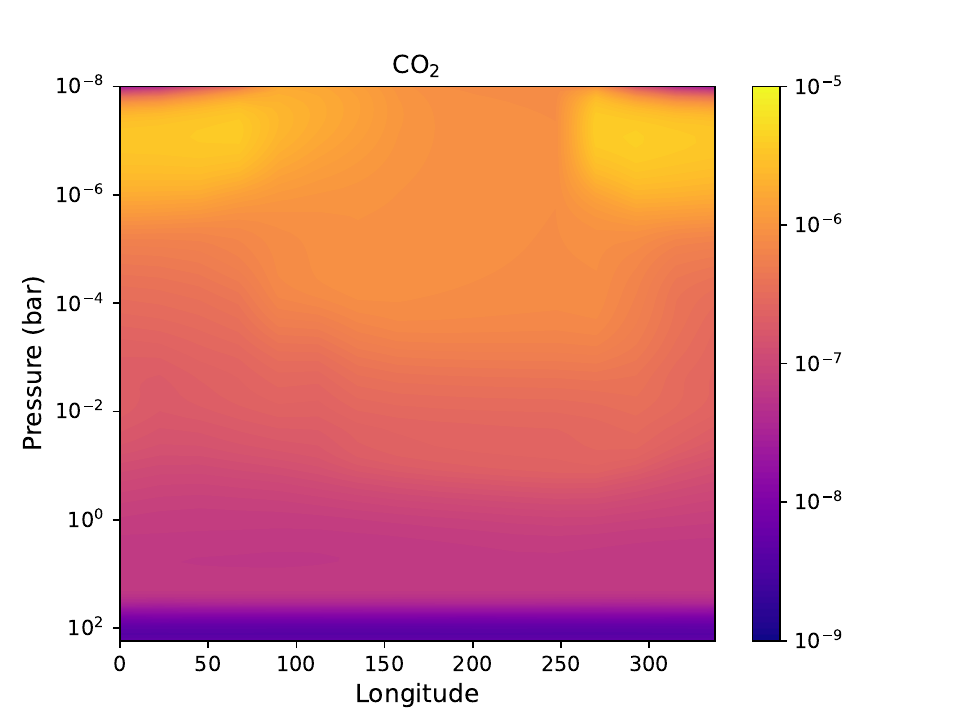}
   \includegraphics[width=0.7\columnwidth]{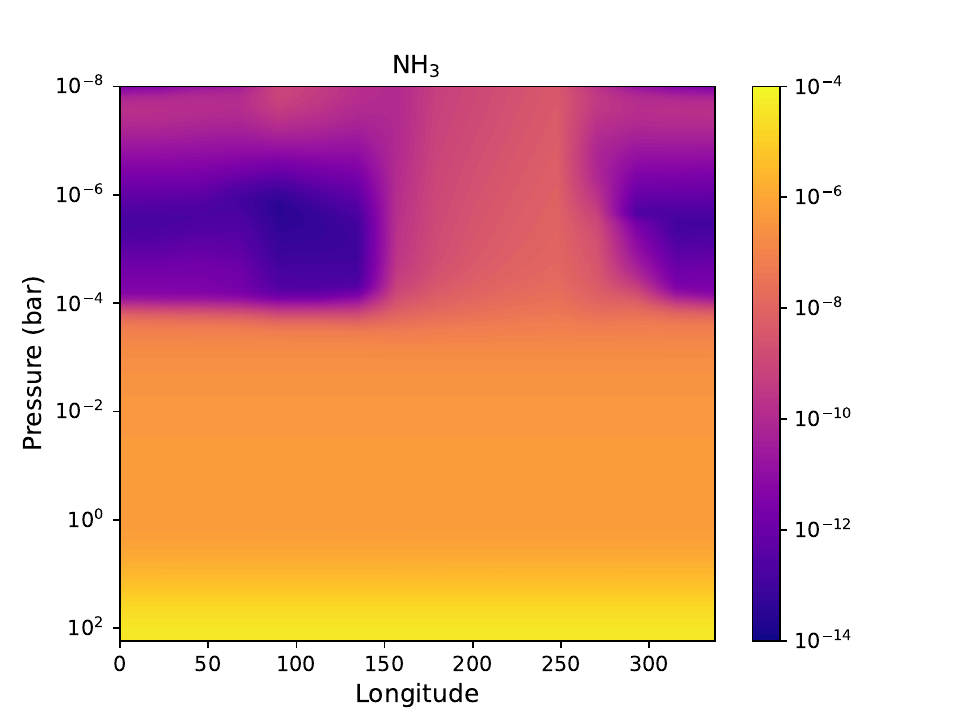}
   \includegraphics[width=0.7\columnwidth]{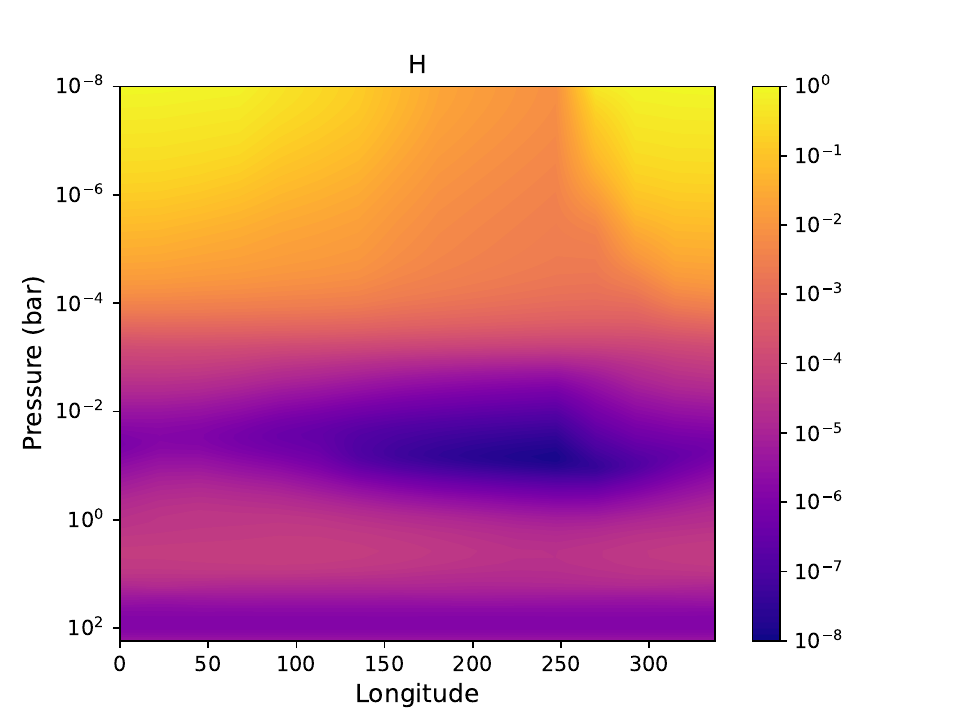}
   \includegraphics[width=0.7\columnwidth]{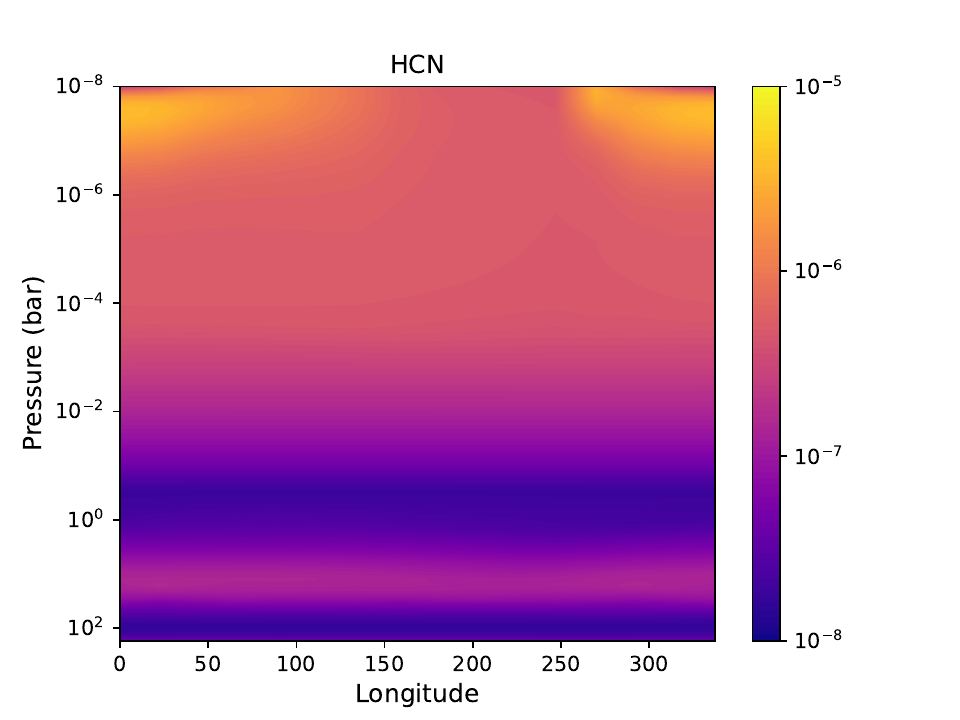}
\caption{The equatorial abundance distribution (in color contours) of several chemical species as a function of longitude and pressure on HD 189733b. The substellar point is located at 0$^{\circ}$ longitude.}
\label{fig:HD189-contours}
\end{figure*}

\begin{figure*}[!ht]
   \centering
   \includegraphics[width=0.7\columnwidth]{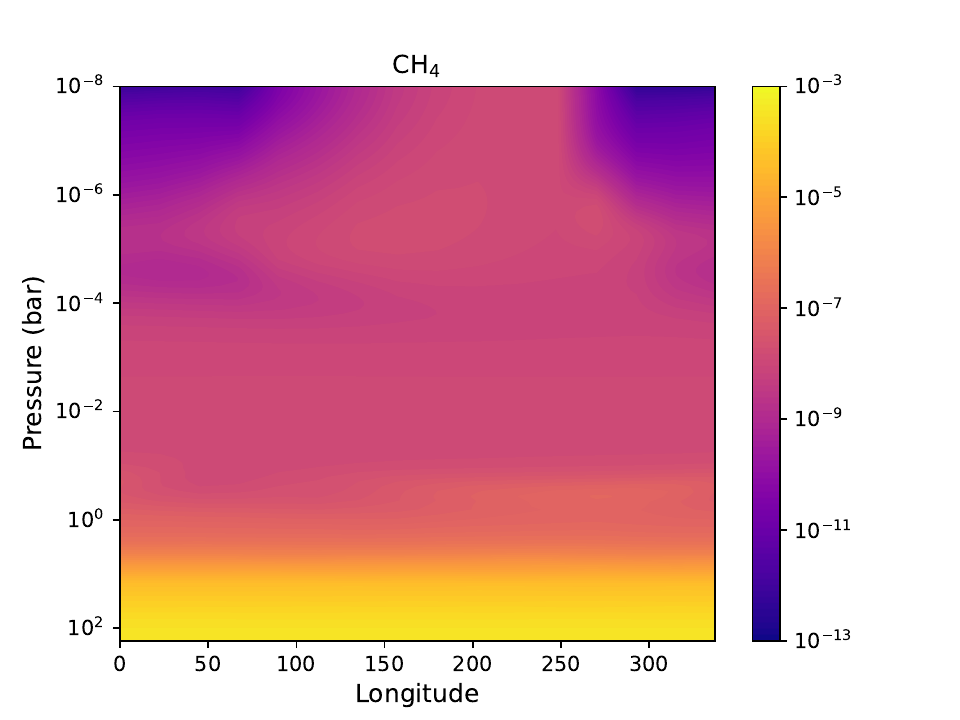}
   \includegraphics[width=0.7\columnwidth]{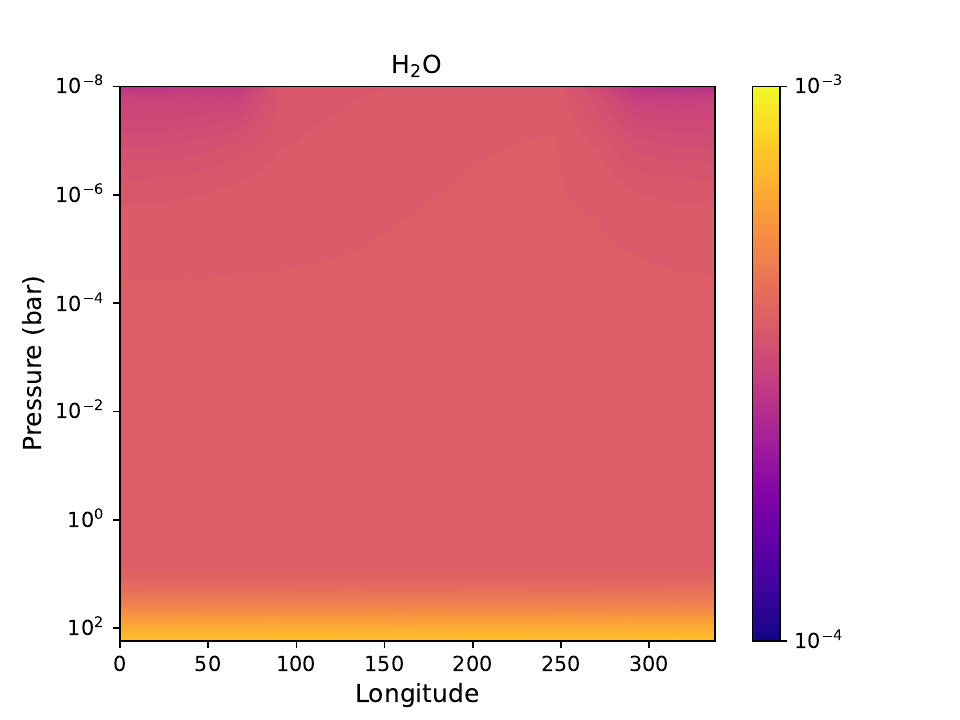}
   \includegraphics[width=0.7\columnwidth]{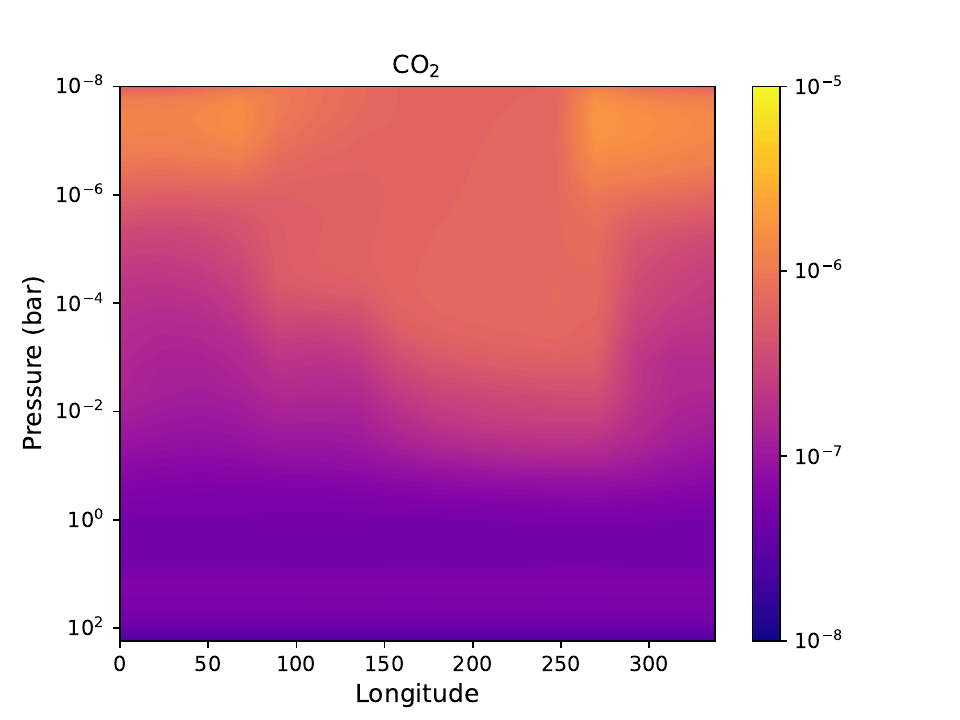}
   \includegraphics[width=0.7\columnwidth]{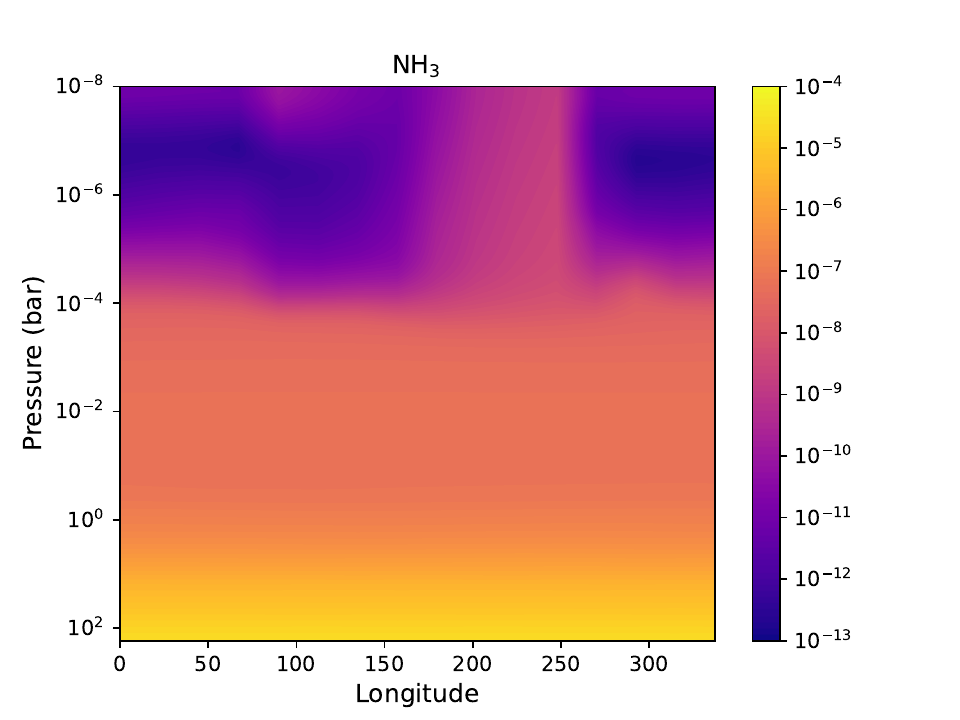}
   \includegraphics[width=0.7\columnwidth]{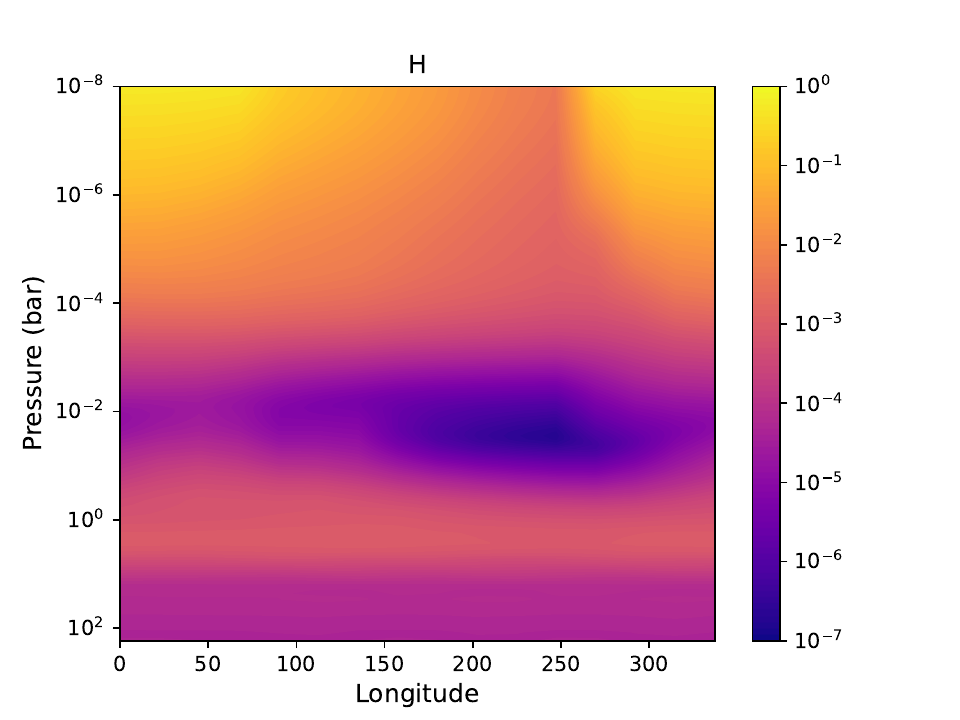}
   \includegraphics[width=0.7\columnwidth]{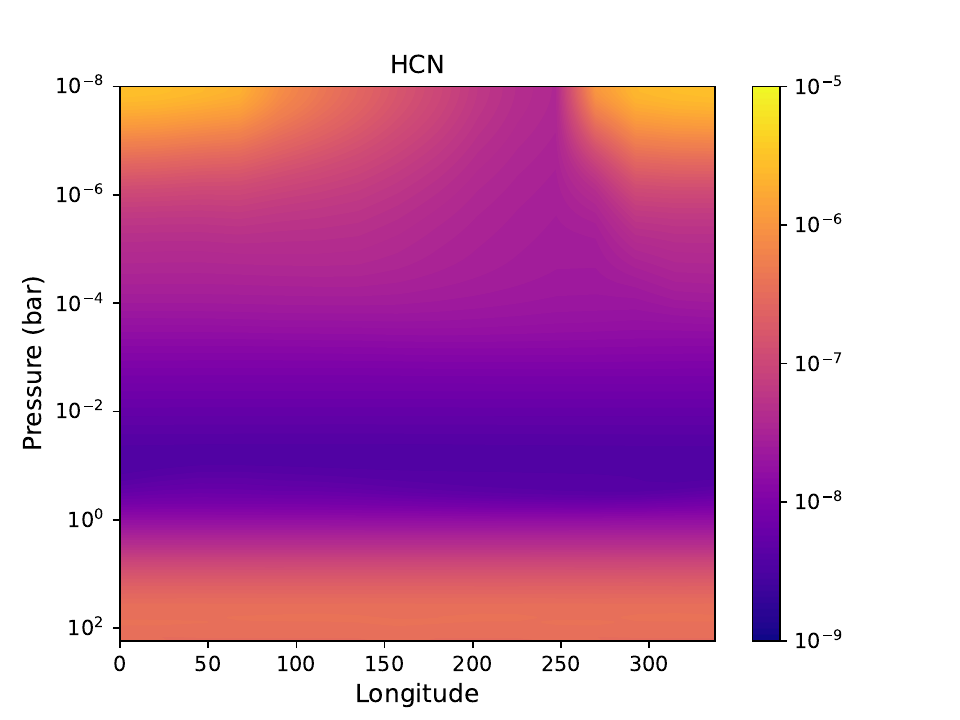}
\caption{Same as Figure \ref{fig:HD189-contours} but for HD 209458 b.}
\label{fig:HD209-contours}
\end{figure*}

\begin{figure*}[!ht]
   \centering
   \includegraphics[width=\columnwidth]{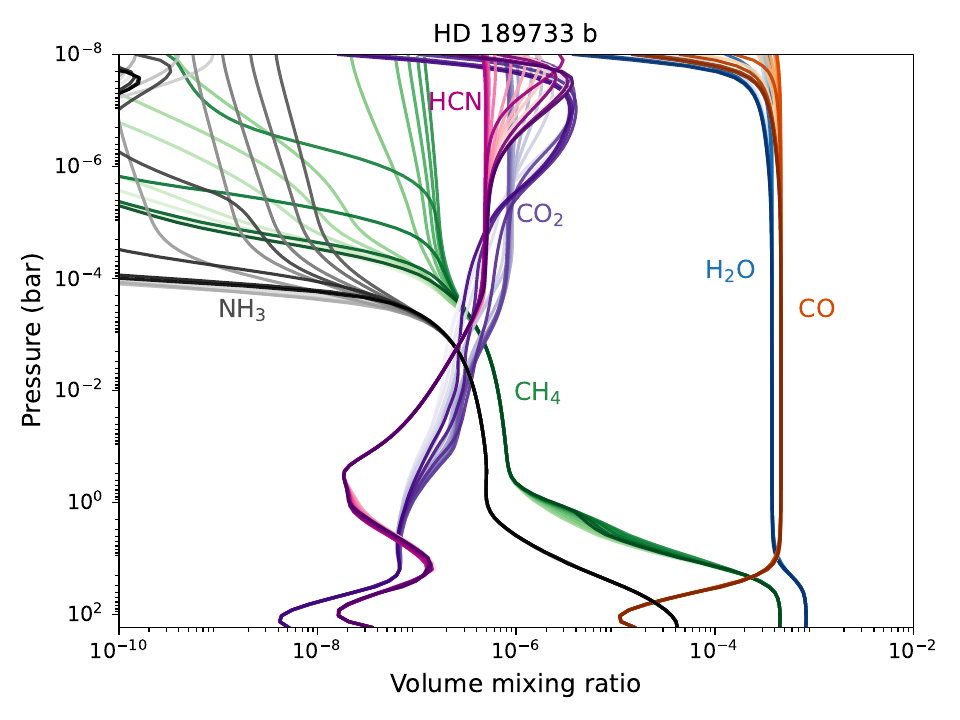}
   \includegraphics[width=\columnwidth]{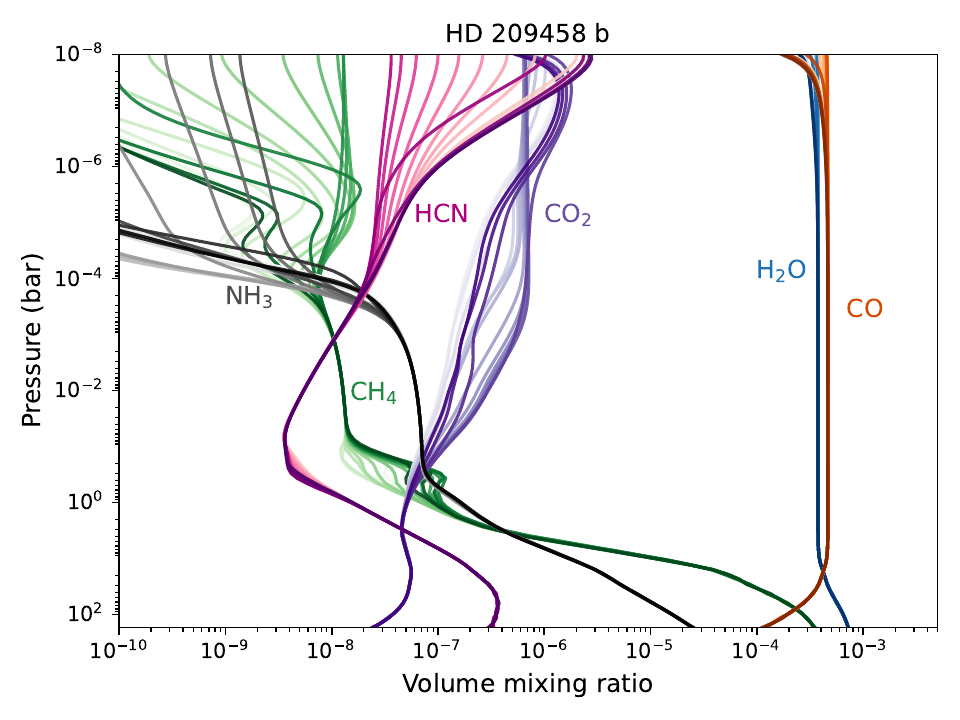}
\caption{The vertical profiles in the equatorial regions of HD 189733b and HD 209458 b simulated by 2D VULCAN. The shades of colors from light to dark correspond to longitudes from the substellar point (0$^{\circ}$) through nightside and back around to the substellar point (360$^{\circ}$).}
\label{fig:2D_pasta}
\end{figure*}

\subsection{Setup and overview}
We adopt 3D GCM output from SPARC/MITgcm \citep{Showman2009,Adcroft2004} for HD 189733 b and HD 209458 b. Chemical equilibrium is assumed when calculating radiative transfer with the correlated-k distribution method. For HD 189733 b, we use the same GCM output as \cite{Agundez2014}, where detailed GCM parameters are further listed in Table 1 of \cite{Steinrueck2019} (Table 1). For HD 209458 b, we do not include TiO and VO as shortwave opacity sources in our simulation, based on the lack of evidence of a thermal inversion from the reanalysis of emission spectra obtained by Spitzer and HST \citep{DiamondLowe2014,Line2016}. The parameters for our HD 209458 b GCM can be found in \citep{Parmentier2013}, with the change that TiO and VO opacities removed. The lower and upper pressure boundaries of the GCM simulations are 200 bar and 2$\times$ 10$^{-6}$ bar, respectively. To extend the model domain for VULCAN 2D, the temperatures and winds at the upper GCM boundary are held constant up to 10$^{-8}$ bar.

The equatorial average temperatures and winds for our HD 189733b and HD 209458 b simulations are shown in Figure \ref{fig:GCMs}. The zonal wind speeds evidently vary across longitudes at pressures above 0.1 bar. Both planets share similar thermal and dynamical structures in the equatorial region, though HD 209458 b exhibits higher temperatures and overall faster zonal wind speeds. We adopt the same global eddy diffusion coefficients as \cite{Agundez2014} for vertical mixing. These $K_{\textrm{zz}}$ profiles are compared with those estimated by the mixing length theory in Figure \ref{fig:w_Kzz}. We will further explore the sensitivity to the adopted eddy diffusion coefficients in Section 4.4. For VULCAN 2D, we apply the N-C-H-O chemical network\footnote{\url{https://github.com/exoclime/VULCAN/blob/master/thermo/NCHO_photo_network.txt}} and omit sulfur throughout this study. 


Figures \ref{fig:HD189-contours} and \ref{fig:HD209-contours} showcase the chemical abundance distributions simulated by VULCAN 2D in the equatorial regions of HD 189733b and HD 209458 b. The vertical abundance distributions along longitudes of several important species are summarized in Figure \ref{fig:2D_pasta}. In the observable regions of the atmospheres, species such as \ce{CH4}, H, \ce{NH3}, and HCN show considerable horizontal (longitudinal) gradient, whereas \ce{H2O} and CO (not shown) are rather uniform everywhere. The uniformity is expected since the equilibrium abundances of \ce{H2O} and CO do not vary with longitude in the first place. Species that are more susceptible to photodissociation, such as \ce{CH4} and \ce{NH3}, are destroyed on the dayside but can partly recover on the nightside. Photochemical products, such as atomic H and HCN, build up on the dayside while being transported horizontally by the zonal jet at the same time. We find that the chemical transport exhibits qualitative similarities between HD 189733b and HD209458 b, with the main difference being that \ce{CH4} and \ce{NH3} are more favored on the cooler HD 189733 b.


This behavior of chemical transport can be understood by comparing the horizontal and vertical dynamical timescales, as illustrated in Figure \ref{fig:2D_tau}: Horizontal mixing dominates at altitudes below $\sim$0.1 mbar, while vertical mixing dominates at altitudes above $\sim$0.1 mbar level. Within the horizontal mixing region, species with longer timescales tend to display more uniform global abundances, whereas compositional gradients in the upper atmosphere are mainly controlled by vertical mixing and photochemistry.

\begin{figure*}[!ht]
   \centering
   \includegraphics[width=\columnwidth]{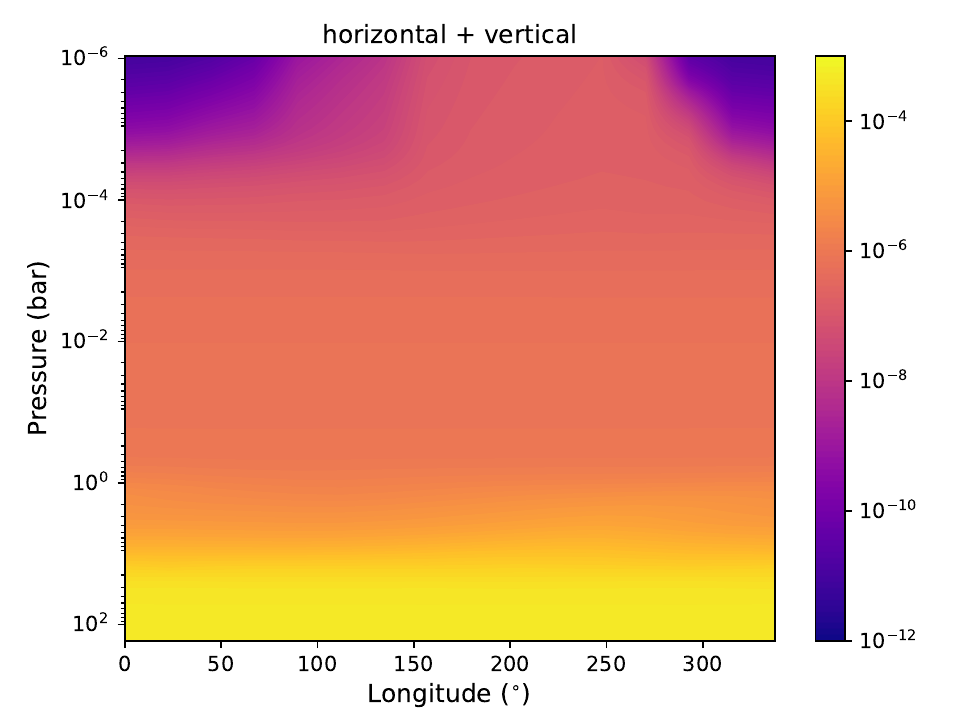}
   \includegraphics[width=\columnwidth]{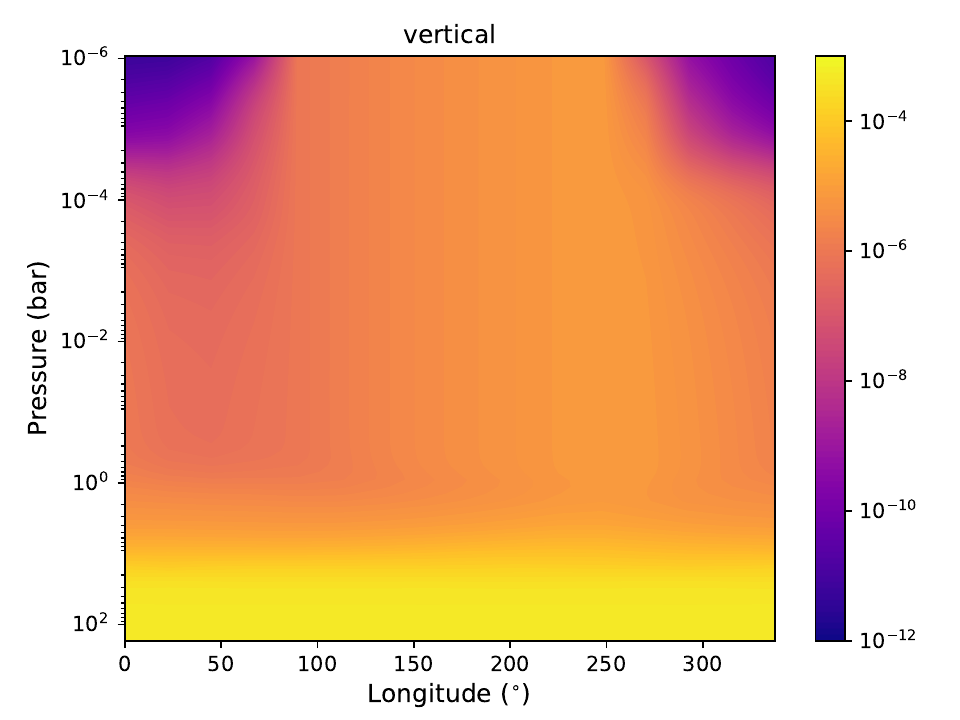}
   \includegraphics[width=\columnwidth]{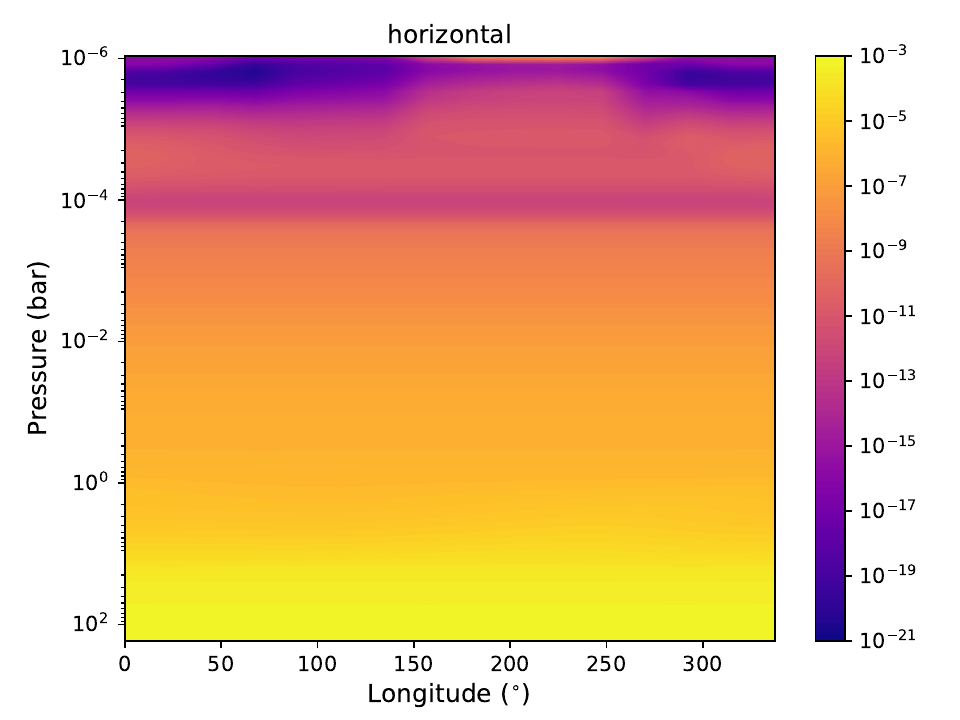}
   \includegraphics[width=\columnwidth]{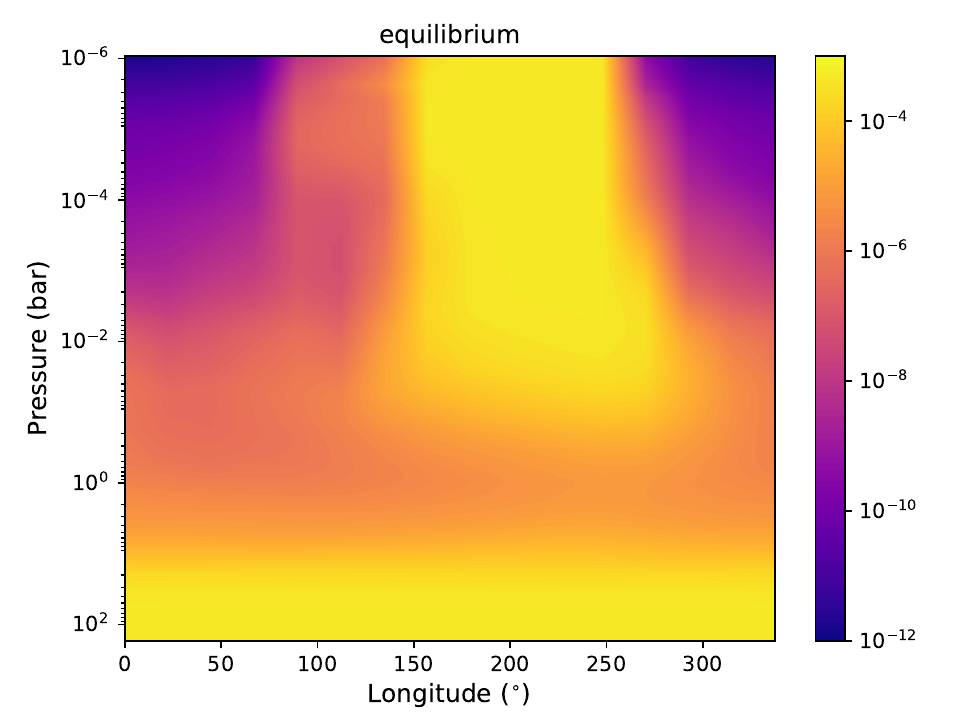}
\caption{The equatorial distribution of \ce{CH4} as a function of longitude and pressure on HD 189733 b, from our nominal model, vertical-mixing model, horizontal-transport model, and chemical equilibrium. Note the colorbars have different scales.}
\label{fig:HD189-2D-CH4-limit}
\end{figure*}

\begin{figure*}[!ht]
   \centering
   \includegraphics[width=\columnwidth]{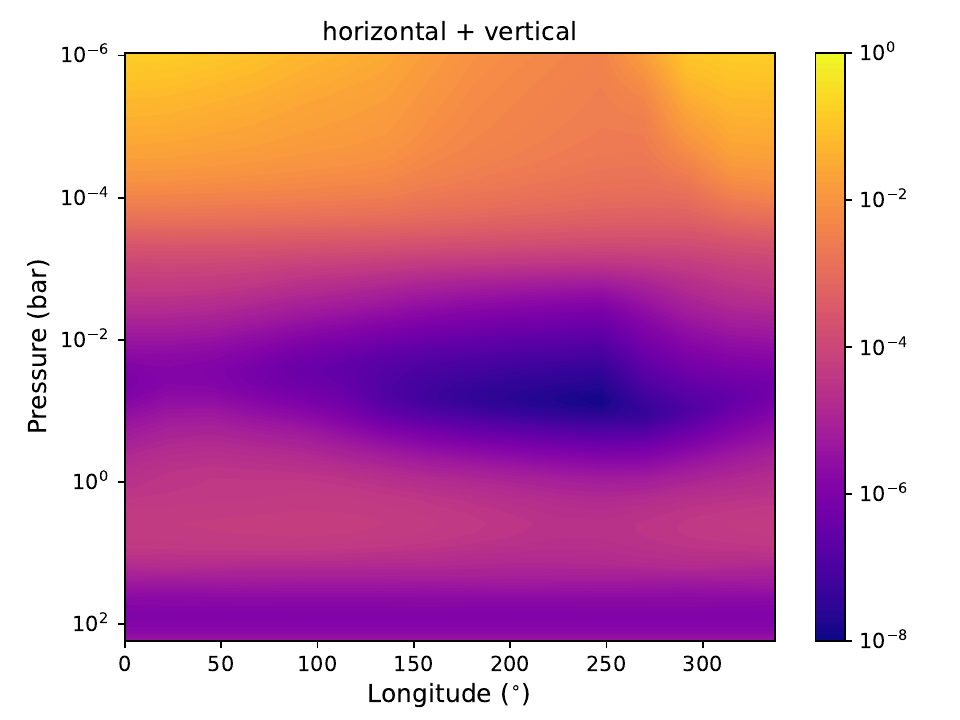}
   \includegraphics[width=\columnwidth]{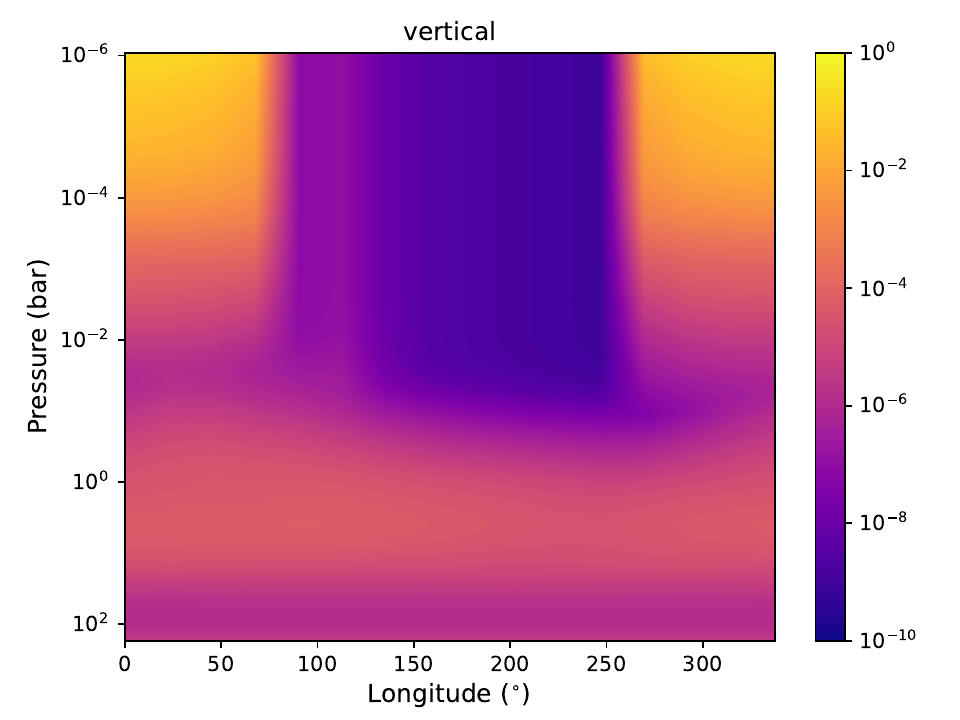}
   \includegraphics[width=\columnwidth]{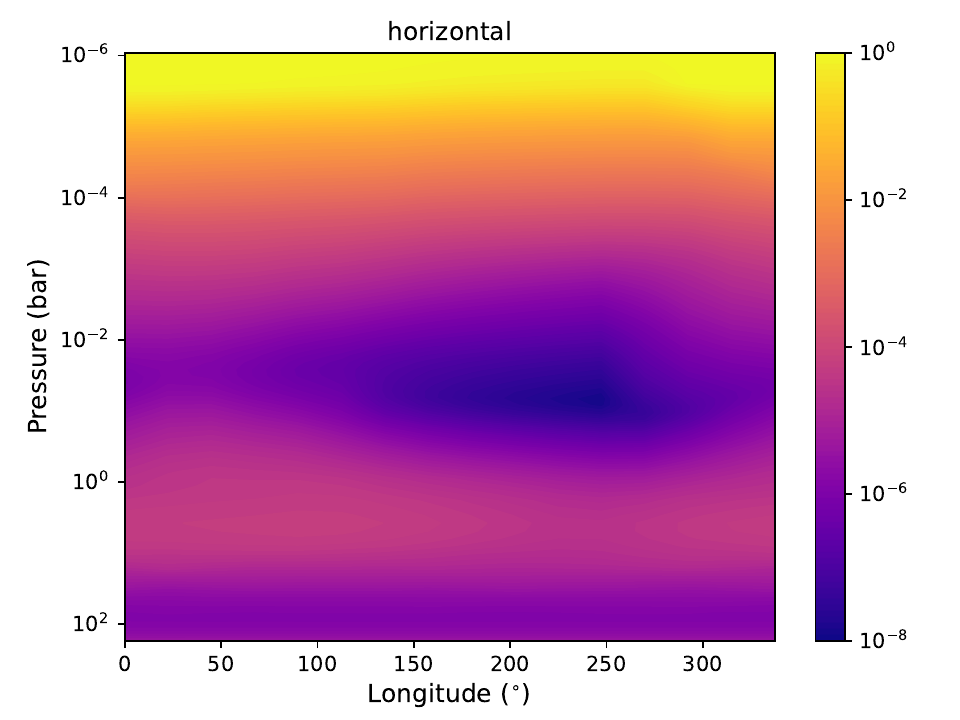}
   \includegraphics[width=\columnwidth]{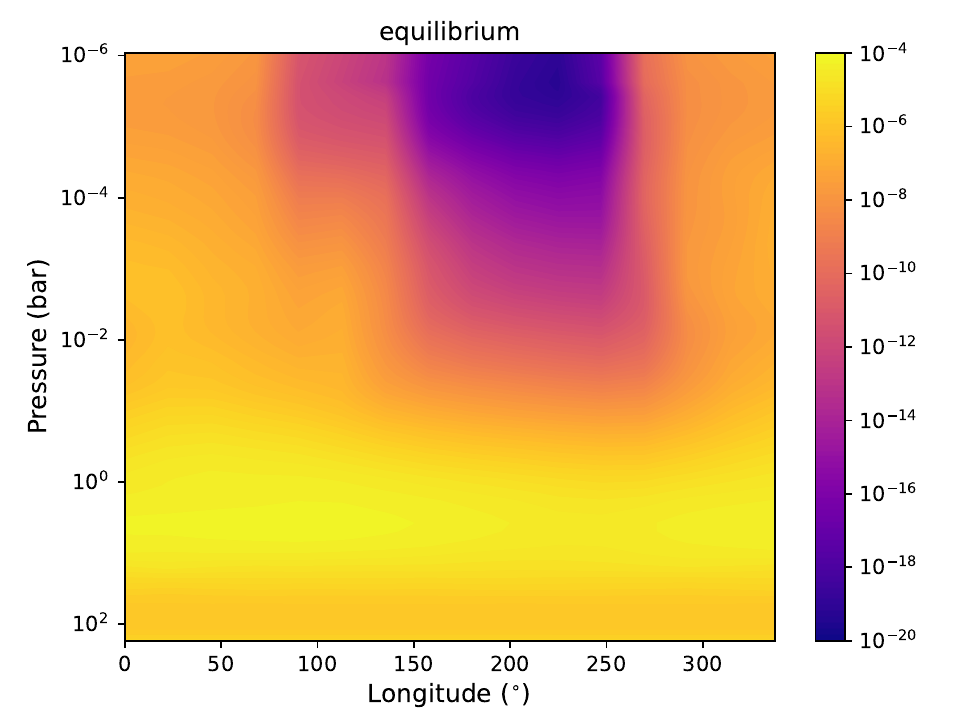}
\caption{Same as Figure \ref{fig:HD189-2D-CH4-limit} but for H.}
\label{fig:HD189-2D-H-limit}
\end{figure*}

\begin{figure*}[ht!]
   \centering
   \includegraphics[width=\columnwidth]{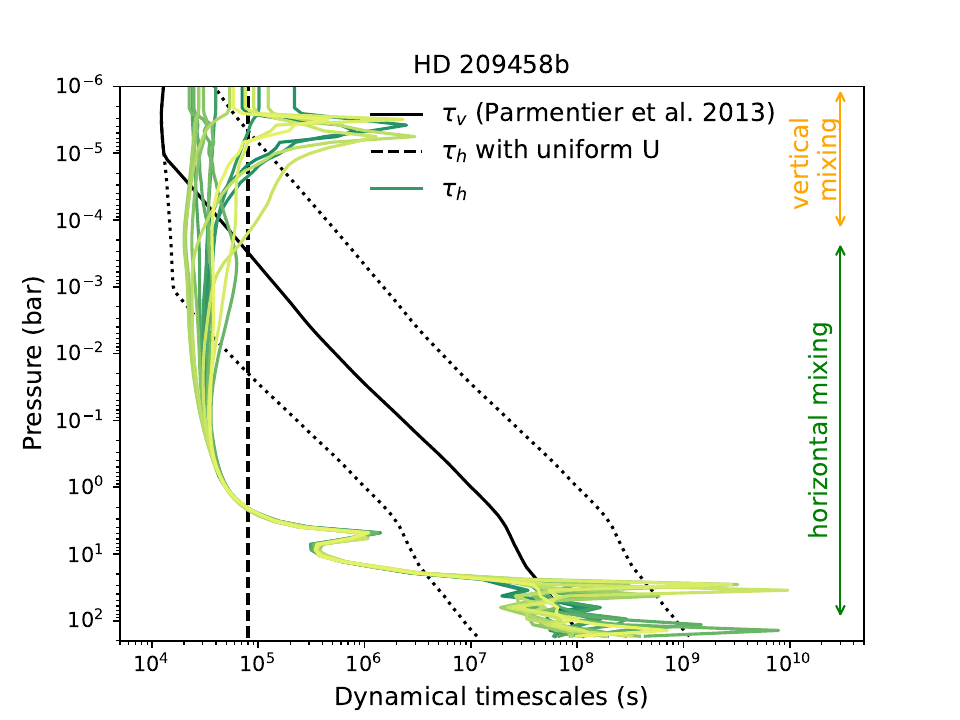}
   \includegraphics[width=\columnwidth]{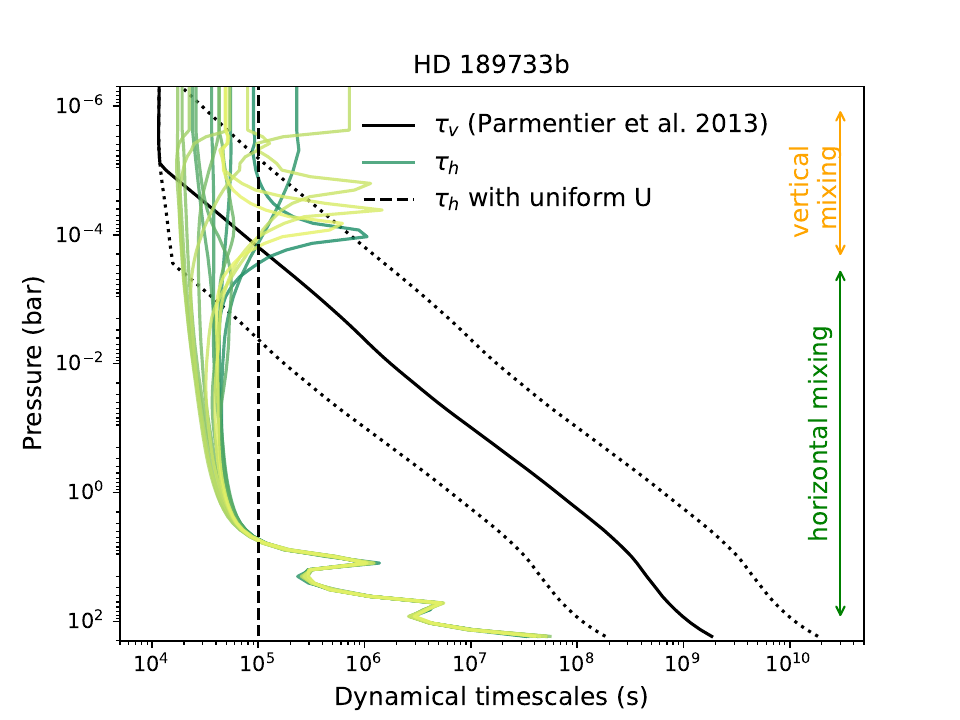}
\caption{The dynamical timescales of vertical transport ($\tau_v$) and horizontal transport ($\tau_h$). The black dotted lines indicate the vertical transport timescale with 10 times and 0.1 times eddy diffusion coefficients. The horizontal transport timescale at different longitudinal locations is depicted in gradient shades of green, while the uniform horizontal wind assumed in \cite{Agundez2014} is shown as the dashed lines. The arrows on the right indicate the regions dominated by horizontal transport and vertical transport processes, respectively.}
\label{fig:2D_tau}
\end{figure*}

\begin{figure*}[!ht]
   \centering
   \includegraphics[width=\columnwidth]{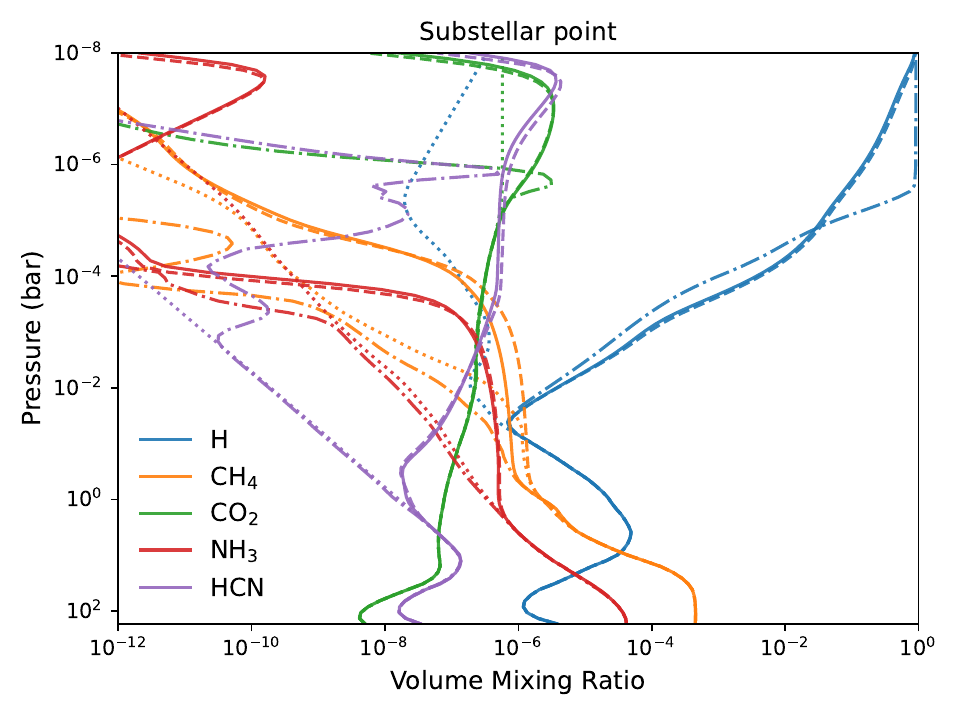}
   \includegraphics[width=\columnwidth]{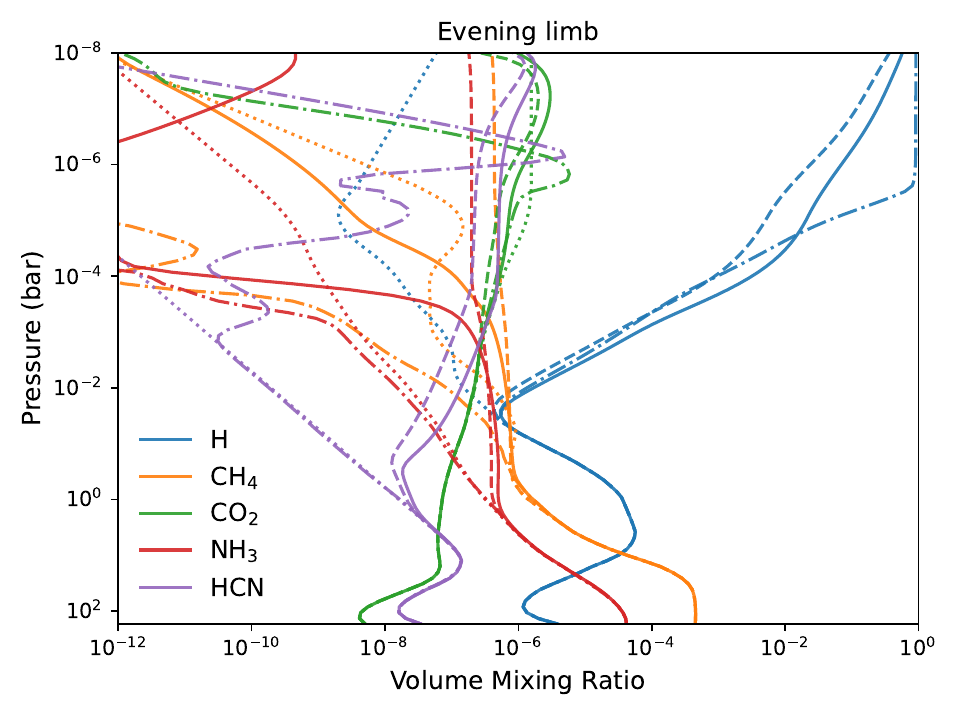}
   \includegraphics[width=\columnwidth]{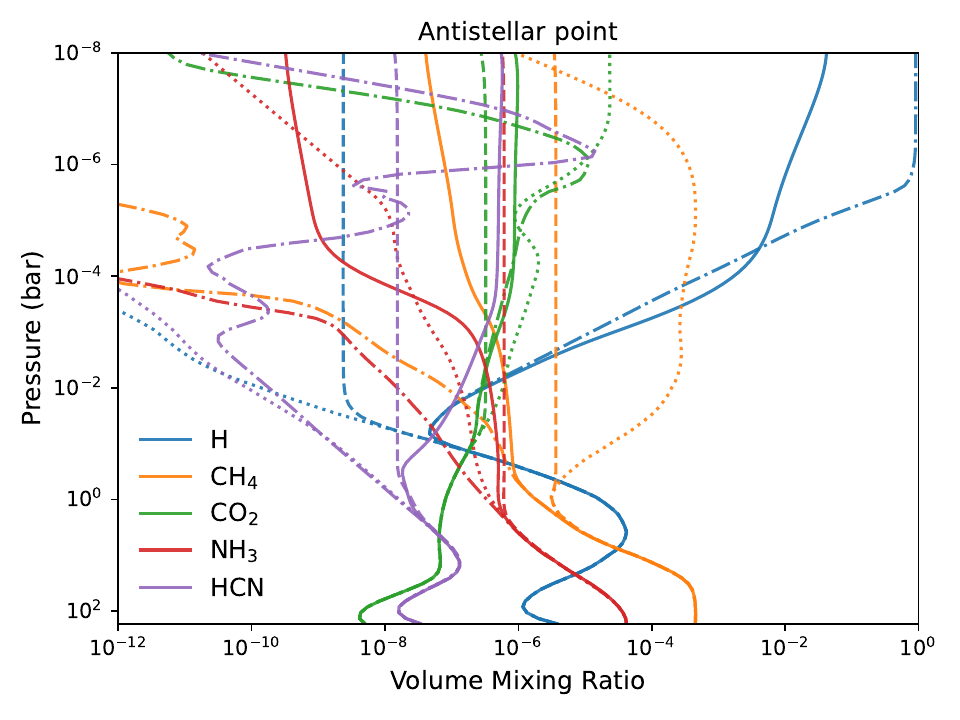}
   \includegraphics[width=\columnwidth]{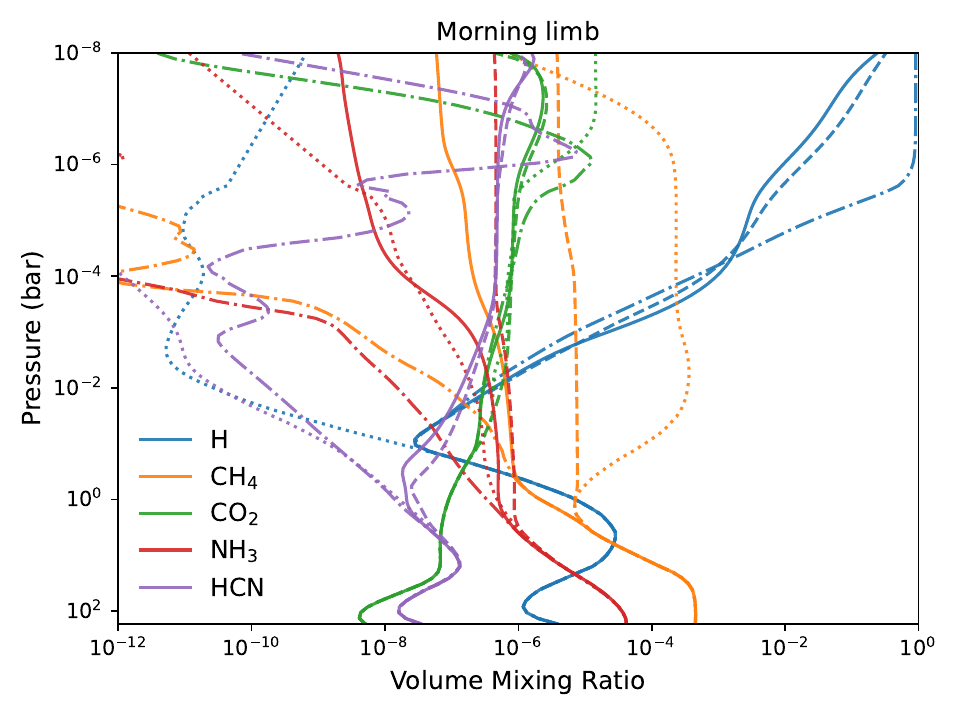}
\caption{The abundance profiles of selected species on HD 189733 b at four longitudes: substellar point, evening limb, antistellar point, and morning limb. The volume mixing ratios computed by the nominal 2D model (solid) are compared to the vertical-mixing case (excluding horizontal transport; dashed), the horizontal-transport case (excluding vertical diffusion; dash-dotted), and thermochemical equilibrium (dotted).}
\label{fig:HD189-4parts}
\end{figure*}

\begin{figure*}[ht]
   \centering
   \includegraphics[width=\columnwidth]{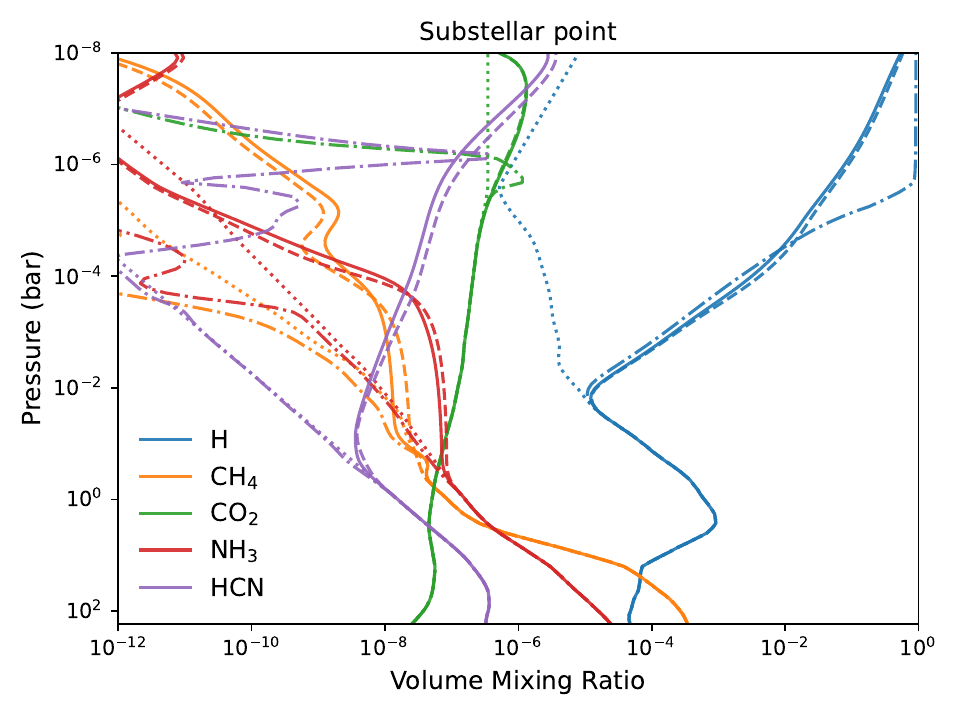}
   \includegraphics[width=\columnwidth]{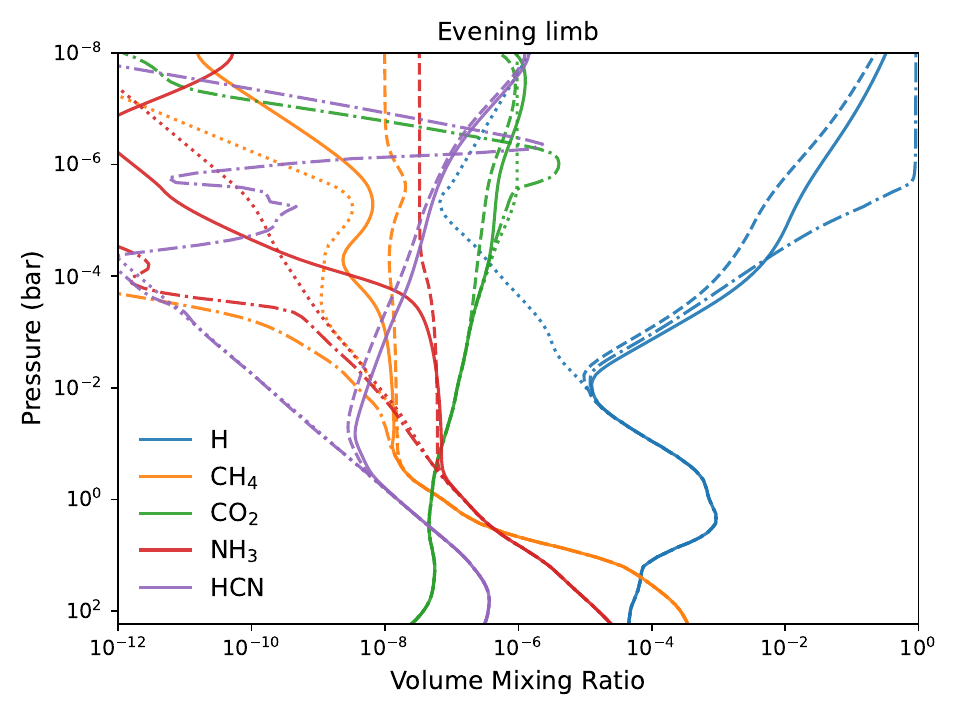}
   \includegraphics[width=\columnwidth]{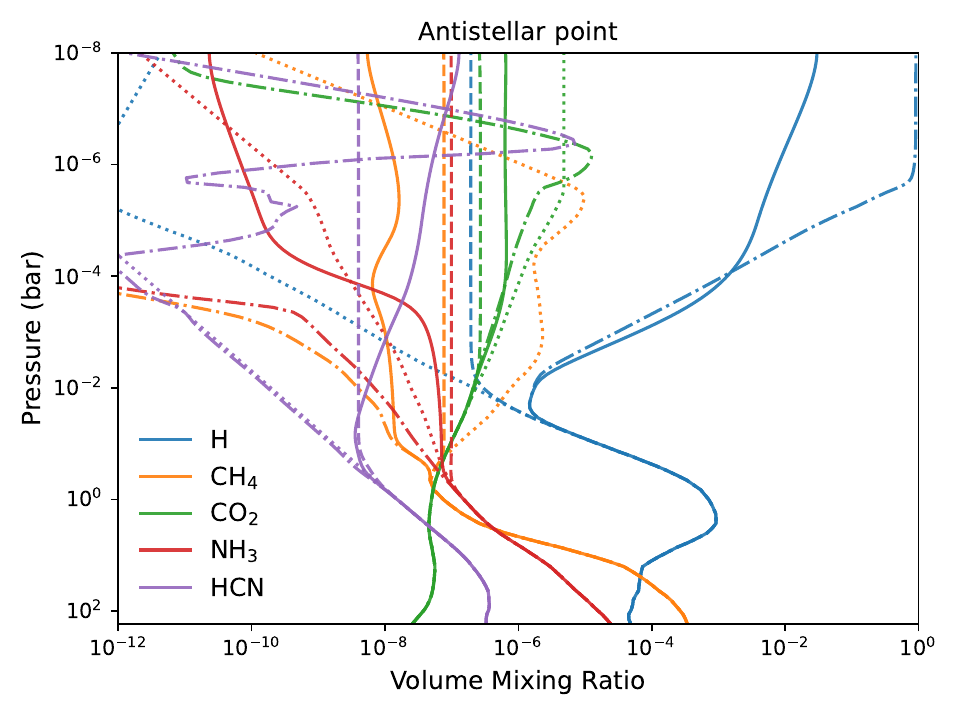}
   \includegraphics[width=\columnwidth]{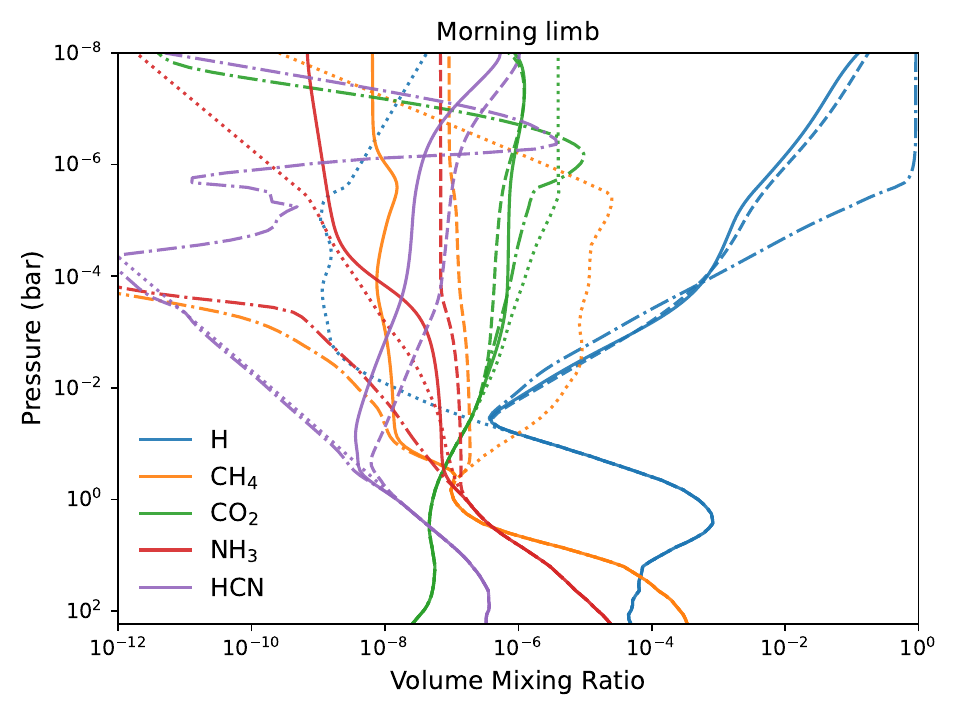}
\caption{The abundance profiles of selected species on HD 209458 b at four longitudes: substellar point, evening limb, antistellar point, and morning limb. The volume mixing ratios computed by the nominal 2D model (solid) are compared to the vertical-mixing case (excluding horizontal transport; dashed), the horizontal-transport case (excluding vertical diffusion; dash-dotted), and thermochemical equilibrium (dotted).}
\label{fig:HD209-4parts}
\end{figure*}

\subsection{Limiting cases with only horizontal transport and with only vertical mixing}
Following the pedagogical exercise in \cite{Agundez2014}, this section presents the simulated vertical abundance profiles of HD 189733 b and HD 209458 b under limiting cases to isolate individual dynamical effects. Specifically, we examine scenarios where we exclude horizontal transport and vertical mixing, respectively. Figures \ref{fig:HD189-2D-CH4-limit} and \ref{fig:HD189-2D-H-limit} first show the distributions of \ce{CH4} and H in these limiting cases, demonstrating the principles of horizontal transport and vertical mixing. As expected, vertical mixing tends to homogenize the vertical gradient of composition, while horizontal transport tends to homogenize the horizontal gradient. However, the influence of transport processes on a species depends on its chemical properties, which can be broadly categorized into two groups: For species like \ce{CH4} that are replenished from the deep thermochemical regions, vertical mixing provides a zeroth order prediction on their abundances. For photochemical products like H that are predominantly produced on the dayside upper atmosphere, horizontal transport is crucial to account for their circulation to the nightside.


Detailed comparisons between the nominal 2D models, the vertical-mixing case, the horizontal-transport case, and chemical equilibrium at different longitudinal locations are presented in \ref{fig:HD189-4parts} and \ref{fig:HD209-4parts}. Examining the quenched species such as \ce{CH4} and \ce{NH3}, we find that the vertical mixing and horizontal transport cases have the same quench levels. This is not surprising, as these quench levels correspond to the transition point associated with the same chemical timescales, regardless of the dynamical process.

However, the abundance distributions above the quench levels differ substantially between the vertical mixing and horizontal transport scenarios. This is because vertical mixing makes chemical abundances quenched from the hot deep layers, while horizontal transport results in quenching from the hot and irradiated dayside, as also noted in \cite{Agundez2014}. Owing to this nature of horizontal transport, the substellar abundance profiles closely resemble those from the vertical mixing case (the close match between the solid and dashed lines) for both planets. 



Taking a closer look at Figures \ref{fig:HD189-4parts} and \ref{fig:HD209-4parts}, species with long chemical timescales like \ce{CH4} and \ce{NH3} closely follow the vertical mixing distribution at hotter longitudes, the substellar point and evening limb, up to about 10$^{-4}$ bar. On the other hand, at the colder antistellar point and morning terminator, the equilibrium abundances differ substantially from the dayside, causing the abundance distribution predicted by the vertical-mixing model to diverge from the nominal 2D model (with both vertical and horizontal transport) at these cooler locations. These results shed light on the limitation of applying 1D models to interpret observations that probe the chemical properties of different regions. 

One crucial feature of horizontal transport is transporting the photochemical products from the dayside to the nightside \citep{Agundez2014,Baeyens2022}. For both planets, atomic H produced by photolysis can penetrate into the nightside, even reaching regions around the antistellar point where no UV photons are available. Transport of atomic H plays a key role in reacting with \ce{CH4} and \ce{NH3} to form HCN on the nightside. The abundances of \ce{CH4} and \ce{NH3} in Figures \ref{fig:HD189-4parts} and \ref{fig:HD209-4parts} fall between those in the vertical mixing and horizontal transport cases, highlighting the importance of considering both mixing processes. 


To conclude our limiting-case analysis, we note that for hot Jupiters similar to HD 189733 b and HD209458 b, 1D model including vertical mixing serves as a fairly good approximation for the dayside. Given that the equatorial jet efficiently transports heat, 1D models generally better capture the hotter evening limb better compared to the cooler morning limb in the pressure regions most relevant for transmission spectroscopy. We will discuss the limb asymmetry as a result of horizontal transport more in Section \ref{sec:HD209_CtoO105} and \ref{sec:spectra}.





\subsection{Comparison with \cite{Agundez2014}} 
There are several differences in modeling assumptions and setups between our models of HD 18933b and HD 209458 b and those presented in \cite{Agundez2014}: (i) \cite{Agundez2014} assume uniform zonal winds, while we adopt longitude- and pressure-dependent wind profiles from the GCM. (ii) Shortwave opacity sources of TiO and VO, which are responsible for generating thermal inversion, are included in \cite{Agundez2014} but are not included in our model for HD 209458 b (iii) \cite{Agundez2014} use the chemical scheme of \cite{Venot2012}, while our study adopts the chemical scheme of \cite{Tsai2021}.

Despite these differences, our nominal 2D VULCAN and those from the limiting cases are qualitatively consistent with the pseudo-2D outcomes presented in  \cite{Agundez2014}. The major difference lies in the thermal inversion of HD 209458 b in \cite{Agundez2014}, making the dayside temperature at low pressures $\sim$ 1000 K higher than that in our model. Consequently, \cite{Agundez2014} predict lower levels of \ce{CH4}, \ce{NH3}, and HCN on HD 209458 b. The hotter dayside also leads to a lower equilibrium abundance of \ce{CO2}, creating the notable \ce{CO2} day-night contrast seen in \cite{Agundez2014}. Overall, apart from the temperature inversion included in \cite{Agundez2014}, we show qualitative agreements in terms of quenching and transport behaviors. We will discuss the implications of assuming uniform zonal winds (i.e. pseudo-2D model) in Section \ref{sec:pseudo2D} in more detail. 





\begin{figure*}[!ht]
   \centering
   \includegraphics[width=\columnwidth]{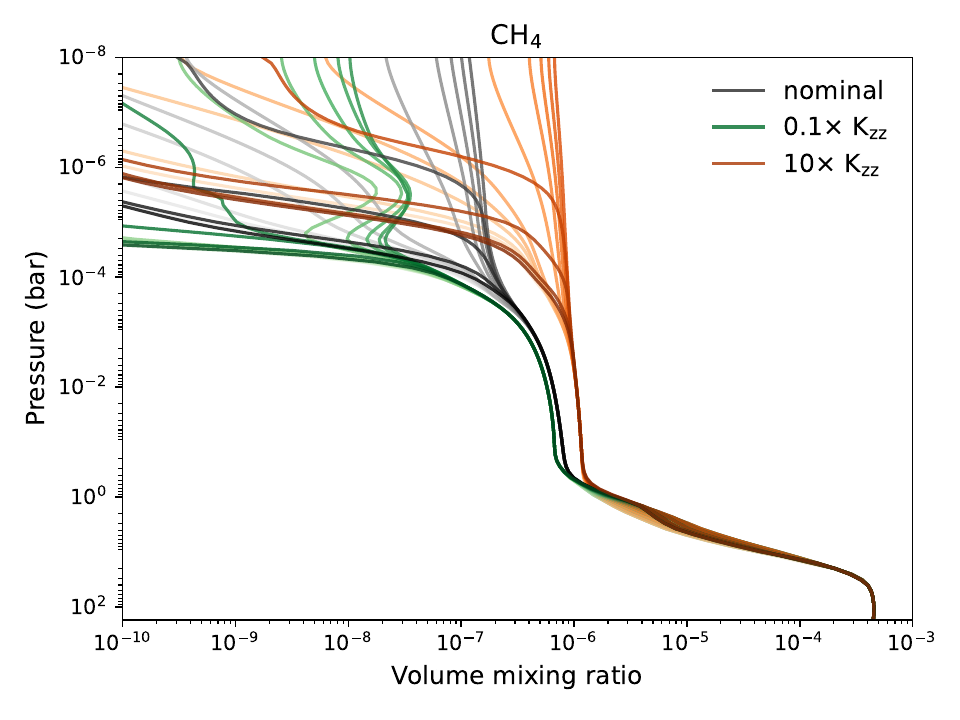}
   \includegraphics[width=\columnwidth]{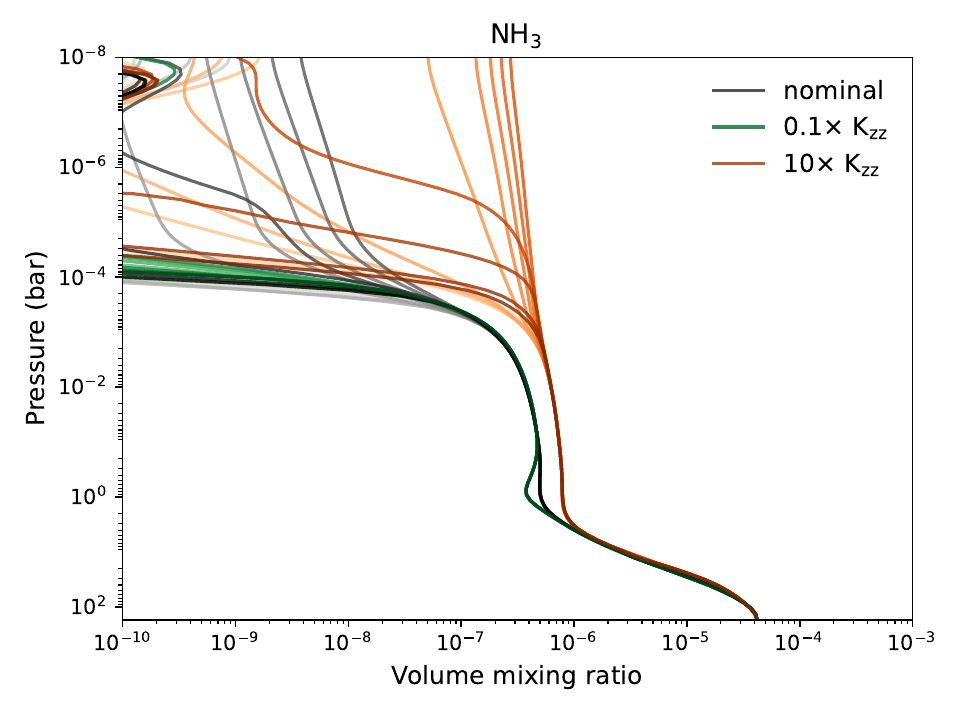}
\caption{\ce{CH4} and \ce{NH3} vertical profiles in the equatorial regions of HD 189733 b, simulated with 10 times stronger (orange) and weaker (green) eddy diffusion. The shades of colors from light to dark correspond to longitudes from the substellar point (0$^{\circ}$) through nightside and back around to the substellar point (360$^{\circ}$).}
\label{fig:2D-pasta-Kzz}
\end{figure*}

\begin{table}
\begin{tabular}{ c|c|c|c } 
species & 0.1$\times$ $K_{\textrm{zz}}$ & nominal & 10$\times$ $K_{\textrm{zz}}$ \\
\hline
\ce{CH4} & 0.067 & 0.14 & 0.22\\ 
\ce{NH3} & 0.14 & 0.28 & 0.57 \\ 
\end{tabular}
\caption{The pressure levels (mbar) where the morning and evening abundances begin to deviate by more than 50\%.}\label{tab:am-pm}
\end{table}

\subsection{Sensitivity to $K_{\textrm{zz}}$}\label{sec:Kzz}
While VULCAN has the capability to employ vertical advection instead of eddy diffusion \citep{Tsai2021}, it is numerically more stable to employ diffusion in the vertical direction in conjunction with molecular diffusion. One apparent caveat is that the parameterization of vertical mixing with eddy diffusion has been a long-standing uncertainty in atmospheric modeling \citep[e.g., ][]{Smith1998,Parmentier2013,Zhang2018a,Komacek2019}. To test the sensitivity to uncertainties in estimating the eddy diffusion coefficient ($K_{\textrm{zz}}$), we vary the eddy diffusion coefficient profile by an order of magnitude in our model of HD 189733 b, both increasing and decreasing it. This explored range roughly spans the uncertain range considered in the previous literature \citep{Smith1998,Parmentier2013}.


We examine the effects of varying vertical eddy diffusion coefficients on \ce{CH4} and \ce{NH3}, the major carbon and nitrogen species that exhibit transport-induced disequilibrium, in Figure \ref{fig:2D-pasta-Kzz}. Stronger vertical mixing makes the transition to a vertical-mixing-dominated region occur at a higher pressures, as expected. This allows morning-evening asymmetries to emerge at somewhat higher pressures when horizontal transport becomes less efficient compared to vertical mixing. The pressure levels where morning-evening asymmetry becomes notable are generally between 1 and 0.05 mbar, as summarized in Table \ref{tab:am-pm}. Since the equilibrium abundances of \ce{CH4} and \ce{NH3} monotonically increase with pressure in our HD 189733 b model (also see the discussion in 4.3.2. in \cite{Tsai2021} and \cite{Fortney2020}), increased vertical mixing also results in slightly higher quenched abundances, while weaker mixing leads to lower abundances from horizontal quenching from the dayside. Other hydrocarbons and HCN also follow similar trends in response to mixing processes as \ce{CH4} and \ce{NH3}. Despite these effects, the global distribution of main species only mildly depends on the choice of $K_{\textrm{zz}}$ within the explored range. For HD 189733 b and HD 209458b, the bulk of the atmosphere below the millibar level still remains in the horizontal-transport-dominated regime, even when considering strong eddy diffusion.




\begin{figure}[!ht]
   \centering
\includegraphics[width=\columnwidth]{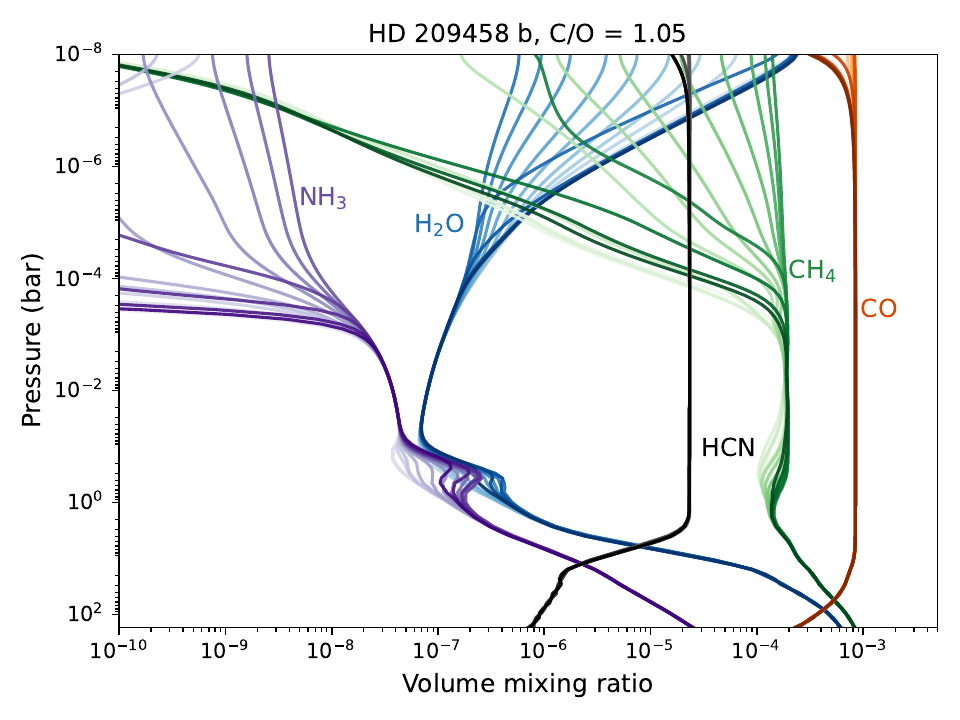}
\includegraphics[width=\columnwidth]{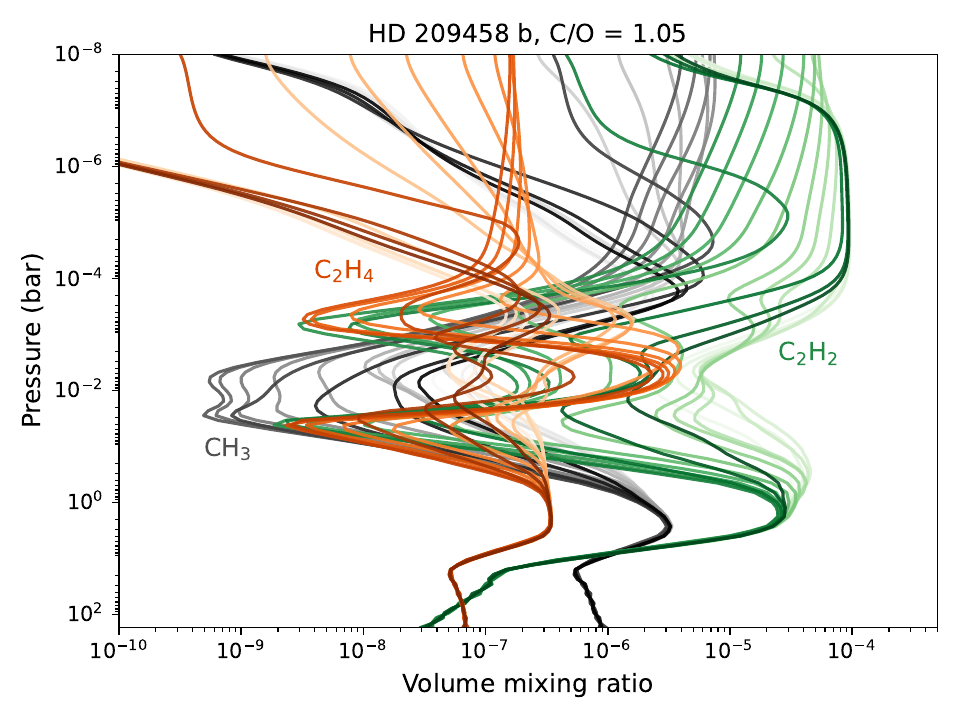}
\includegraphics[width=\columnwidth]{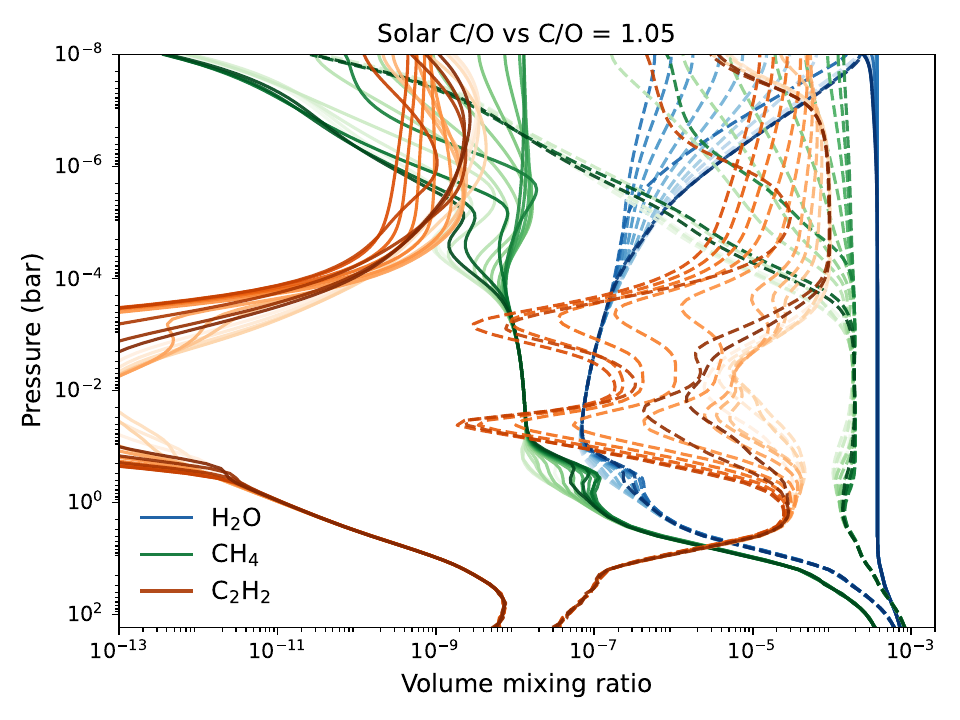}
\caption{The top and middle panels show the vertical profiles in the equatorial regions as Figure \ref{fig:2D_pasta} but for HD 209458 b with with C/O = 1.05. The bottom panel compares the key abundance distributions between solar C/O (solid; with C/O $\simeq$ 0.45) and C/O = 1.05 (dashed).}
\label{fig:HD209-solar-CtoO105}
\end{figure}

\begin{figure}[!ht]
   \centering
\includegraphics[width=\columnwidth]{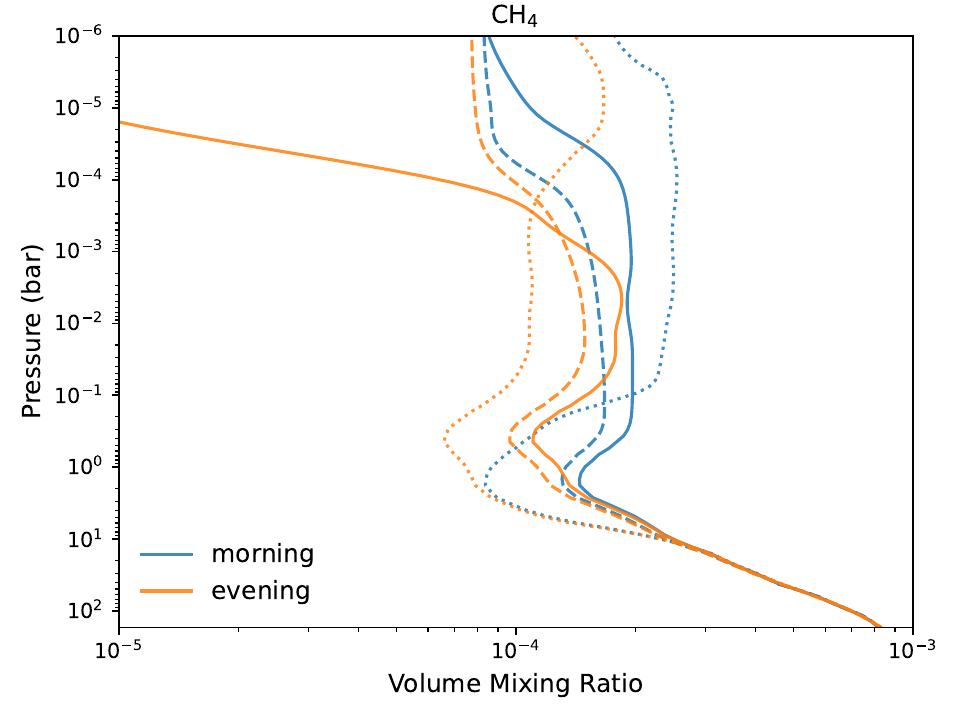}
\includegraphics[width=\columnwidth]{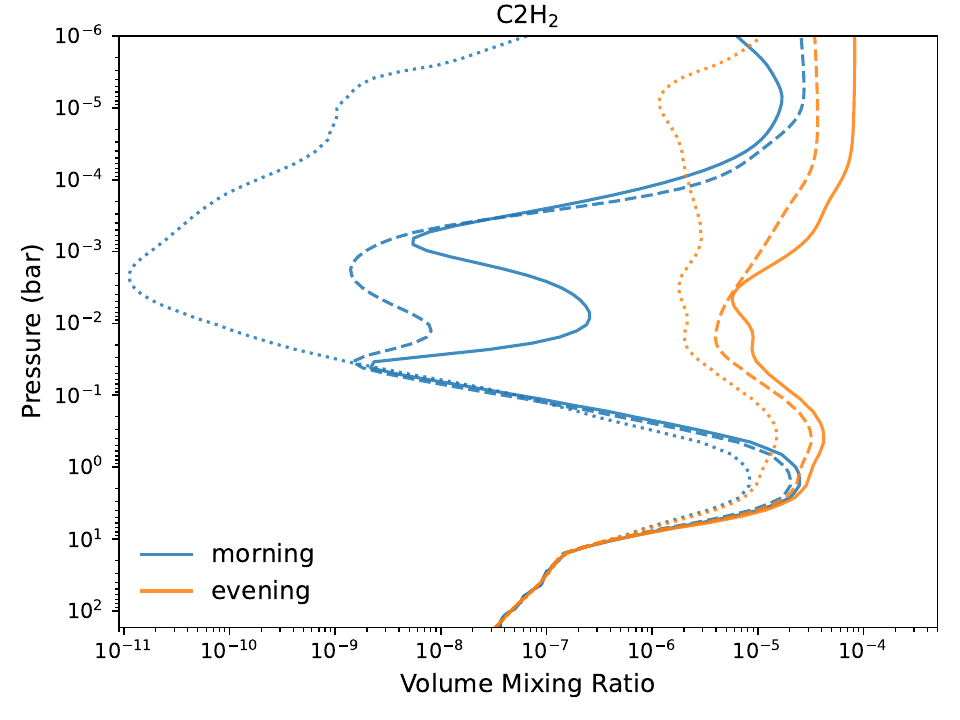}
\includegraphics[width=\columnwidth]{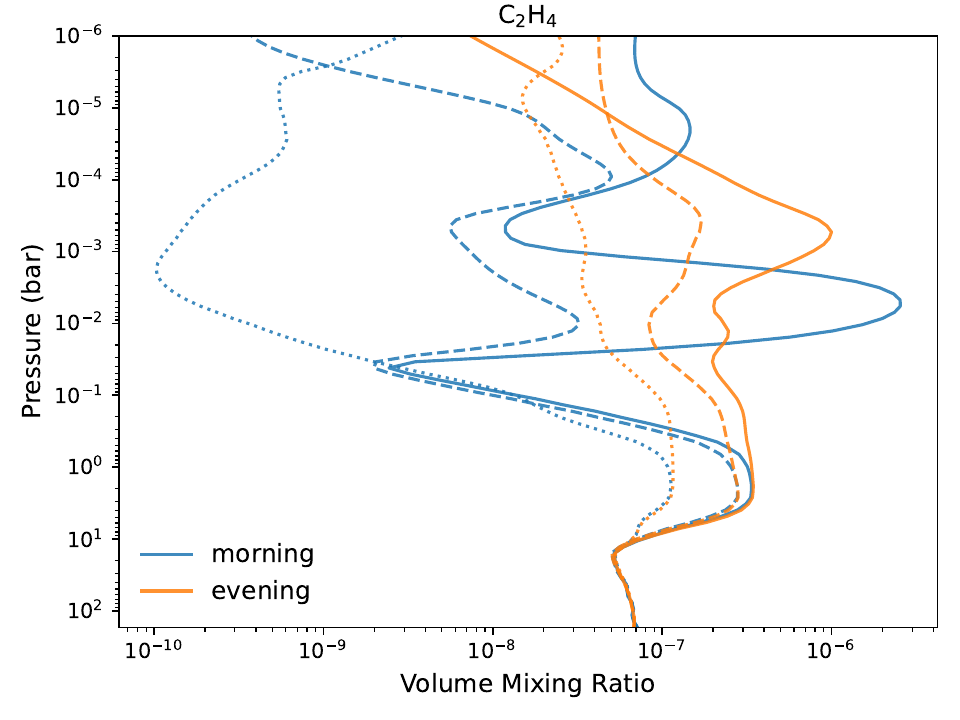}
\caption{The vertical profiles of \ce{CH4}, \ce{C2H2}, and \ce{C2H4} on the morning and evening limbs in our 2D model of HD 209458 b with C/O = 1.05. The distributions without zonal wind are shown in dashed lines, while chemical equilibrium abundances are indicated by dotted lines.}
\label{fig:HD209-CtoO105-am-pm}
\end{figure}

\subsection{Super-solar C/O of HD 209458 b}
\subsubsection{background}
The transit observations using high-resolution spectroscopy of HD209458 b reported detection of \ce{H2O}, CO, HCN, \ce{CH4}, \ce{NH3}, and \ce{C2H2} molecules at statistically significant levels \citep{Giacobbe2021}. Notably, the presence of \ce{CH4}, \ce{C2H2} and HCN are indicative of super-solar carbon-to-oxygen ratio (C/O), consistent with the previous high-resolution analysis that also detected HCN \citep{Hawker2018}. In contrast, recent JWST/NIRCam transmission spectrum of this planet reported non-detections of \ce{CH4}, HCN, and \ce{C2H2} with upper limits provided \citep{Xue2023}, interpreted to have sub-solar C/O based on chemical equilibrium. 

While the interpretation by \cite{Giacobbe2021} is driven by cross-correlating with grids of models also assuming thermochemical equilibrium, \ce{CH4} is influenced by quenching while \ce{C2H2} and HCN can be efficiently produced by photochemistry out of equilibrium. In addition to the nominal HD 209458 b model with solar metallicity and C/O presented in this section, we run a 2D VULCAN model for HD 209458 b with super-solar C/O to explore the chemical transport under the alternative scenario of a carbon-rich atmosphere. We follow the best-fit C/O value of 1.05 as determined by \cite{Giacobbe2021} and keep the same solar metallicity for comparison with our nominal model, since metallicity is not well constrained in \cite{Giacobbe2021}.


\subsubsection{Enhanced hydrocarbon and horizontal transport induced limb asymmetry}\label{sec:HD209_CtoO105}
The top and middle panels of Figure \ref{fig:HD209-solar-CtoO105} depict the same mixing-ratio distributions as in Figure \ref{fig:2D_pasta} but for C/O = 1.05. When C/O exceeds unity, \ce{H2O} loses its dominance to \ce{CH4} due to the lack of available oxygen after CO \citep{Madhusudhan2012,Moses2013,Heng2016}. What is intriguing is the production of water resulting from the photolysis of CO in the dayside upper atmosphere above 0.1 mbar. The same process also operates in a solar C/O condition \citep{Moses2011}, but the increase of \ce{H2O} is more pronounced here, owing to its lower equilibrium composition. Compared to solar C/O, \ce{C2H2} abundances are significantly enhanced. Moreover, \ce{C2H2} exhibits substantial zonal variations, spanning several orders of magnitude across the equator between 1 bar and 0.1 mbar level (bottom panel of Figure \ref{fig:HD209-solar-CtoO105}).   

Although \ce{CH4} is rather uniform below 10$^{-4}$ bar across the planet, several hydrocarbons show notable compositional gradients in the zonal direction, as seen in Figures \ref{fig:HD209-solar-CtoO105} and \ref{fig:HD209-CtoO105-am-pm}. \ce{C2H2} abundance on the evening limb is about 100--1000 times higher than that on the morning limb, whereas \ce{C2H4} peaks between 0.1 and 10$^{-3}$ bar on the morning limb with strong vertical variations. The morning-evening limb asymmetry in \ce{C2H2} is driven by the temperature difference and manifested due to its relatively short chemical timescale. \ce{C2H2} is favored at the hotter evening limb, where the temperature is about 200--300 K higher than the morning limb, as evident in the profiles without zonal wind in Figure \ref{fig:HD209-CtoO105-am-pm}. In this region, the main destruction pathway for \ce{C2H2} proceeds through a sequence of unsaturated hydrocarbons:
\begin{eqnarray}
\begin{aligned}
\ce{C2H2 + H &->[M] C2H3}\\
\ce{C2H3 + H2  &-> C2H4 + H}\\
\ce{C2H4 + H &->[M] C2H5}\\
\ce{C2H5 + H &-> CH3 + CH3}\\
2(\ce{CH3 + H2 &-> CH4 + H})\\
\hline 
\noalign{\vglue 3pt}
\mbox{net} : \ce{C2H2 + 3H2 &-> 2CH4}.
\end{aligned}
\label{re:c2h2-path}
\end{eqnarray}
The timescale of \ce{C2H2} can be estimated from the rate-limiting steps, either the formation of \ce{CH3} or \ce{C2H3} in the above pathway. At 1 mbar pressure level, the lifetime of \ce{C2H2} on the nightside ranges from 10$^3$ to 10$^5$ seconds. This timescale is comparable to that of horizontal transport (Figure \ref{fig:2D_tau}), explaining the longitudinal compositional gradient \ce{C2H2} displays.

Although \ce{C2H4} also has a higher equilibrium abundance on the warming evening limb, photochemistry and horizontal transport lead to an accumulation of \ce{C2H4} on the cooler morning limb instead. The combining effects result in a peak \ce{C2H4} distribution on the morning limb greater than the abundance on the evening limb, in contrast with the lower morning \ce{C2H4} abundance predicted in the absence of zonal transport (bottom panel of Figure \ref{fig:HD209-CtoO105-am-pm}). The photolysis of \ce{CH4} on the dayside produced methyl radical (\ce{CH3}), the precursor to produce other hydrocarbons. \ce{CH3} flow into the nightside where the lower temperature promotes the formation of \ce{C2H6}, \ce{C2H5}, and \ce{C2H4}, since the combination of \ce{CH3} into \ce{C2H6} is exothermic and kinetically favored at cooler temperatures. With the aid of horizontal transport, the photochemically produced \ce{CH3} is able to transport to the nightside to initiate the production of hydrocarbon species. This hydrocarbon production on the nightside is then carried to the morning limb by the zonal wind, leading to the peak shown in \ce{C2H4}. Similar behavior is found in \ce{C2H6} distribution as well, but at significantly lower abundances. For completeness, the equatorial abundance distribution as a function of the longitude and pressure of several key species can be found in Figure \ref{fig:HD209-CtoO105-contours}.

\section{When does the assumption of a uniform zonal jet in pseudo-2D models break down?}\label{sec:pseudo2D}
The equatorial jet is a robust feature for tidally-locked planets that receive steady day-night thermal forcing \citep{Heng2015,Showman2020}. However, the equatorial jet transitions to a day-to-night flow when the radiative timescale or drag timescale becomes short \citep{Showman2015,Komacek2019}. In the case of cooler sub-Neptunes or nonsynchronously rotating planets, the zonal wind in the equatorial region can also develop a more complex structure with winds changing directions across varying pressure levels \citep{Showman2015,Carone2020,Charnay2020,Innes2022}. Here, we examine how horizontal transport changes on a hot Jupiter with strong frictional drag dominated by a day-to-night flow. We adopted the output of T$_{\textrm{eq}}$ = 1600 K with strong drag ($\tau$ = 10$^4$ s) from \cite{Tan2019} as our fiducial atmosphere with a day-to-night flow. The equatorial thermal and wind structures of our fiducial day-to-night circulation and those of HD 189733 b are compared in Figure \ref{fig:T1600-HD189-eq}. Our goal is to determine when a pseudo-2D approach with uniform flow remains valid and when it breaks down.


\begin{figure}[ht!]
   \centering
   \includegraphics[width=\columnwidth]{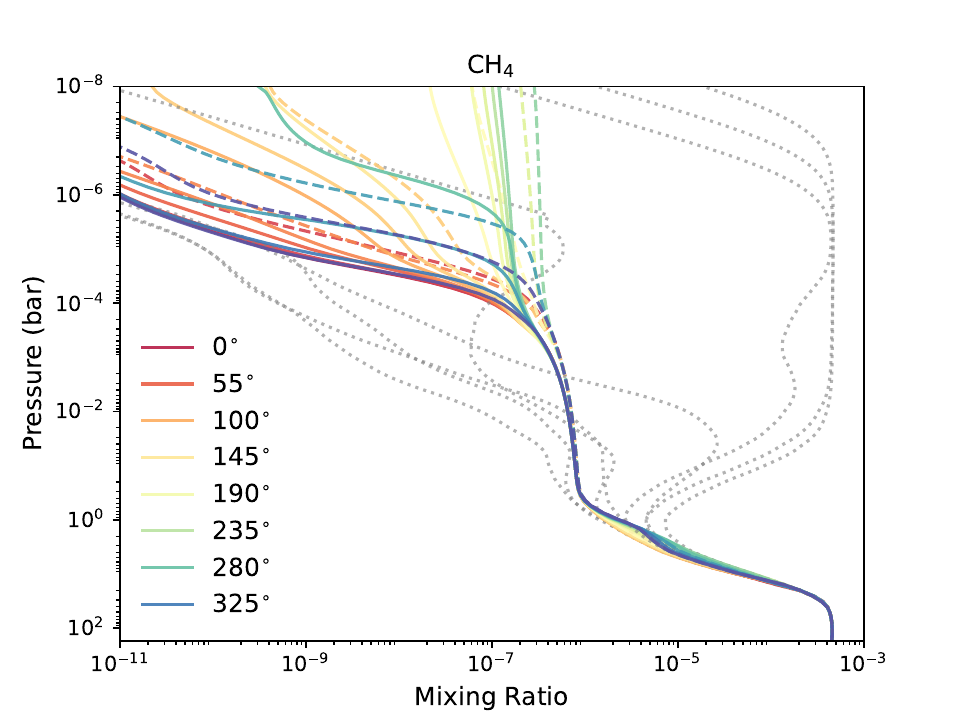}
        \caption{The mixing ratio profiles of \ce{CH4} across different longitudes computed from our nominal HD 189733 b model with pressure- and longitude- dependent zonal winds (solid) and from the pseudo-2D model with uniform zonal winds (dashed). We adopt a mean zonal wind velocity of 1736 m/s at 1 bar in the nominal model for the uniform wind speed in the pseudo-2D model, following the same approach as in \cite{Agundez2014}.}
\label{fig:HD189-2D-pseudo}
\end{figure}

\subsection{Comparisons within the superrotating regime} 
The equatorial jets induced by the stationary day-night heating in our HD 189733 b and HD 209458 b models resemble Gaussian-like vertical profiles, gradually diminishing towards zero in both the upper and deeper layers of the atmosphere, as depicted in Figure \ref{fig:GCMs}. We use HD 189733 b as an example of circulation characterized by equatorial superrotation for the comparison between 2D and pseudo-2D approaches. For the uniform zonal wind, the pseudo-2D model needs to adopt either the peak jet speed \citep{Agundez2014} or the averaged wind velocity over the jet region \citep{Baeyens2021}. For our pseudo-2D model of HD 189733b, we adopt a mean zonal wind of 1736 m/s at 1 bar from the GCM as the uniform zonal wind.

Figure \ref{fig:HD189-2D-pseudo} shows the comparison of \ce{CH4} distributions, demonstrating differences caused by the simplified transport in the pseudo-2D approach verse the 2D model. In the deep region where thermochemical equilibrium holds, the actual wind pattern is not effectively relevant. The \ce{CH4} distribution transitions from a horizontal homogenized regime to a vertical mixing-dominated regime at the same pressure level around 10$^{-4}$ bar in both models. In the upper atmosphere above 10$^{-4}$ bar, the pseudo-2D model has slightly slower horizontal transport (Figure \ref{fig:2D_tau}) and predicts slightly more vertically mixed profiles compared to those in our nominal 2D model, where stronger winds around 1 mbar level at certain longitudes can alter the distribution. Similar trends are seen in other species as well. Despite these minor differences, we find the pseudo-2D approach to be a valid assumption when a broad superrotation jet is present.



\begin{figure*}[ht!]
   \centering
   \includegraphics[width=\columnwidth]{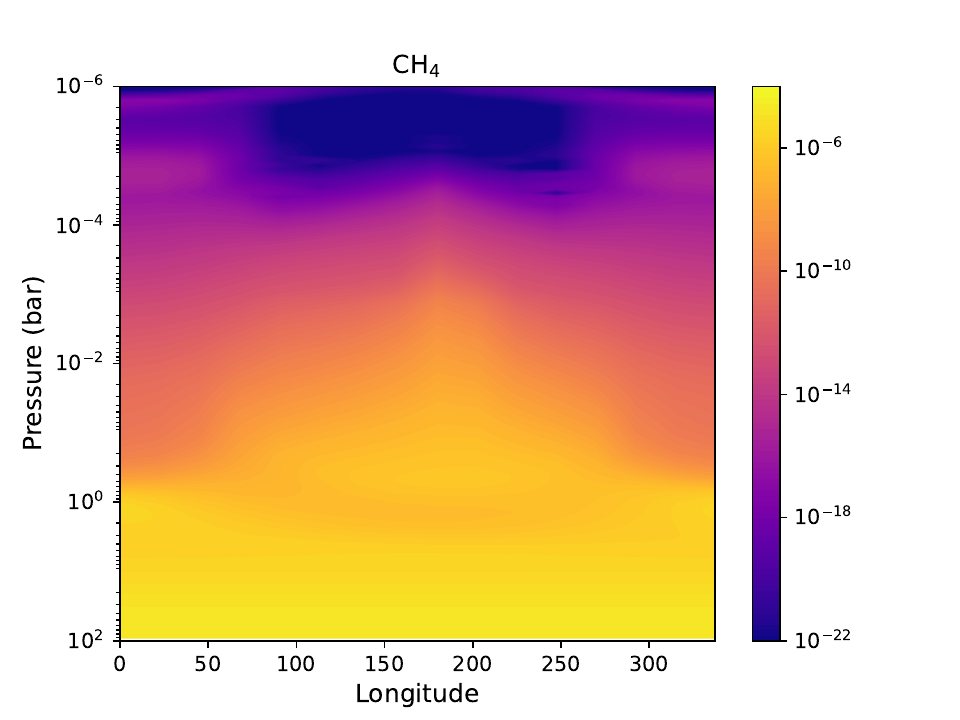}
   \includegraphics[width=\columnwidth]{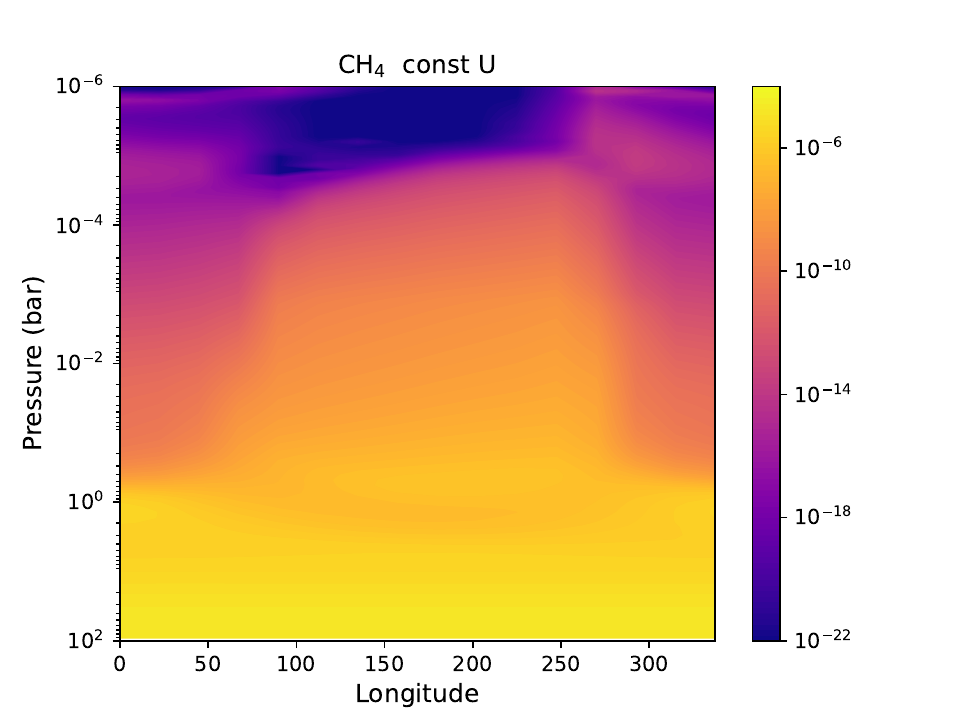}
   \includegraphics[width=\columnwidth]{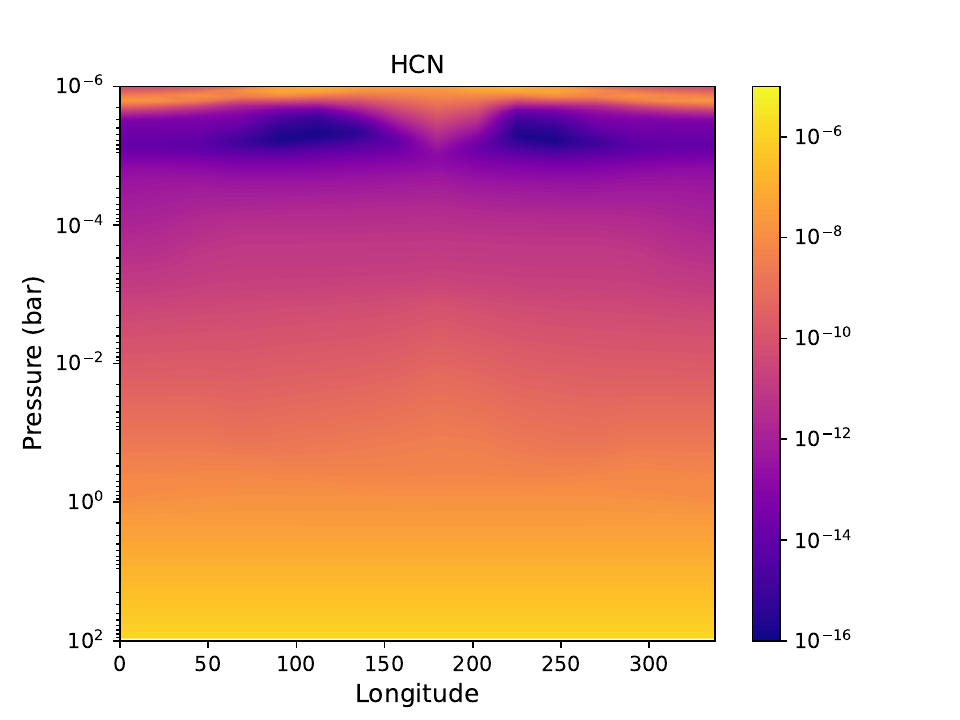}
   \includegraphics[width=\columnwidth]{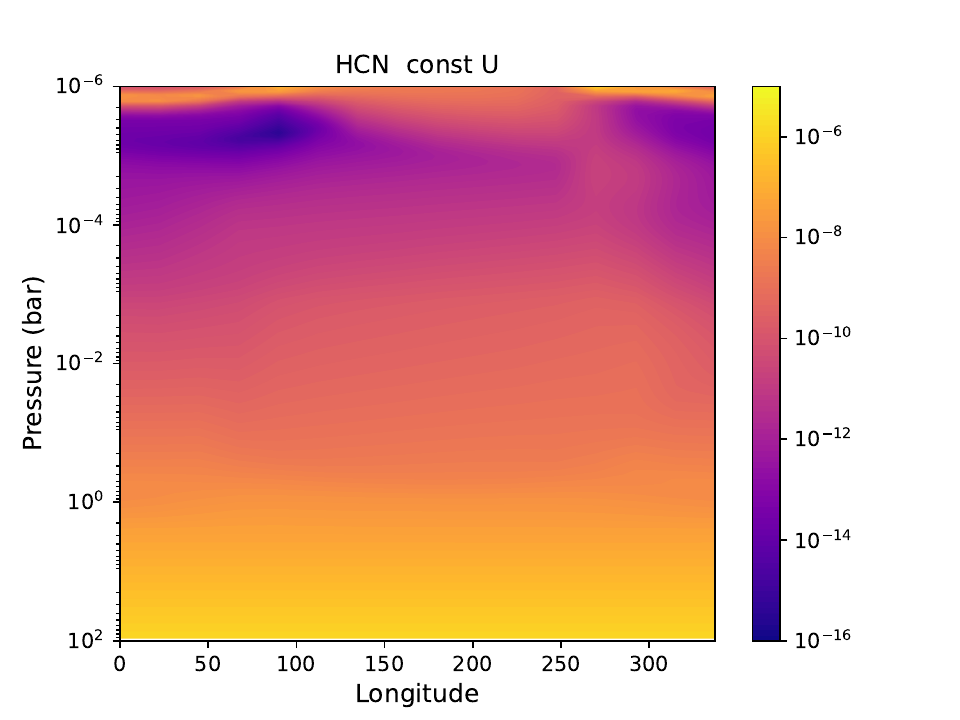}
        \caption{The equatorial abundance distribution of \ce{CH4} and HCN. The left column shows the nominal model and the right column shows the model assuming uniform zonal to simulate the pseudo-2D method. The substellar point is located at 0$^{\circ}$ longitude.}
\label{fig:Teq1600-2D}
\end{figure*}

\begin{figure*}[ht!]
   \centering
   \includegraphics[width=\columnwidth]{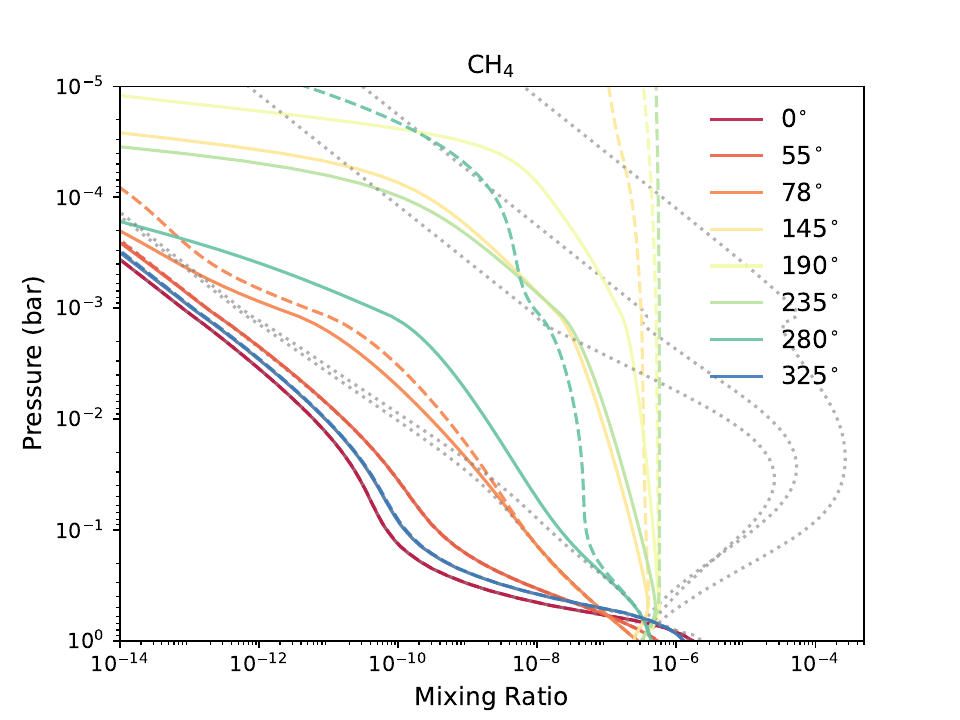}
   \includegraphics[width=\columnwidth]{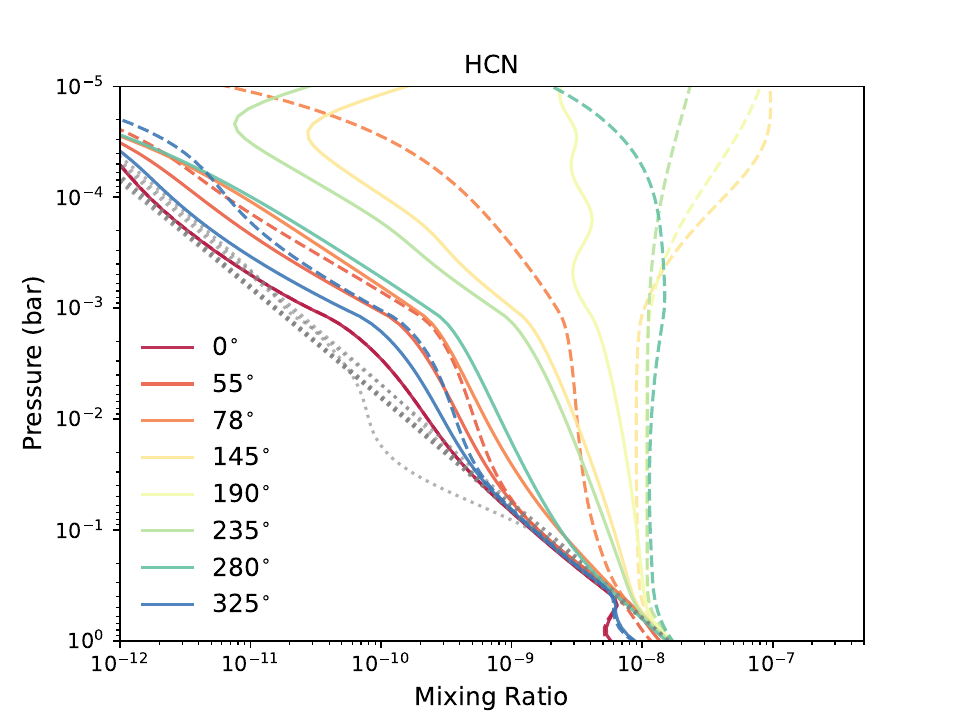}
        \caption{Mixing ratio profiles at several different longitudes in our fiducial ultra-hot Jupiter models, with the nominal day-to-night flow (solid), uniform zonal winds (dashed), and chemical equilibrium (grey dotted lines). The pressure domain is zoomed in to the most observable region between 1 and 10$^{-5}$ bar for clarity.}
\label{fig:Teq1600-pasta}
\end{figure*}

\subsection{comparisons within the day-to-night flow regime}
Next, we explore the scenario of a hotter hot Jupiter with strong radiative \citep[e.g., ][]{Komacek2016} or magnetic drag \citep[e.g., ][]{Perna2010}, where the global circulation has changed from equatorial superrotation to day-to-night flow \citep{Kempton2012,Showman2015,Tan2019}. Figure \ref{fig:Teq1600-2D} compares the abundance distributions of \ce{CH4} and HCN, the two main molecules that display compositional gradients, in the equatorial region of the fiducial strong-drag hot Jupiter atmosphere with $T_{\textrm{eq}}$ = 1600 K simulated by our nominal 2D model and pseudo-2D model. It is evident that the day-night flow leads to a symmetrical distribution, differing from the distribution governed by uniform zonal winds. This discrepancy is most pronounced around the morning limb, owing to the distinctive transport dynamics at play. In the nominal model, nightside-to-morning-limb advection occurs, contrasting with the morning-limb-to-nightside advection in the model with uniform zonal winds. This circulation pattern also enables more efficient transport of atomic H produced photochemically on the dayside to the morning limb. In the nominal 2D model, the \ce{CH4} and HCN abundances remain low on both morning and evening limbs. Conversely, when assuming uniform eastward winds, \ce{CH4} and HCN exhibit higher abundances around the morning limb due to nightside transport. Similar to \ce{CH4}, species favored on the nightside over the dayside, such as \ce{NH3}, follow similar trends showing higher morning limb abundances in the uniform winds model as well.

The disparity in abundance profiles across longitudes from the nominal model and pseudo-2D model is further illustrated in Figure \ref{fig:Teq1600-pasta}. Except near the substellar point and at high pressures (P $\gtrsim$ 1 bar), this comparison demonstrates that for species susceptible to photochemistry, the assumption of uniform zonal winds in the pseudo-2D framework can yield orders of magnitude differences in the observable region (approximately 0.1 to 10$^{-5}$ bar) when the dominant circulation is featured by a day-to-night flow. The pseudo-2D approach is only suitable for a tidally locked atmosphere with moderate drag such that the circulation is still within the equatorial superrotation regime. 







\begin{figure}[ht!]
   \centering
   \includegraphics[width=\columnwidth]{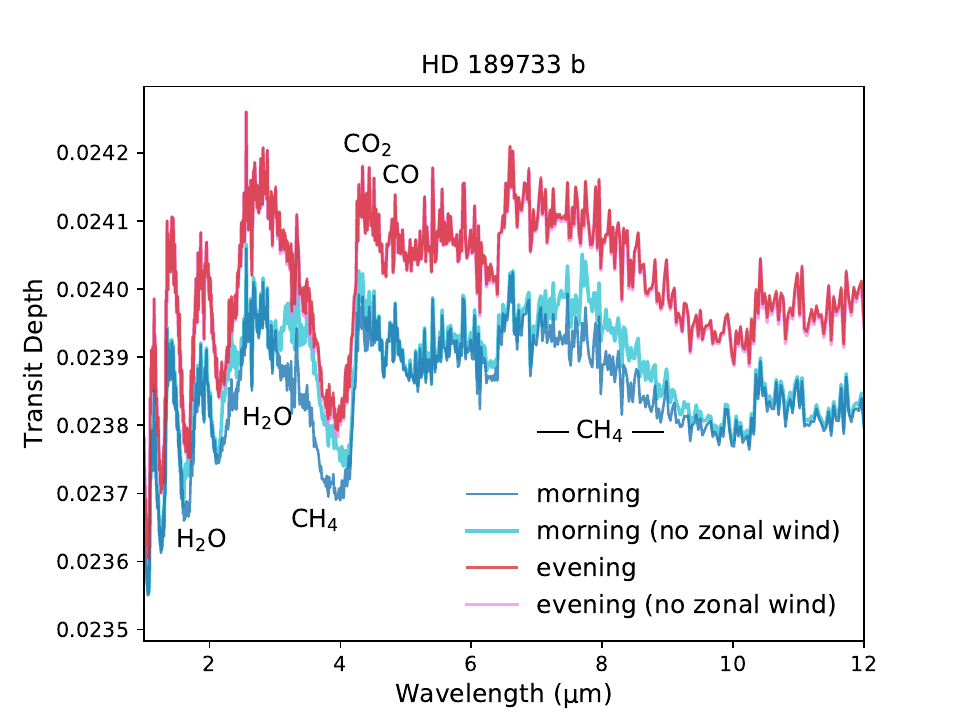}
   \includegraphics[width=\columnwidth]{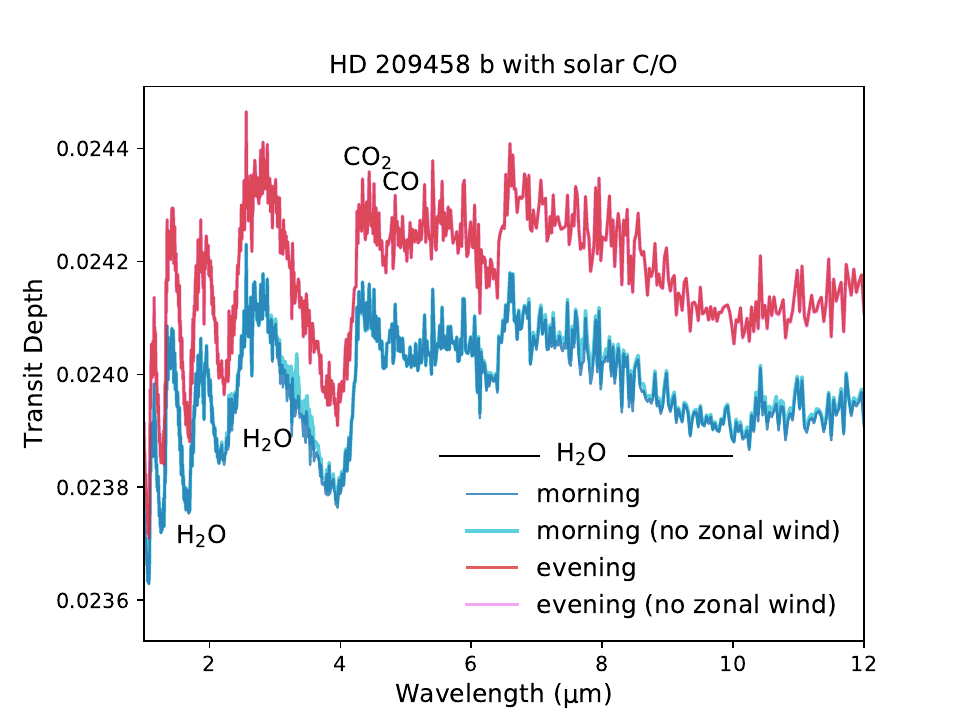}
   \includegraphics[width=\columnwidth]{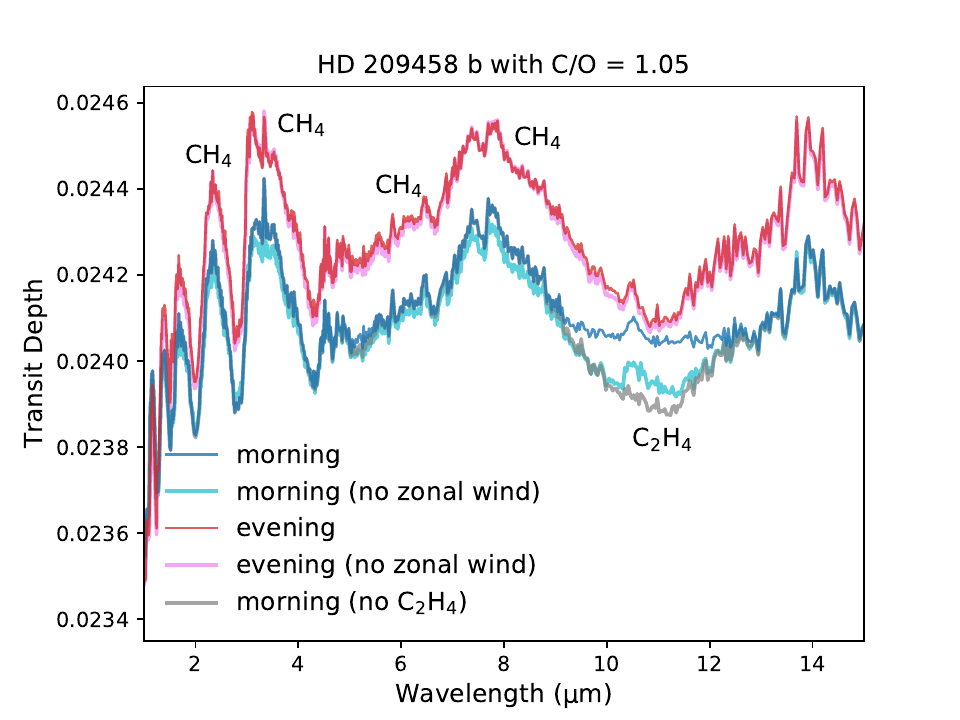}
        \caption{The synthetic transmission spectra of the evening and morning limbs of HD 189733 b and HD209458 b (for both solar and super-solar C/O). The atmospheric structure is derived from MITgcm simulation \citep{Parmentier2013} with the chemical composition computed by 2D VULCAN. The models excluding zonal transport are also shown for comparison.}
\label{fig:spectra}
\end{figure}

\section{Synthetic spectra (for the evening and morning limbs)}\label{sec:spectra}
We present synthetic transmission spectra for the morning and evening limbs based on our 2D model results. The transmission spectra are computed using PLATON \citep{Zhang2019,Zhang2020}. including opacity sources of \ce{CH4}, \ce{CO}, \ce{CO2}, \ce{C2H2}, \ce{H2O}, HCN, \ce{NH3}, \ce{O2}, \ce{NO}, \ce{OH}, \ce{C2H4}, \ce{C2H6}, \ce{H2CO}, \ce{NO2}, and collision-induced absorption (CIA) of \ce{H2}-\ce{H2} and \ce{H2}-He. We will focus on the observational impact of horizontal transport and the differences between the evening and morning limbs. It is worth noting that the distribution of clouds and hazes can also significantly influence limb asymmetry \citep{Kempton2017,Powell2019,Steinrueck2021,Savel2023}. For the scope of this study, we will leave the effects of clouds to future work and will solely delve into the chemical transport within our cloud-free models.

Figure \ref{fig:spectra} illustrates the synthetic transmission spectra with and without horizontal transport for HD 189733 b and HD 209458 b (including solar and super-solar C/O).
For HD 189733 b, 1D models (i.e. without horizontal transport) produce methane features on the cooler morning limb that are absent on the evening limb (Figure \ref{fig:HD189-4parts}). However, this compositional gradient is readily homogenized once the zonal transport is included. For HD 209458 b (solar C/O), methane abundance remains too low in both 1D and 2D models (below ppm level; see Figure \ref{fig:HD209-4parts}). Consequently, the influence of horizontal transport on the spectra of HD 209458 b with solar C/O is negligible. Instead, the dominant molecules that show up in the spectra are \ce{H2O}, \ce{CO}, and \ce{CO2}, all of which exhibit relatively uniform equilibrium abundances throughout the planet. As a result, these gases do not contribute to limb asymmetry, whether horizontal transport is in play or not. In the case of HD 209458 b with super-solar C/O, \ce{CH4} took over \ce{H2O} to make the strongest spectral features at 2.3, 3.3-3.9, 5.5-6.6, and 7--8.5 $\mu$m. As \ce{CH4} becomes the predominant carbon-bearing molecule, it also reaches a uniform distribution across the entire planet. The major limb asymmetry is the presence of \ce{C2H4} absorption on the morning limb but not on the evening limb, due to the horizontal transport of \ce{C2H4} from the nightside to the morning limb.









\section{Discussions and Conclusions}
In this paper, we present the 2D version of the photochemical model VULCAN. We first validate VULCAN 2D with analytical solutions. VULCAN 2D successfully reproduces the special case equivalent to the pseudo-2D approach with uniform winds and also demonstrates consistent results with a 3D GCM \citep{Mendonca2018a}. We use limiting cases to demonstrate the distinct effects of the vertical and horizontal mixing processes. For canonical hot Jupiters, such as HD 189733 b and HD 209458 b, we find most of the atmosphere below 1--0.1 mbar is within the horizontal transport-dominated region, where zonal advection prevails over vertical mixing. In the upper atmosphere above this region, photochemistry and vertical mixing control the composition. We explore the sensitivity to the parametrization of vertical mixing and find a mild dependence in the abundance distribution of our hot Jupiter models. We note that stronger vertical mixing can, in principle, promote morning-evening asymmetry.  

For HD 189733 b, the morning-evening limb asymmetry in \ce{CH4} predicted by 1D models is readily homogenized when horizontal transport is included. For HD 209458 b with solar C/O, the transmission spectra exhibit no limb asymmetry attributed to the composition due to the paucity of \ce{CH4}. However, with super-solar C/O, horizontal transport results in notable limb asymmetries in hydrocarbons (\ce{C2H4} in this case). For atmospheres with circulation dominated by an equatorial jet, we show that the pseudo-2D (rotation 1D column) approach can reasonably capture the transport, but the assumption of uniform flow breaks down for day-to-night circulations under stronger drag and pseudo-2D models can yield observable differences (0.1 to 10$^{-5}$ bar) orders of magnitude apart, particularly for species susceptible to photochemistry or with inherent compositional gradient in equilibrium.

The 2D modeling framework highlights the need to consider both horizontal and vertical transport when interpreting the compositions from transmission observations probing the limbs. The 2D framework developed here bridges the gap between traditional 1D photochemical kinetics models and 3D general circulation models that typically exclude chemical kinetics. Future directions include incorporating sulfur chemistry and applying the model to the meridional plane or tidally locked coordinate \citep{Koll2015} to explore the role of overturning circulation. 

In this work, we did not explore the effects of vertical advection. The upward or downward transport can lead to significantly different distributions, as already demonstrated in the ammonia distribution on Jupiter by a 1D model \citep{Tsai2021}. \citet{Hammond2021} also showed how slow overturning winds could transport much more heat than fast zonal jets on tidally locked planets with weak temperature gradients. A future model development should represent both of these processes.


By elucidating how the atmosphere composition is regulated by global circulation, this 2D modeling approach will pave the way for self-consistent and more comprehensive models and provide a useful tool to enhance the capacity of 3D GCMs for interpreting observations.



\medskip
Part of this work is supported by the European community through the ERC advanced grant EXOCONDENSE (\#740963; PI: R.T. Pierrehumbert). S.-M.T. acknowledges support from NASA Exobiology Grant No. 80NSSC20K1437 and the University of California at Riverside. X.Z. acknowledges support from the NASA Exoplanet Research Grant 80NSSC22K0236 and the NASA Interdisciplinary Consortia for Astrobiology Research (ICAR) grant 80NSSC21K0597. Financial support to R.D. was provided by a Natural Sciences and Engineering Research Council of Canada (NSERC) Discovery Grant to C.Goldblatt. 

%
%
\bibliographystyle{aa}
\bibliography{master_bib.bib}

\begin{appendix}

\begin{figure}[ht!]
   \centering
\includegraphics[width=0.49\columnwidth]{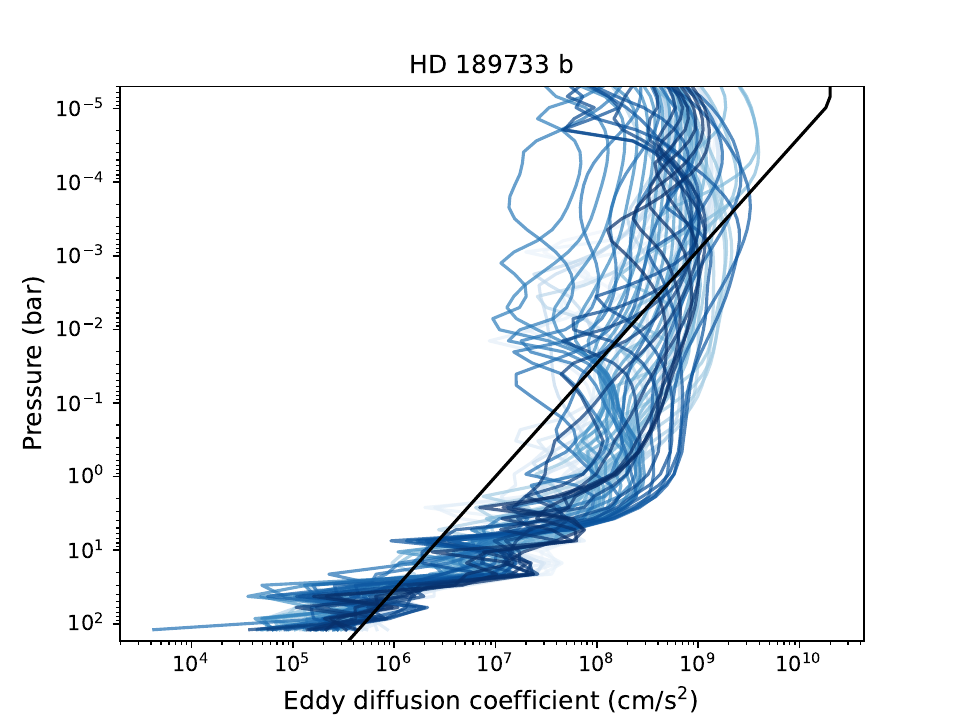}
\includegraphics[width=0.49\columnwidth]{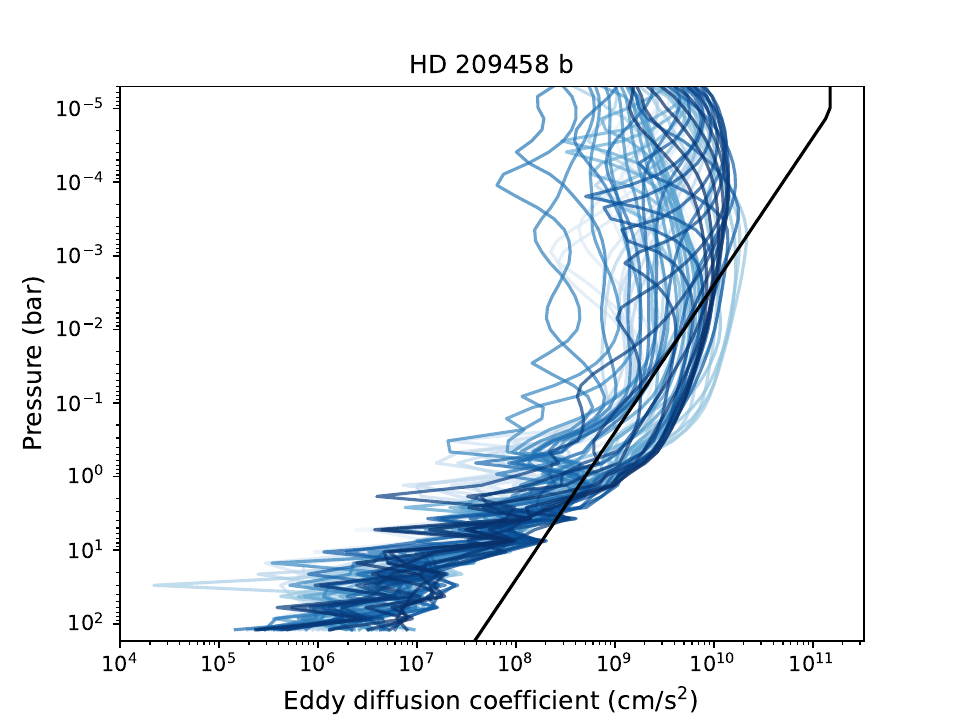}
\caption{The eddy diffusion coefficient profiles derived from \cite{Agundez2014} (black) of HD 189733 b and HD 209458 b compared with the root-mean-squared vertical wind multiplied by 0.1 scale height from the GCM in this study (blue).}
\label{fig:w_Kzz}
\end{figure}

\begin{figure}[!htp]
   \centering
   \includegraphics[width=0.495\columnwidth]{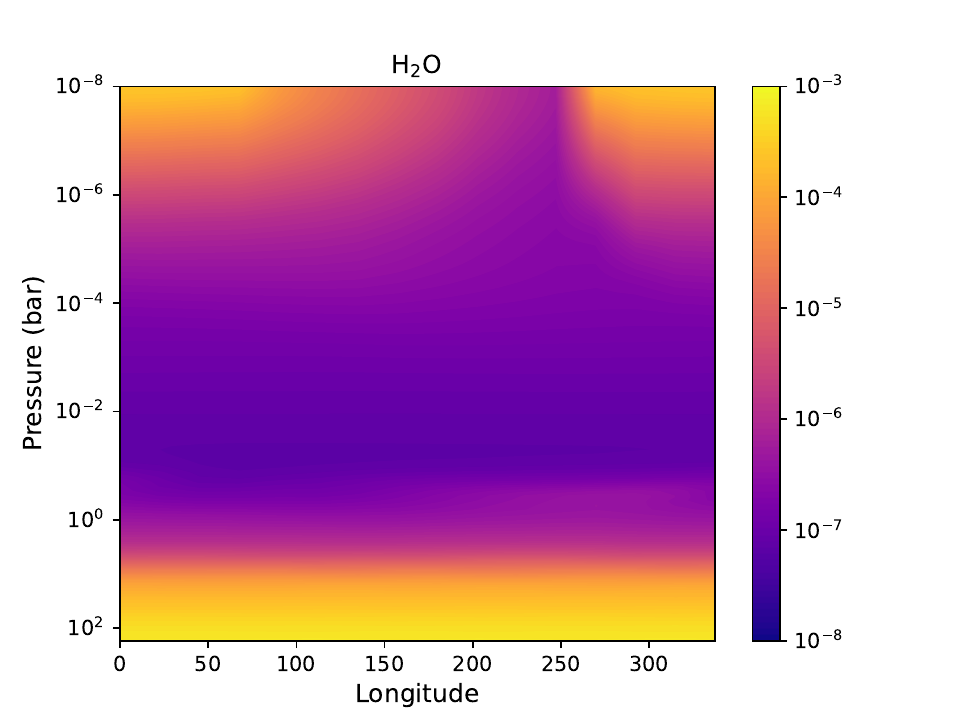}
   \includegraphics[width=0.495\columnwidth]{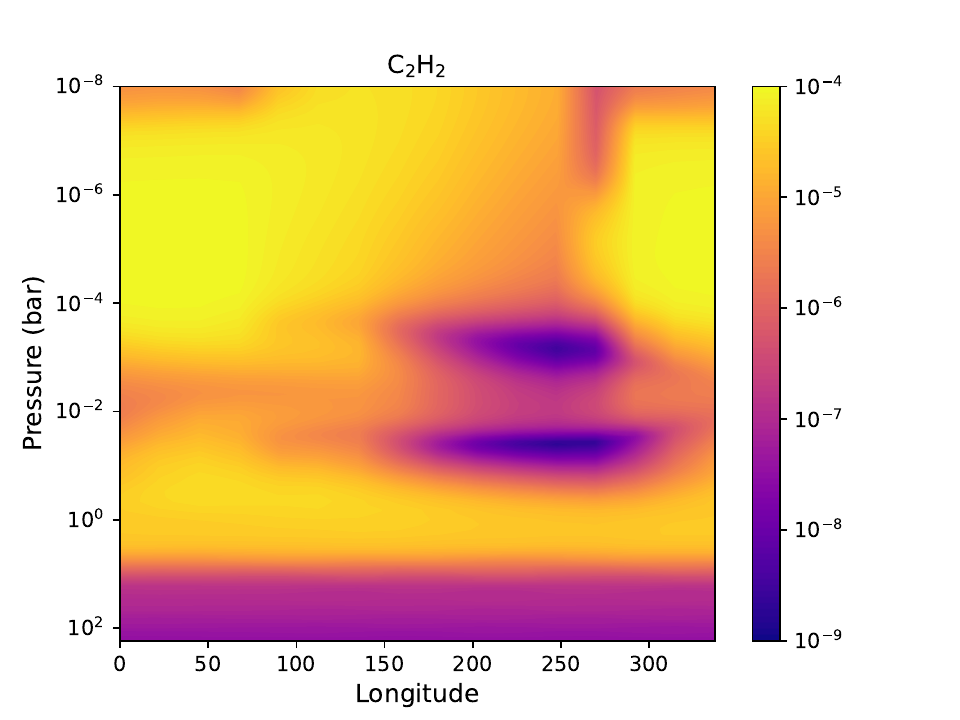}
   \includegraphics[width=0.495\columnwidth]{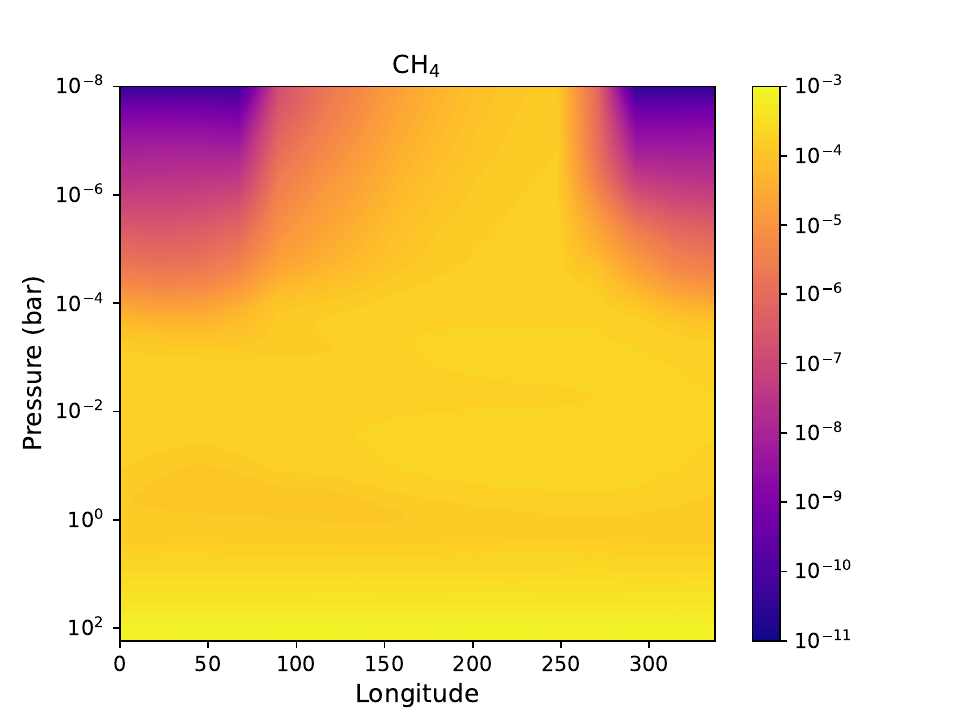}
   \includegraphics[width=0.495\columnwidth]{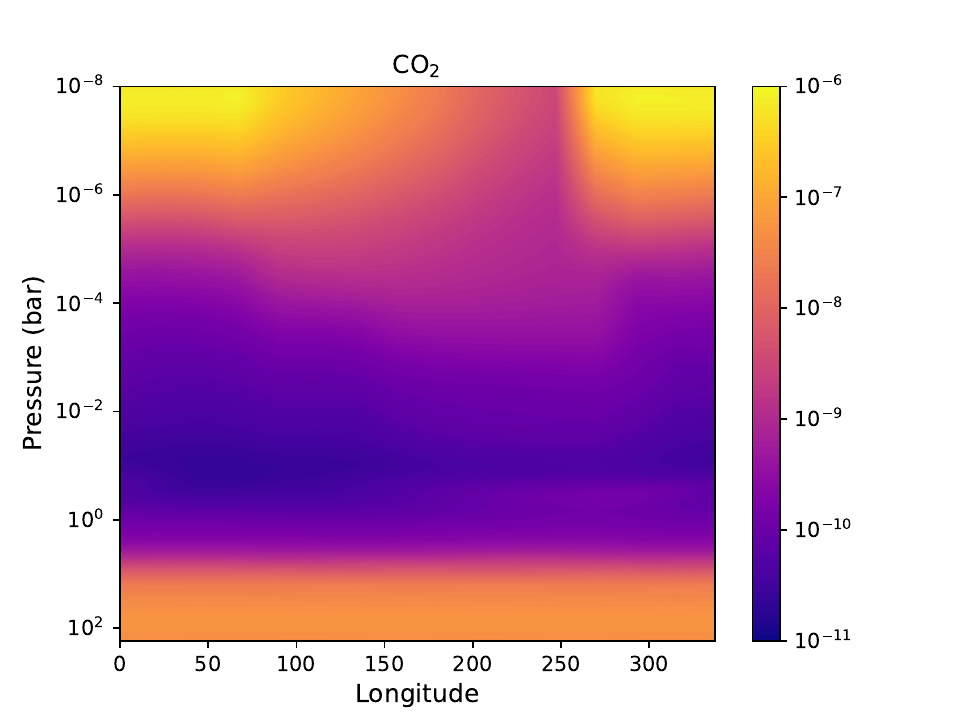}
   \includegraphics[width=0.495\columnwidth]{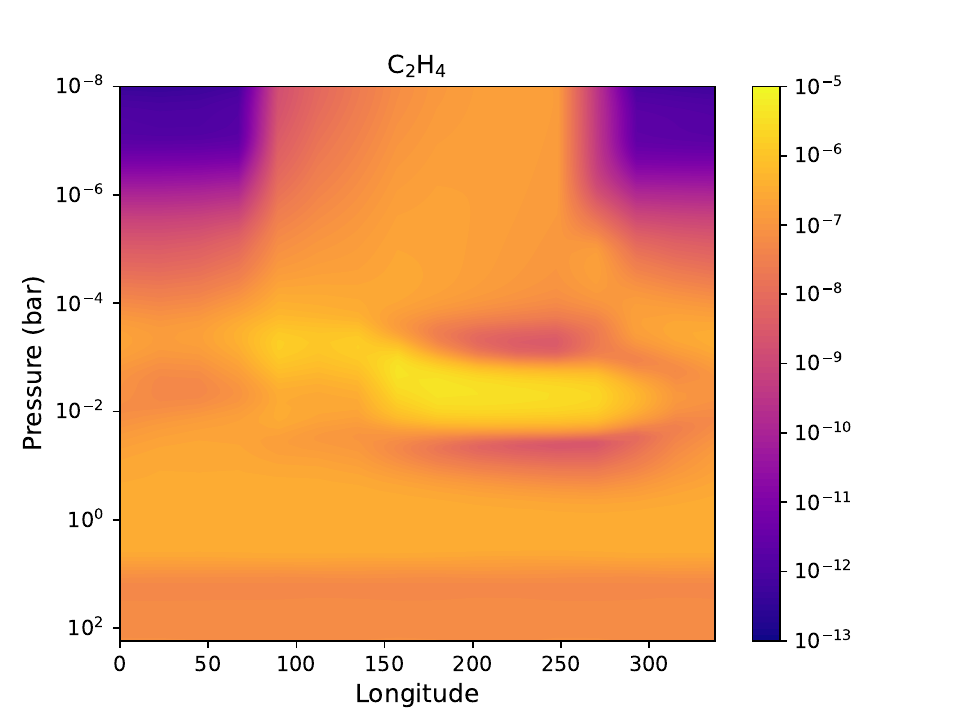}
   \includegraphics[width=0.495\columnwidth]{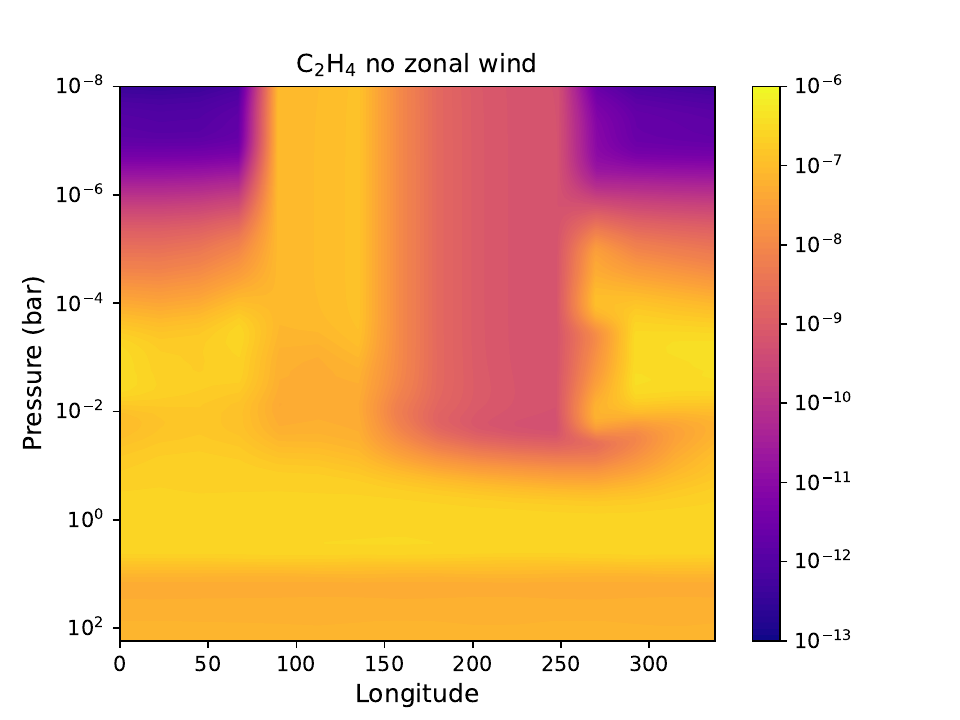}
\caption{The equatorial abundance distribution of several chemical species as a function of longitude and pressure on HD 209458 b with C/O = 1.05, with the\ce{C2H4} distribution when horizontal transport is omitted also included for comparison. The substellar point is located at 0$^{\circ}$ longitude.}
\label{fig:HD209-CtoO105-contours}
\end{figure}

\begin{figure}[!htp]
   \centering
   \includegraphics[width=0.495\columnwidth]{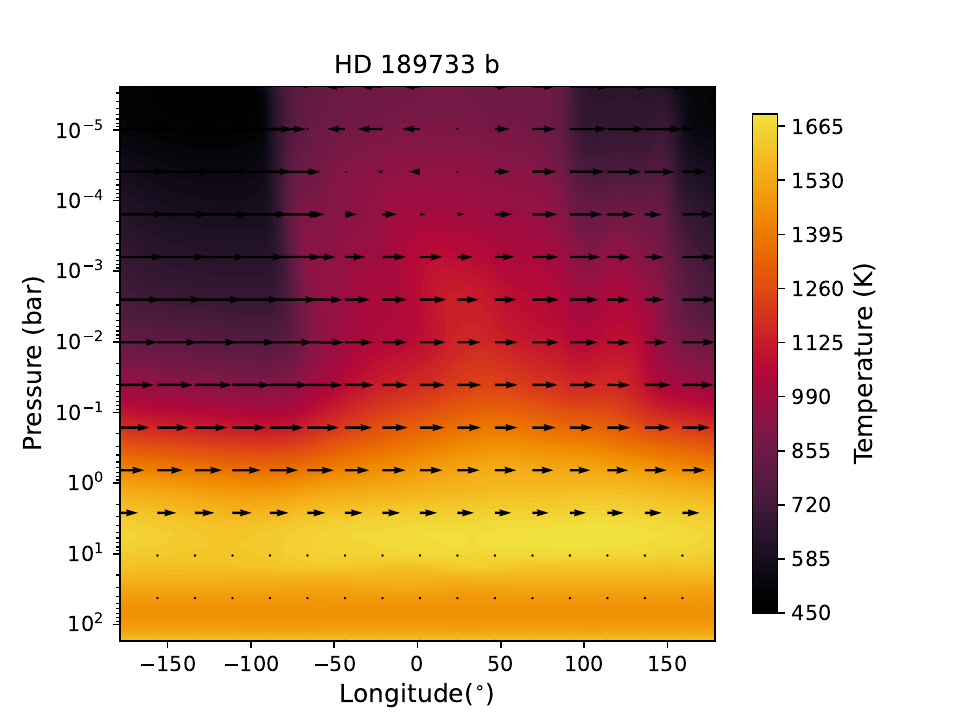}
   \includegraphics[width=0.495\columnwidth]{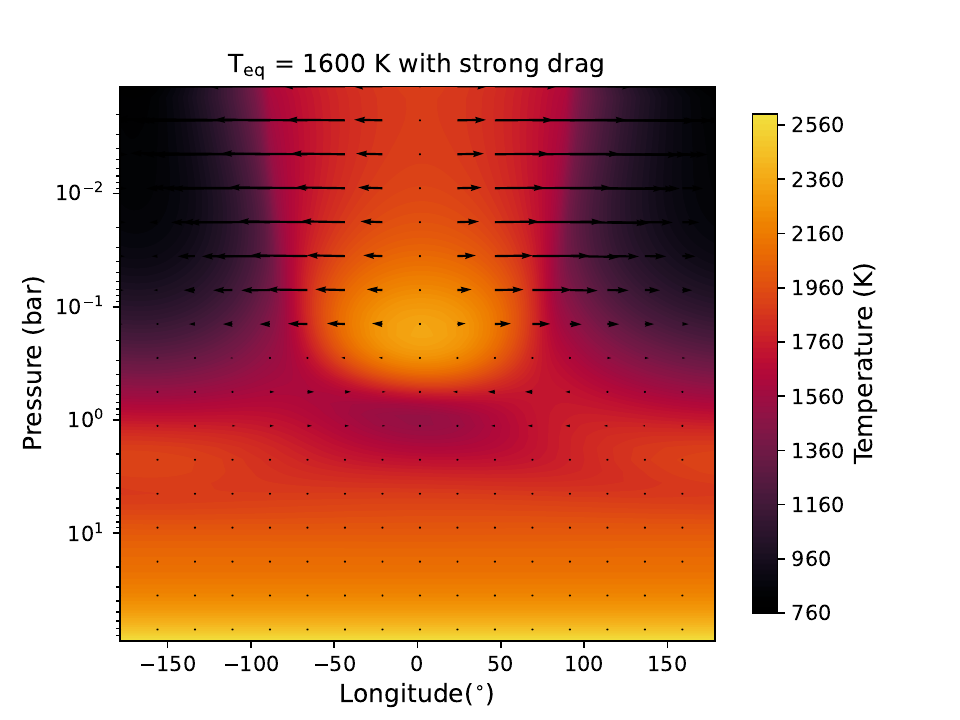}
\caption{The temperatures (color scale) and winds (arrows) on the meridionally averaged equatorial plane of HD 189733 b (left) and those of our fiducial hot Jupiter atmosphere with strong drag from \cite{Tan2019} (right). The substellar point is located at 0$^{\circ}$ longitude. A superrotating jet extends from approximately a few bar to 0.1 mbar level on HD 189733 b, whereas the strong-drag fiducial circulation is dominated by the day-to-night flow above 1 bar.} 
\label{fig:T1600-HD189-eq}
\end{figure}



\end{appendix}

\end{document}